\documentstyle[11pt, leqno]{article}
\input amssym.def
\input amssym.tex

\def\utw{\smash{\rlap{\lower5pt\hbox{$\sim$}}}}
\def\udtw{\smash{\rlap{\lower6pt\hbox{$\approx$}}}}

\def\bbbr{{\rm I\!R}} 
\def\bbbn{{\rm I\!N}} 

\def\bbbone{{\mathchoice {\rm 1\mskip-4mu l} {\rm 1\mskip-4mu l}
{\rm 1\mskip-4.5mu l} {\rm 1\mskip-5mu l}}}
\def\bbbc{{\mathchoice {\setbox0=\hbox{$\displaystyle\rm C$}\hbox{\hbox
to0pt{\kern0.4\wd0\vrule height0.9\ht0\hss}\box0}}
{\setbox0=\hbox{$\textstyle\rm C$}\hbox{\hbox
to0pt{\kern0.4\wd0\vrule height0.9\ht0\hss}\box0}}
{\setbox0=\hbox{$\scriptstyle\rm C$}\hbox{\hbox
to0pt{\kern0.4\wd0\vrule height0.9\ht0\hss}\box0}}
{\setbox0=\hbox{$\scriptscriptstyle\rm C$}\hbox{\hbox
to0pt{\kern0.4\wd0\vrule height0.9\ht0\hss}\box0}}}}
\def\bbbe{{\mathchoice {\setbox0=\hbox{\smalletextfont e}\hbox{\raise
0.1\ht0\hbox to0pt{\kern0.4\wd0\vrule width0.3pt
height0.7\ht0\hss}\box0}}
{\setbox0=\hbox{\smalletextfont e}\hbox{\raise
0.1\ht0\hbox to0pt{\kern0.4\wd0\vrule width0.3pt
height0.7\ht0\hss}\box0}}
{\setbox0=\hbox{\smallescriptfont e}\hbox{\raise
0.1\ht0\hbox to0pt{\kern0.5\wd0\vrule width0.2pt
height0.7\ht0\hss}\box0}}
{\setbox0=\hbox{\smallescriptscriptfont e}\hbox{\raise
0.1\ht0\hbox to0pt{\kern0.4\wd0\vrule width0.2pt
height0.7\ht0\hss}\box0}}}}
\def\bbbq{{\mathchoice {\setbox0=\hbox{$\displaystyle\rm Q$}\hbox{\raise
0.15\ht0\hbox to0pt{\kern0.4\wd0\vrule height0.8\ht0\hss}\box0}}
{\setbox0=\hbox{$\textstyle\rm Q$}\hbox{\raise
0.15\ht0\hbox to0pt{\kern0.4\wd0\vrule height0.8\ht0\hss}\box0}}
{\setbox0=\hbox{$\scriptstyle\rm Q$}\hbox{\raise
0.15\ht0\hbox to0pt{\kern0.4\wd0\vrule height0.7\ht0\hss}\box0}}
{\setbox0=\hbox{$\scriptscriptstyle\rm Q$}\hbox{\raise
0.15\ht0\hbox to0pt{\kern0.4\wd0\vrule height0.7\ht0\hss}\box0}}}}
\def\bbbt{{\mathchoice {\setbox0=\hbox{$\displaystyle\rm
T$}\hbox{\hbox to0pt{\kern0.3\wd0\vrule height0.9\ht0\hss}\box0}}
{\setbox0=\hbox{$\textstyle\rm T$}\hbox{\hbox
to0pt{\kern0.3\wd0\vrule height0.9\ht0\hss}\box0}}
{\setbox0=\hbox{$\scriptstyle\rm T$}\hbox{\hbox
to0pt{\kern0.3\wd0\vrule height0.9\ht0\hss}\box0}}
{\setbox0=\hbox{$\scriptscriptstyle\rm T$}\hbox{\hbox
to0pt{\kern0.3\wd0\vrule height0.9\ht0\hss}\box0}}}}
\def\bbbs{{\mathchoice
{\setbox0=\hbox{$\displaystyle     \rm S$}\hbox{\raise0.5\ht0\hbox
to0pt{\kern0.35\wd0\vrule height0.45\ht0\hss}\hbox
to0pt{\kern0.55\wd0\vrule height0.5\ht0\hss}\box0}}
{\setbox0=\hbox{$\textstyle        \rm S$}\hbox{\raise0.5\ht0\hbox
to0pt{\kern0.35\wd0\vrule height0.45\ht0\hss}\hbox
to0pt{\kern0.55\wd0\vrule height0.5\ht0\hss}\box0}}
{\setbox0=\hbox{$\scriptstyle      \rm S$}\hbox{\raise0.5\ht0\hbox
to0pt{\kern0.35\wd0\vrule height0.45\ht0\hss}\raise0.05\ht0\hbox
to0pt{\kern0.5\wd0\vrule height0.45\ht0\hss}\box0}}
{\setbox0=\hbox{$\scriptscriptstyle\rm S$}\hbox{\raise0.5\ht0\hbox
to0pt{\kern0.4\wd0\vrule height0.45\ht0\hss}\raise0.05\ht0\hbox
to0pt{\kern0.55\wd0\vrule height0.45\ht0\hss}\box0}}}}

\def\bbbz{{\mathchoice {\hbox{$\sf\textstyle Z\kern-0.4em Z$}}
{\hbox{$\sf\textstyle Z\kern-0.4em Z$}}
{\hbox{$\sf\scriptstyle Z\kern-0.3em Z$}}
{\hbox{$\sf\scriptscriptstyle Z\kern-0.2em Z$}}}}

\def\diameter{{\ifmmode\oslash\else$\oslash$\fi}}

\def\init{\setcounter{equation}{0}}

\newtheorem{theoreme}{Theorem }[section]
\newtheorem{proposition}[theoreme]{Proposition}
\newtheorem{lemma}[theoreme]{Lemma}
\newtheorem{definition}[theoreme]{Definition}

\newtheorem{remark}[theoreme]{Remark}

\def\rr{\bbbr}
\def\cc{\bbbc}
\def\nn{\bbbn}
\def\zz{\bbbz}
\def\one{\bbbone}

\def\cBc{{\cal T}_\alpha}
\def\cUc{\cU_{0}}

\def\e{{\rm e}}

\def\i{{\rm i}}
\def\d{{\rm d}}
\def\12{\frac{1}{2}}

\def\proof{{\bf  Proof. }}
\def\slim{\hbox{\rm s-}\lim}
\def\coinf{C_{0}^{\infty}}

\def\qed{$\Box$}

\def\supp{{\rm supp\,}}
\def\cH{{\cal H}}

\def\cU{{\cal U}}
\def\cB{{\cal B}}
\def\G{\Gamma}

\def\cR{{\cal R}}
\def\cRb{{\cal R}_{\scriptscriptstyle AW}}

\def\ch{{\frak h}}
\def\p{\partial}

\def\wlim{{\rm w}-\lim}

\def\dG{\d\Gamma}

\def\cD{{\cal D}}

\def\cV{{\cal V}}

\def\e{{\rm e}}

\def\pfi2{P(\varphi)_{2}}

\newcommand{\beq}{\begin{equation}}
\newcommand{\eeq}{\end{equation}}
\newcommand{\bet}{\begin{theoreme}}
\newcommand{\eet}{\end{theoreme}}
\newcommand{\bel}{\begin{lemma}}
\newcommand{\eel}{\end{lemma}}
\newcommand{\bep}{\begin{proposition}}
\newcommand{\eep}{\end{proposition}}

\newcommand{\bear}[1]{\begin{array}{#1}}
\newcommand{\ear}{\end{array}}

\begin{document}
\def\j{{\rm j}}
\def\tC{c}
\def\bc{{\rm c}}
\def\bp{{\rm p}}
\def\bt{{\rm t}}
\def\q{{\rm q}}
\def\chbar{\overline{\ch}}
\def\cO{{\cal O}}
\def\circle{\tt C}
\def\stp{ stochastic process }
\def\stps{stochastic processes }
\def\cUbar{\overline{\cal U}}
\def\stpos{ stochastically positive }
\def\kms{ KMS system}
\def\Xbar{\overline{X}}\def\cBbar{{\overline \cB}}
\def\cS{{\cal S}}
\def\cA{{\cal A}}

\title{Thermal Quantum Fields without Cut-offs  
\linebreak
in 1+1 Space-time Dimensions \protect\footnotetext{AMS 1991 {\it{Subject 
Classification}}. 81T08, 82B21, 82B31, 46L55} \protect}          

\author{Christian G\'{e}rard\footnote{ christian.gerard@math.u-psud.fr, Universit\'e Paris Sud XI, F-91405 Orsay, France} \ and Christian D.\ J\"akel\footnote{
christian.jaekel@mathematik.uni-muenchen.de, Math.\ Inst.\ der LMU,
Theresienstr.~39, 80333~M\"unchen}}        
\date{March 2004}
\maketitle
\abstract{We construct  interacting quantum fields in
1+1 dimensional Minkowski space, representing neutral scalar bosons at positive temperature. 
Our work is based on prior work by Klein and Landau and H\o egh-Krohn.}

\tableofcontents

\section{Introduction}
\init\label{introd}
\noindent
Constructive thermal field theory allows one  to
circumvent (at least in 1+1 space-time dimensions) the
severe infrared problems (see e.g.~\cite{St}) of thermal perturbation
theory. 
A class of models representing scalar neutral bosons with polynomial interactions
was constructed by H\o egh-Krohn \cite{H-K} 
more than twenty years ago. 
Shortly afterwards,
several related results on the construction of self-interacting 
thermal fields were
announced by Fr\"ohlich \cite{Fr2}.


Our first paper was devoted to the construction of neutral and charged
thermal fields with {\em spatially cutoff} interactions in 1+1 space--time
dimensions, using the notion of {\em stochastically positive} KMS
systems due to Klein and Landau \cite{KL1}.

The construction of interacting thermal quantum fields without cutoffs  
presented here includes several of the original ideas of H\o egh-Krohn \cite{H-K}, but instead
of starting from  the  interacting system in a box
we start from the {\em Araki-Woods representation} 
for the free thermal system in infinite volume.
This `algebraic' approach eliminates some cumbersome 
limiting procedures present in H\o egh-Krohn's work due to the introduction of boxes.
We provide complete proofs for a number of statements which where only
touched upon in H\o egh-Krohn's work. The list of `new' contributions contains 
the Wick (re-)ordering with respect to different
covariance functions, the existence of interacting sharp-time fields, the identification of local algebras,
the existence and uniqueness of the solution of H\o egh-Krohn's time dependent heat-equation, local normality of the 
interacting KMS state, uniqueness of the weak$^*$ accumulation point of the sequence of 
approximating KMS states, and a number of inequalities that enter into a rigorous construction at several points.
Although some of our results were probably already known by the experts (most of our work is based on results by Glimm and Jaffe, H\o egh-Krohn, Fr\"ohlich, Klein and Landau, and Simon) more than twenty years ago, we 
feel that it is worth while to present the arguments in full detail.

We will provide a detailed description of the content of this paper in the next subsection. But before we do so, we 
give a rough outline of the main ideas.

Let $\ch$ and $\epsilon$ denote the one-particle Hilbert space and
the one-particle energy for a single neutral scalar boson.
On the Weyl algebra $\cal {W}(\ch)$ we define a quasi-free
$(\tau^{\circ}, \beta)$-KMS  
state~$\omega^{\circ}_\beta$  for the time evolution~$\{\tau^{\circ}_{t}\}_{t\in \rr}$
by
\[
\omega^{\circ}_{\beta} \bigl(W(h) \bigr):= \e^{-\frac{1}{4}(h, (1+ 2\rho)h)}, \quad 
\tau^{\circ}_{t}\bigl(W(h) \bigr)= W(\e^{\i t\epsilon}h), \:h\in \ch,\: t\in \rr,
\]
where $\rho:= ( \e^{\beta {\rm \epsilon}}-1)^{-1}$, $\beta>0$. 

\goodbreak

A convenient realization of the  GNS representation associated to the pair~$\bigl( {\cal W}(\ch),
\omega^{\circ}_{\beta} \bigr)$ is the {\em Araki-Woods representation}
defined by:
\[
\begin{array}{l}
\cH_{\scriptscriptstyle AW}:= \G(\ch\oplus \chbar),\\[3mm]
\Omega_{\scriptscriptstyle AW}:= \Omega, \\[3mm]
\pi_{\scriptscriptstyle AW}(W(h))= W_{\scriptscriptstyle AW}(h):= W_{\rm F} \bigl( (1+ \rho)^{\frac{1}{2}}h\oplus
\overline{\rho}^{\frac{1}{2}}\overline{h} \bigr),\quad h\in 
\ch.\\[3mm]
\end{array}
\] 
Here  $\overline{\ch}$ is the conjugate Hilbert space to $\ch$, $W_{F}(.)$ denotes the Fock Weyl operator  on
$\G(\ch\oplus\chbar)$ and $\Omega\in \G(\ch\oplus \chbar)$ is the Fock
vacuum. The von Neumann algebra generated by $\{ \pi_{\scriptscriptstyle AW}(W(h)) \mid h \in \ch \}$
is denoted by $\cRb$. The {\em local von Neumann algebra} generated by $\{ \pi_{\scriptscriptstyle AW}(W(h)) \mid h \in \ch_I \}$
is denoted by $\cRb (I)$. Here $I \subset \rr$ is an open and bounded interval and $\ch_I$ will be defined in (\ref{ekg4}).

Since $\omega^{\circ}_\beta$
is $\tau^\circ$-invariant, there exists a standard implementation 
(see \cite{DJP})
of the time evolution 
in the  representation $\pi_{\scriptscriptstyle AW}$:
\[ {\rm e}^{i L_{\scriptscriptstyle AW} t} \pi_{\scriptscriptstyle AW} (A) \Omega_{\scriptscriptstyle AW} :=
\pi_{\scriptscriptstyle AW} \bigl(\tau_t^\circ(A)\bigr) \Omega_{\scriptscriptstyle AW} \quad \hbox{\rm and} \quad
L_{\scriptscriptstyle AW} \Omega_{\scriptscriptstyle AW} =0. \]
The generator $L_{\scriptscriptstyle AW}$ of the free time evolution  is called 
the (free) Liouvillean.

{\em Euclidean techniques} were used in our first paper
to define the operator sum 
\[ H_l := L_{\scriptscriptstyle AW} +  \int_{-l}^{l}:\!P(\phi(0, x))\!:_{C_{0}}\d x \]
and to show that $H_l$ is essentially selfadjoint. 

Using Trotter's product formula as in \cite{GJ2}, a
finite propagation speed argument shows that 
\[ \tau^l_t (A) = {\rm e}^{i H_l t} A {\rm e}^{-i H_l t}\] 
is independent of $l$ for $t \in \rr$ and $ A \in \cRb (I)$ fixed, if $I$ is bounded  and $l$ is sufficiently large.
Thus there exists a limiting dynamics $\tau$ such that 
\beq  \label{nc}
\lim_{l \to \infty} \bigl\| \tau_t^l (A) - \tau_t (A) \bigr\| = 0 \eeq
for all $A \in \cRb (I)$, $I$ bounded. This norm convergence extends to the {\em norm closure}
\[
\cA:= \overline{ \bigcup_{I \subset \rr} \cRb (I)}^{(*)}  \]
of the {\em local von Neumann algebras}. The $C^*$-algebra $\cA$ is called the 
{\em algebra of local observables}.

It follows from general results of \cite{KL1} that the vector $\Omega_{l} \in \cH_{\scriptscriptstyle AW}$,
\beq \label{bperturbations}
\Omega_{l} := \frac{\e^{-\frac{\beta}{2}H_{l}}\Omega_{\scriptscriptstyle AW}}
{\|\e^{-\frac{\beta}{2}H_{l}}\Omega_{\scriptscriptstyle AW}\|} ,\eeq
induces a $(\tau^l, \beta)$-KMS state $\omega_{l}$ for the $W^*$-dynamical system $(\cA, \tau^l)$. Equation (\ref{bperturbations}) should be compared with similar expressions which are well-known 
(see e.g.~\cite[Theorem 5.4.4]{BR}) for
bounded perturbations and which have recently been derived  for a class of unbounded
perturbations in \cite[Theorem 5.6]{DJP}.

The existence of weak limit points (which are states) of the net $\{ \omega_l \}_{l>0}$ is a consequence of 
the Banach-Alaoglu theorem (see \cite[Theorem 2.3.15]{BR}). 

That fact that all limit states satisfy the {\em KMS condition} w.r.t.\ the pair $(\cA, \tau)$ follows
from (\ref{nc}), which itself is a consequence of   finite propagation speed.

Since $\cA$ is the norm closure
of the {\em weakly closed\/} local  algebras,
all limit points are {\em locally normal} KMS states w.r.t.\ the Araki-Woods
representation \cite{TW}.

To prove that there is only one accumulation point is more delicate. Following
H\o egh-Krohn \cite{H-K} we will use {\em Nelson symmetry}
to  relate the interacting vacuum theory on the circle 
to the interacting thermal model on the real line.

\vskip 1cm

\goodbreak
\subsection{Content of this paper}

In Section \ref{sec0} we recall the notions of {\em stochastically
positive KMS systems} and associated {\em generalized path spaces}, due to Klein and
Landau \cite{KL1}.  The property corresponding to stochastic positivity in the $0$-temperature case is 
called {\em Nelson-Symanzik positivity}.

In Subsection \ref{sec0.1} we recall the characterization of the thermal equilibrium states of a
dynamical system $(\cB, \tau)$ by the
{\em KMS condition} and the definition of  {\em Euclidean
Green's functions}. The notion of  a {\em stochastically
positive KMS systems}  $(\cB, \cU, \tau, \omega)$ rests on the introduction of a
distinguished abelian sub-algebra $\cU$ of the observable algebra
$\cB$. In our case this algebra will be the algebra generated by
the  time-zero fields.

In Subsection \ref{sec0.2}  we recall the notion of a {\em generalized path
space} $(Q, \Sigma, \Sigma_{0}, U(t), R, \mu)$. It consists of a probability space $(Q, \Sigma, \mu)$,
a distinguished sub $\sigma$-algebra $\Sigma_{0}$, a one-parameter group $t\mapsto U(t)$ of automorphisms
of $L^{\infty}(Q, \Sigma, \mu)$ such that $\Sigma=\bigvee_{t\in \rr}U(t)\Sigma_{0}$
and a reflection~$R$, acting as an automorphism on~$L^{\infty}(Q, \Sigma,
\mu)$ such that $R^{2}=\one$, $RU(t)= U(-t)R$.

Klein and Landau (see \cite{KL1}) have shown that  for $\beta>0$ there is a one to one correspondence 
between stochastically positive $\beta$-KMS systems and
$\beta$-periodic OS-positive path spaces (for $\beta=\infty$ the
object associated to an OS-positive path space is called a  {\em
positive semigroup structure}, see \cite{K}).  The role of
OS-positivity is to ensure the positivity of the inner product in the Hilbert space $\cH$
on which the real time quantum fields act.

The {\em reconstruction theorem} provides a concrete realization of the {\em GNS  triple} $(\cH_\omega, \pi_\omega, \Omega_\omega)$ associated to
the pair $(\cB, \omega)$. The {\em Liouvillean} $L$  implements the 
time evolution in the GNS representation $\pi_\omega$. 

In Subsection \ref{sec0.3} we recall some results from \cite{KL1}
(with some improvements in \cite{GeJ}) concerning  perturbations of generalized path spaces obtained
from  Feynman-Kac-Nelson kernels. The main examples of FKN kernels are
those obtained from a selfadjoint operator $V$ on the physical Hilbert
space $\cH_{\omega}$, which  is  affiliated to $\cU\cong L^\infty (K, \nu_\omega)$.
If $\e^{-\beta V}\in L^{1}(K, \nu_\omega)$ and
\[
V\in L^{p}(K, \nu_\omega),\quad
\e^{-\frac{\beta}{2}V}\in L^{q}(K, \nu_\omega) \quad \hbox{for} \quad p^{-1}+ q^{-1}=
\frac{1}{2} , \quad 2\leq p, q\leq \infty,
\]
then the operator sum $L+V$ is essentially selfadjoint on $\cD (L) \cap \cD(V)$
and  the perturbed time-evolution~$\tau_{V}$ on
${\cal B}$ is given by 
\[ \tau_{V, t}(B)=\e^{\i t \overline{L+V}}B\e^{-\i t \overline{L+V} }  .\]
The KMS state $\omega_{V}$ for the pair $(\cB, \tau_{V})$
is  the vector state    induced by 
\[ \Omega_{V}:=\frac{\e^{-\frac{\beta}{2}\overline{L+V}}\Omega_\omega}{ 
\|\e^{-\frac{\beta}{2}\overline{L+V}}\Omega_\omega\| } .\]
The {\em Liouvillean}~$L_{V}$ for the perturbed $\beta$-KMS system
$({\cal B}_{V}, \tau_{V}, \omega_{V})$ equals
$\overline{L+V - JVJ}$. ($J$~denotes the modular conjugation associated to the pair $({\cal B},\Omega_\omega)$). 
It satisfies
\[ {\rm e}^{it L_{V}} A \Omega_{V} = \tau_{V, t} (A) \Omega_{V} \hbox{ and }
L_{V} \Omega_{V} = 0 .\]

In Section \ref{sec1} we recall some standard facts about Gaussian measures
on distribution spaces and fix some notation.
{\em Gaussian measures} are reviewed in Subsection
\ref{sec1.2}.
{\em Sharp-time free fields} are introduced in Subsection
\ref{sec1.3}. If  the space dimension $d$ is $1$, then it is possible to define similarly {\em sharp-space free
fields}. This is done in Subsection~\ref{sec1.4}.

In Section \ref{sec2} we recall two well known path spaces supported by
$(\cS'_{\rr}(S_{\beta}\times \rr),
\d\phi_{C})$, where $S_{\beta}$ is the circle of length $\beta$. 
In Subsection \ref{sec2.1} we identify the generalized path space on
$(\cS'_{\rr}(S_{\beta}\times \rr),\d\phi_{C})$ corresponding to the free
massive scalar field on the circle $S_{\beta}$ at temperature $0$.

In Subsection \ref{sec2.2a}
we  identify the  generalized path space on
$(\cS_{\rr}'(S_{\beta}\times \rr),\d\phi_{C})$ corresponding to the free
massive scalar field on the real line $\rr$ at temperature $\beta^{-1}$.
The physical Hilbert space associated to this path space  can be
unitarily identified with the Fock space $\G(\ch\oplus\overline{\ch})$. The KMS
vector $\Omega_{\scriptscriptstyle AW}$ is identified with the Fock vacuum vector~$\Omega$ 
in~$\G(\ch\oplus\overline{\ch})$. The dynamics $\tau^{\circ}$ can be unitarily implemented in
$\pi_{\scriptscriptstyle AW}$: The (free) Liouvillean~$L_{\scriptscriptstyle AW}$ is
identified with  $\d\Gamma(\epsilon\oplus
-\overline{\epsilon})$.

In Section \ref{perturbations} we describe perturbations of the two path spaces
defined in Subsects.~\ref{sec2.1} and~\ref{sec2.2a}. The perturbed path spaces are  obtained from FKN
kernels corresponding to $P(\phi)_{2}$ interactions. 

In Subsection \ref{sec2.2} we recall some well known facts concerning  the Wick ordering of
Gaussian random variables.  In 1+1 space-time dimensions 
Wick ordering is sufficient to eliminate the UV divergences of polynomial interactions. As it turns out, the leading order 
in the UV  divergences is independent of the temperature. Thus it is a matter of convenience
whether one uses the thermal covariance function $C_0$  or the
vacuum covariance function $C_{\rm vac}$ to define the Wick ordering.

In Subsection \ref{sec2.4a} the
$P(\phi)_{2}$ model on the circle $S_{\beta}$ at  temperature $0$ is discussed.
It is specified by the formal interaction  
\[
V_{\tt C}=\int_{S_{\beta}}:\!P(\phi(t, 0))\!:_{C_{\beta}}\d t.
\]
Here $P(\lambda)$ is a real valued polynomial, which is bounded from below. 
The time-evolution $x \mapsto {\rm e}^{ix H_{\tt C}^{\rm ren}}$ is generated by $H_{\tt C}^{\rm ren}:=
H_{\tt C}-E_{\tt C}$, where  $E_{\tt C}:=\inf(\sigma(H_{\tt C}))$
and
\[
H_{\tt C}=\overline{\d\Gamma \bigl( (D_{t}^{2}+ m^{2})^{\12} \bigr)+ V_{\tt C}} .
\] 
The operator  $H_{\tt C}$ is bounded from below
and has a  unique vacuum state $\omega_{\tt C} ( \: . \: )= (\Omega_{\tt C}, \: . \: \Omega_{\tt C})$ such that
$(\Omega_{\tt C}, \Omega)>0$ and $H_{\tt C}^{\rm ren} \Omega_{\tt C}= 0$. 
The renormalized energy operator $H_{\tt C}^{\rm ren}$ is called the~{\em $P(\phi)_{2}$ Hamiltonian on the circle
$S_{\beta}$}. 

Some  bounds are provided in Proposition 5.4, which are used in the sequel to prove the existence of interacting
sharp-time fields.

The
{\em spatially cutoff $P(\phi)_{2}$  model on the real line $\rr$ at temperature
$\beta^{-1}$} is discussed in Subsection \ref{sec2.4}.
It is  specified by the formal interaction  
\[
V_{l} =\int_{-l}^{l} :\! P(\phi(0, x))\!:_{C_{0}}\d x.
\]
Here $P(\lambda)$ is once again a real valued polynomial, which is bounded from below, and $l\in
\rr^{+}$ is a spatial cutoff parameter. The perturbed KMS state $\omega_{l}$ turns out to be normal w.r.t.\ the Araki-Woods representation $\pi_{\scriptscriptstyle AW}$.
In fact, it
is the vector state induced by
\[ \Omega_{l} := \frac{\e^{-\frac{\beta}{2}H_{l}}\Omega_{\scriptscriptstyle AW}}
{\|\e^{-\frac{\beta}{2}H_{l}}\Omega_{\scriptscriptstyle AW}\|} ,
\]
where $H_{l}$ is the selfadjoint operator 
$H_{l}:=\overline{L_{\scriptscriptstyle AW} +V_{l}}$. The perturbed time-evolution on
${\cal B}$ is given by 
\[ \tau^{l}_{t}(B):=\e^{\i t
H_{l}}B\e^{-\i tH_{l}}, \quad B \in \cB .
\]
The following consequence of Lemma \ref{2.2} will be important in Section~\ref{sec3}:
\beq
\e^{-\int_{-\beta/2}^{\beta/2} U(t) \int_{-l}^{l} :  P(\phi(0, x)) :_{C_{0}}\d x \d t} 
= \e^{-\int_{-l}^{l}U_{\tt C}(x) \int_{S_{\beta}}: P(\phi(t, 0)) :_{C_{\beta}}\d t \d x} .
\label{ns}
\eeq
The analog of (\ref{ns}) in the zero temperature case is called {\it Nelson symmetry} (see e.g.~\cite{Si1}).

\goodbreak

The thermodynamic limit is discussed in Section \ref{sec4}.
We prove that the limits
\[
\lim_{l\to +\infty}\tau^{l}_{t}(A)=: \tau_{t}(A) \quad \hbox{and} \quad \lim_{l\to
+\infty}\omega_{l}(A)=: \omega_\beta(A)
\]
exist for $A$ in the  $C^{*}$-algebra  of local observables $\cA$ and that
$(\cA, \tau , \omega_\beta)$ is a $\beta$-KMS system, describing
the {\em translation invariant $P(\phi)_{2}$ model at temperature
$\beta^{-1}$}. 

\goodbreak 

In Subsection \ref{sec4.1} we recall  localization properties of the classical solutions of the 
Klein-Gordon equation.

In Subsection \ref{sec4.1a}
we introduce the net of local algebras $I \to \cRb(I)$
for the free thermal field: for a bounded open interval $I\subset \rr$, the 
symbol
$ \cRb(I)$ denotes the von Neumann algebra generated by $ \{W_{\scriptscriptstyle AW}(h) \mid h\in \ch_{I}\}$. 
By a result of  Araki \cite{Ar}, 
the local von Neumann algebras for the free thermal scalar field are regular from the inside and from the outside:
\[ \bigcap_{J\supset \overline{I}}\cRb(J)=\cRb(I)= \bigvee_{\overline{J}\subset
I}\cRb(J). \]
Moreover,  if $I$ is bounded, then the local algebra $\cRb(I)$ is $*$-isomorphic to the
unique hyper-finite factor of type~III$_1$.

In Subsection \ref{sec4.2} the existence of the limiting dynamics is discussed.
For $t\in \rr$ fixed,
the norm limit
\[
\lim_{l\to \infty}\tau_{t}^{l}(B)=: \tau_{t}(B)\]
exists
for all $B$ in
\[ \cA := \overline { \bigcup_{I \subset \rr} \cRb (I) }^{(*)},\]
where the $I$'s are open and bounded. 
{\em Finite propagation speed}
is used to  show that
$\tau^{l}_{t}(B)$, for~$B\in \cRb(I)$ and $|t|\leq T$,
is independent of $l$ for $l>|I|+ T$. The proof uses {\em Trotter's product formula}, which requires that
$L_{\scriptscriptstyle AW} + V_l$ is essentially self-adjoint on $\cD(L_{\scriptscriptstyle AW}) \cap \cD(V_l)$.

In order to apply the results of Section~\ref{sec3} to the
$C^{*}$-algebra $\cA$, it is necessary to identify the  local von
Neumann algebra $\cRb(I)$ with the von Neumann algebra obtained by 
applying the interacting dynamics $\tau$ to the local abelian
algebra of time-zero fields. This is done in Subsection \ref{ila}:
for $I\subset \rr$ a bounded open interval, we denote by $ \cU_{\scriptscriptstyle AW} (I)$ the
abelian von Neumann algebra generated by 
$\{W_{\scriptscriptstyle AW}(h) \mid h\in \ch_{I},\: h\hbox{ real
valued}\}$.
We denote by ${\cal B}_{\alpha}(I)$ the
von Neumann algebra generated by 
\[ \bigl\{\tau_{t}(A) \mid A\in \cU_{\scriptscriptstyle AW}(I), \;
|t|<\alpha \bigr\} . \]
We set
${\cal B}(I):=\bigcap_{\alpha>0} {\cal B}_{\alpha}(I)$ and show that ${\cal B}(I)= \cRb(I)$.

Taking the existence of the interacting path space (which we will
construct in Section \ref{sec3}) for granted, we show that the net $\{\omega_{l}\}_{l>0}$
has a unique accumulation point. This is done in 
Subsection \ref{sec4.3}, using the identification of algebras established in the previous subsection.
Thus
\[
\wlim_{l\to +\infty}\omega_{l}=: \omega_{\beta}  \hbox{ exists on }\cA.
\]
The state $\omega_{\beta}$ is a $(\tau, \beta)$-KMS state on $\cA$. It follows from
a result of Takesaki and Winnink~\cite{TW} that $\omega_{\beta}$ is
{\em locally normal}, i.e., if $I$ is an open and bounded interval, then $\omega_{\beta |\cRb(I)}$ is normal 
w.r.t.~the Araki-Woods representation; thus $\omega_{\beta |\cRb(I)}$
is also normal with respect to the Fock representation.
Moreover, $\omega_\beta$ is invariant under spatial translations and
satisfies the {\em space-clustering property}:
\[\lim_{x\to\infty}\omega_\beta (A\alpha_{x}(B))=
\omega_\beta(A)\omega_\beta(B),\:  A,
B\in \cA.
\]

Finally, the main results of this paper, namely the  explicit construction
of the translation invariant
$P(\phi)_{2}$ model at positive temperature 
is given in Section \ref{sec3}.
Following ideas of H\o egh-Krohn \cite{H-K}, {\em Nelson symmetry} is used
to establish the existence of the model in the thermodynamic limit.

The first step is to construct  the {\em interacting path space} supported by
${\cal S}_{\rr}' (S_{\beta}\times \rr)$ describing the translation
invariant $P(\phi)_{2}$ model at temperature $\beta^{-1}$.

Following H\o egh-Krohn~\cite{H-K} we consider 
the operator $W_{[-\infty, \infty]}(f)$ solving the time-dependent heat equation
\[
\frac{\d}{ \d b}W_{[a,b]}(f)= W_{[a,b]}(f)\bigl(-H_{\tt C}^{\rm
ren}+\i \phi(f_{b}) \bigr),\quad a\leq b,
\]
where  $ f_{b} (\cdot):= f(\cdot,b) \in \cS_{\rr}(S_{\beta})$ for $f\in \cS_{\rr} (S_{\beta}\times \rr)$.
We show that for $f\in
C^{\infty}_{0\:\rr}(S_{\beta}\times \rr)$ 
\[
\lim_{l\to +\infty}\int \e^{\i
\phi(f)}\d\mu_{l}= (\Omega_{\tt C},
W_{[-\infty, \infty]}(f)\Omega_{\tt C})
\]
exists and that the map
\[
\matrix{ & \cS_{\rr}(S_{\beta}\times\rr)& \to & \rr^+
\cr
&f & \mapsto & \bigl(\Omega_{\tt C},
W_{[-\infty, \infty]}(f)\Omega_{\tt C}\bigr) \cr} 
\]
is the generating functional of a {\em Borel probability measure} $\mu$ on $(Q,
\Sigma)$. The measure $\mu$ is invariant under space
translations, time translations
and time-reflection.

In Subsection \ref{secsec} we prove the {\em existence of interacting sharp-time fields}.  
(Note that the necessary bounds (\ref{e3.1d}) depend on the dimension of space-time.)
This result allows us to equip the
probability space $(Q, \Sigma, \mu)$ with an {\em OS-positive
$\beta$-periodic path space structure\/}: 

\begin{itemize}
\item[--] $U(t)$ is the group of transformations
generated by the time translations~${\scriptstyle \frak T}_s$ induced on~$Q$ by the map $(t,x) \mapsto (t +s, x)$;
\item[--] $R$ is the 
transformation generated by the (euclidean) time reflection at $t=0$;
\item[--] $\Sigma_{0}$ is the sub$-\sigma$-algebra of
$\Sigma$ generated by the functions $\{\phi(0, h) \mid h\in \cS_{\rr}(\rr) \}$.
\end{itemize}
 
In Subsection \ref{sec7.3} some properties of the {\em associated interacting $\beta$-KMS system} 
$( \cB, \cU, \tilde \tau, \tilde{\omega})$ are discussed.
We prove the {\em convergence of sharp-time Schwinger functions} and show that
\[\tilde{\omega} \bigl(\alpha_{x}(W_{\scriptscriptstyle AW}(h)) \bigr)= \tilde{\omega}\bigl(W_{\scriptscriptstyle AW}(h)\bigr)\]
for all $x\in \rr$ and
\[ \lim_{x\to \infty}\tilde{\omega}\bigl(W_{\scriptscriptstyle AW}(h)\alpha_{x}(W_{\scriptscriptstyle AW}(g))\bigr)
=\tilde{\omega} \bigl(W_{\scriptscriptstyle AW}(h) \bigr)\tilde{\omega} \bigl(W_{\scriptscriptstyle AW}(g)\bigr)\]
for $h, g \in C_{0\:\rr}^{\infty}(\rr)$.

In Appendix \ref{td} we discuss 
the abstract time-dependent heat equation
\beq 
\label{heat}
\left\{
\begin{array}{l}
\frac{\d}{ \d t}U(t,s)= - \bigl(H+\i R(t) \bigr)U(t,s),\; s\leq t, \\
U(s,s)=\one.
\end{array}\right.
\eeq
Here $H\geq 0$ is a selfadjoint operator on a Hilbert space $\cH$ and $R(t)$, $t\in \rr$, is a
family of  closed operators with $\cD(H^{\gamma})\subset \cD(R(t))$
for some $0\leq \gamma<1$. 
We show that there exists a unique solution $U(t,s)$ 
such that $U(s,s)=\one $ and 
\[U(t,r)U(r,s)= U(t,s) \hbox{ for }s\leq r\leq t .\]

In Subsection \ref{td3} we consider the dissipative case when $R(t)$
is selfadjoint for $t\in \rr$. We establish an
approximation of $U(t,s)$ by time-ordered products and
prove some bounds on $U(t,s)$, which are used in the main text 
to show the existence of interacting sharp-time fields and the convergence of 
sharp-time Schwinger functions.

Finally we establish  a lemma which is used in the main text to prove spatial clustering
for the  translation
invariant $P(\phi)_{2}$ model at temperature $\beta^{-1}$.

\medskip
\noindent
{\bf Acknowledgments}.   The authors would like to thank Jan
Derezi\'nski  for useful discussions.  C. J\"akel wants to thank Hanno 
Gottschalk for discussing H\o egh-Krohn's original paper.
The second author was supported under the FP5 TMR program 
of the European 
Union by the Marie
Curie fellowship HPMF-CT-2000-00881 and is currently supported by the~IQN network of the DAAD.
Both authors benefited from the IHP network HPRN-CT-2002-00277
of the European Union.

\section{Stochastically positive KMS systems and generalized path
spaces}
\init\label{sec0}
In this section we briefly recall the notions of {\em stochastically
positive KMS systems} and associated  {\em  generalized path spaces}, due to Klein and
Landau \cite{KL1}. We will also need the corresponding notions at
$0$-temperature, which can be found in \cite{K}.
\subsection{Stochastically positive KMS systems}
\label{sec0.1}
Let $\cB$ be a $C^{*}$-algebra and let  $\{\tau_{t}\}_{t\in \rr}$ be a one
parameter group of $*$-automorphisms of $\cB$. We recall that a state
$\omega$ on $\cB$ is a $(\tau, \beta)$-{\em KMS state} or $(\cB, \tau,
\omega)$ is a $\beta-${\em KMS system}, if for each pair $A,
B\in \cB$ there exists a  function~$F_{A,B}(z)$ holomorphic in the
strip $I_{\beta}^{+}=\{z\in \cc \mid 0<{\rm Im}z<\beta\}$ and
continuous on~$\overline{I_{\beta}^{+}}$ such
that
\[
F_{A, B}(t)= \omega(A\tau_{t}(B)) \quad \hbox {and} \quad F_{A, B}(t+\i \beta)=
\omega(\tau_{t}(B)A) \quad \forall  t\in \rr.
\]
For $A_{i}\in \cB$ and $t_{i}\in \rr$, $1\leq
i\leq n$, the {\em Green's functions} are defined as follows:
\[
 G(t_{1}, \dots, t_{n}; A_{1}, \dots, A_{n}):= \omega \bigl(\prod_{i=1}^{n}
\tau_{t_{i}}(A_{i}) \bigr) .
\]
It is well known (see \cite{Ar1, Ar2}) that the Green's functions 
are holomorphic in
\[
I_{\beta}^{n+}:=\{(z_{1}, \dots, z_{n})\in \cc^{n} \mid {\rm Im}z_{i}<
{\rm Im}z_{i+1}, \; {\rm Im}z_{n}- {\rm Im}z_{1}<\beta\},
\]
 continuous on $\overline{I_{\beta}^{n+}}$ and bounded there by
$\prod_{1}^{n}\|A_{i}\|$. Therefore one can define the {\em Euclidean
Green's functions}:
\[
^{E}G(s_{1}, \dots, s_{n}; A_{1}, \dots, A_{n}):=G(\i s_{1},\dots, \i s_{n}; A_{1}, \dots, A_{n})\hbox{ for }s_{1}\leq
\cdots\leq s_{n}, \: s_{n}-s_{1}\leq \beta.
\]
The following class of $\beta$-KMS systems has been introduced by
Klein and Landau \cite{KL1}.
\begin{definition}
Let $(\cB, \tau, \omega)$ be a $\beta$-KMS system and let
$\cU\subset \cB$ be an abelian $^{*}$-sub-algebra. The KMS
system $(\cB, \cU,  \tau, \omega)$ is {\em stochastically
positive}  if 
\vskip .3cm
\halign{ \indent \indent \indent #  \hfil & \vtop { 
\parindent =0pt 
\hsize=12cm
                            \strut # \strut} \cr 
{\rm (i)}  & the $C^{*}$-algebra generated by $\bigcup_{t\in
\rr}\tau_{t}(\cU)$ is equal to $\cB$;
\cr
{\rm (ii)} &  the Euclidean Green's functions $^{E}G(s_{1}, \dots, s_{n};
A_{1}, \cdots, A_{n})$ are positive for all $A_{1}, \dots, A_{n}$ 
in~$\cU^{+}=\{A\in \cU \mid A\geq 0\}$.
\cr}
\end{definition}
In applications it is  more convenient  to use a
version of stochastic positivity, which is adapted to von Neumann algebras.

\begin{definition}
Let  $\cB\subset \cB(\cH)$ be a von Neumann
algebra and let $\cU\subset \cB(\cH)$ be a weakly closed abelian sub-algebra of $\cB$. Assume that the dynamics
$\tau \colon \cB \to \cB$ is given by  
\[ \tau_t (B) := \e^{\i tL} B \e^{- \i tL}, \: \: B \in \cB , \]
where $L$ is a selfadjoint operator on $\cH$.
Moreover, assume that $\omega$ is a $\beta$-KMS state 
for the $W^*$-dynamical system $(\cB,\tau)$.
Then the KMS
system $(\cB, \cU,  \tau, \omega)$ is {\em stochastically
positive}  if 
\vskip .3cm
\halign{ \indent \indent \indent #  \hfil & \vtop { 
\parindent =0pt 
\hsize=12cm
                            \strut # \strut} \cr 
{\rm (i)}  & the von Neumann algebra generated by $\bigcup_{t\in
\rr}\tau_{t}(\cU)$ is equal to $\cB$;
\cr
{\rm (ii)} &  the Euclidean Green's functions $^{E}G(s_{1}, \dots, s_{n};
A_{1}, \cdots, A_{n})$ are positive for all $A_{1}, \dots, A_{n}$ 
in~$\cU^{+}=\{A\in \cU \mid A\geq 0\}$.
\cr}
\end{definition}

\subsection{Generalized path spaces}
\label{sec0.2}
Stochastically positive $\beta$-KMS systems can be associated to 
 {\em generalized path spaces} (see \cite{KL1}, \cite{K}).
Let us first recall some terminology.

If $\Xi_{i}$, for $i$ in an index set $I$, is a family of subsets of a set
$Q$, then we denote by~$\bigvee_{i\in I}\Xi_{i}$ the $\sigma$-algebra
generated by  the sets $\bigcup_{i\in J}\Xi_{i}$ where $J$ runs over all
countable subsets of $I$.
\begin{definition}
 A {\em generalized path
space} $(Q, \Sigma, \Sigma_{0}, U(t), R, \mu)$ consists of 
\vskip .3cm
\halign{  \indent \indent \indent #  \hfil & \vtop { 
\parindent =0pt 
\hsize=12cm
                            \strut # \strut} \cr 
{\rm (i)}  & a probability space $(Q, \Sigma, \mu)$;
\cr
{\rm (ii)} & a distinguished sub $\sigma$-algebra $\Sigma_{0}\subset
\Sigma$;
\cr
{\rm (iii)} & a one-parameter group $\rr\ni t\mapsto U(t)$ of measure preserving automorphisms
of $L^{\infty}(Q, \Sigma, \mu)$, strongly continuous in
measure, such that $\Sigma=\bigvee_{t\in \rr}U(t)\Sigma_{0}$;
\cr
{\rm (iv)}  & a measure preserving automorphism $R$ of $L^{\infty}(Q, \Sigma,
\mu)$ such that $R^{2}=\one$, $RU(t)= U(-t)R$ and $RE_{0}= E_{0}R$, where
$E_{0}$ is the conditional expectation with respect to $\Sigma_{0}$.  \cr}
\end{definition}
A path space $(Q, \Sigma, \Sigma_{0}, U(t), R, \mu)$ is said to be
{\em supported }by the probability space $(Q, \Sigma, \mu)$.

It follows from (iii) and (iv) that $U(t)$ extends to a strongly
continuous group of isometries of $L^{p}(Q, \Sigma, \mu)$  
and $R$ extends to an isometry of $L^{p}(Q, \Sigma, \mu)$ for $1\leq
p<\infty$.

We say that the path space $(Q, \Sigma, \Sigma_{0}, U(t), R, \mu)$ is
$\beta$-{\em periodic} for $\beta>0$ if $U(\beta)=\one$. On 
a~$\beta$-periodic path space one can consider the one-parameter group
$U(t)$ as being indexed by the circle~$S_{\beta}=[-\beta/2, \beta/2]$.

For $I\subset \rr$, we denote by $E_{I}$ the conditional expectation
with respect to the $\sigma$-algebra $\Sigma_{I}:=\bigvee_{t\in I}\Sigma_{t}$.

\begin{definition}
{\rm ($0$-temperature case):}  A generalized path space $(Q, \Sigma, \Sigma_{0}, U(t), R, \mu)$ is {\em
OS-positive}, if  
$E_{[0, +\infty[}RE_{[0, +\infty[}\geq 0$ as an operator on~$L^{2}(Q,
\Sigma, \mu)$.

{\rm (Positive temperature case):} A $\beta$-periodic path space $(Q, \Sigma, \Sigma_{0}, U(t), R, \mu)$ is {\em
OS-positive}, if $E_{[0,\beta/2]}RE_{[0, \beta/2]}\geq 0$ as an operator on $L^{2}(Q,
\Sigma, \mu)$.
\end{definition}
For simplicity of notation we will consider $\beta$ as a parameter in $]0,
+\infty]$, the case $\beta=+\infty$ corresponding to the
$0$-temperature case. 

It is shown in \cite{KL1} that  for $\beta>0$ there is a one to one correspondence
between stochastically positive $\beta$-KMS systems and
$\beta$-periodic OS-positive path spaces. For $\beta=\infty$ the
object associated to an OS-positive path space is called a  {\em
positive semigroup structure} (see \cite{K}). 

Let us describe in more
details one part of this correspondence, which is an example of a {\em
reconstruction theorem}. 
Let $(Q, \Sigma, \Sigma_{0}, U(t), R, \mu)$ be an OS-positive
 path space, $\beta$-periodic if $\beta<\infty$. We set 
 \[
\cH_{\rm OS}:=L^{2}(Q, \Sigma_{[0,
\beta/2]}, \mu).
\] Let ${\cal N}\subset \cH_{\rm OS}$ be the kernel of
the positive quadratic form
\[
(\psi,\psi):=\int_{Q}\overline{\psi}R\psi\d \mu.
\] 
Then the {\em physical Hilbert space} is
\[
 \cH:=\hbox{ completion of }\cH_{\rm OS}/{\cal N},
\]
where the completion is done with respect to the positive definite scalar
product $(.,.)$. Let us denote by ${\cal V}$ the canonical map ${\cal
V}\colon \cH_{\rm OS}\to \cH_{\rm OS}/{\cal N}$. Then in $\cH$
there is the {\em distinguished unit vector}
\[
 \Omega:={\cal V}1,
\]
where $1\in \cH_{\rm OS}$ is the constant function equal to $1$ on
$Q$.

\goodbreak
For $A\in L^{\infty}(Q, \Sigma_{0}, \mu)$ one defines
$\tilde{A}\in \cB(\cH)$ by
\beq
 \tilde{A}{\cal V}\psi:= {\cal V}A\psi.
\label{identi}
\eeq
(Note that multiplication by $A$ preserves ${\cal N}$, since 
$A$ is by assumption $\Sigma_0$ measurable). One denotes by ${\cal U}\subset \cB(\cH)$ the abelian von Neumann
algebra ${\cal U}:=\{\tilde{A} \mid A\in L^{\infty}(Q, \Sigma_{0},
\mu)\}$. It is shown in \cite{KL1, K} that the map $A\mapsto
\tilde{A}$ is a weakly continuous $^{*}$-isomorphism between $L^{\infty}(Q, \Sigma_{0},
\mu)$ and $\cU$.

Finally, setting ${\cal M}_{t}=L^{2}(Q, \Sigma_{[0, \beta/2 -t]},
\mu)$ for $0\leq t\leq \beta/2$ and $\cD_{t}={\cal V}{\cal M}_{t}$,
one can define $P(s) \colon {\cal D}_{t}\to \cH$ for $0\leq s\leq
t$ by
\[
P(s){\cal V}\psi:= {\cal V}U(s)\psi, \: \psi\in {\cal M}_{t}.
\]
The triple  $(P(t), \cD_{t}, \beta/2)$ forms a {\em local symmetric
semigroup} (see \cite{Fr, KL3}), and there exists a unique selfadjoint
operator $L$ on $\cH$ such that $P(s)u= \e^{-s L}u$  for
$u\in \cD_{t}$ and $0\leq s\leq t$. The selfadjoint operator
constructed in this way is said to be {\em associated }to the local
symmetric semigroup $(P(t), \cD_{t}, \beta/2)$.

Next one defines:
\begin{itemize}
\item[--] ${\cal B}\subset \cB (\cH)$ as the von  Neumann algebra
generated by $\{ \e^{\i tL}A\e^{-\i tL} \mid t\in \rr,
A\in\cU \}$;
\item[--] $\tau \colon t \mapsto \tau_{t}$ as the weakly continuous 
group of
$*$-automorphisms of $\cB$, which is  given by 
\[ \tau_{t}(B)= \e^{\i
tL}B\e^{-\i tL}\]
for $t\in \rr$ and $B\in {\cal B}$;
\item[--] $\omega$ as the vector  state on ${\cal B}$ given by $\omega(B)=
(\Omega, B\Omega)$ for $B\in {\cal B}$.
\end{itemize}

\noindent
It is shown in \cite{KL1} that  $(\cB, \cU, \tau, \omega)$ is a stochastically positive
$\beta$-KMS system. The relationship between the two objects is fixed
by the following identity:
\beq
^{E}G(s_{1}, \dots, s_{n}; \tilde{A}_{1}, \dots, \tilde{A}_{n})=
\int_{Q} \bigl(\prod_{i=1}^{n}U(s_{i}) A_{i} \bigr) \d \mu 
\label{e0.0}
\eeq 
for 
$A_{i}\in L^{\infty}(Q, \Sigma_{0}, \mu)$, $1\leq i\leq
n$, and $s_{1}\leq \cdots\leq s_n$, $s_{n}-s_{1}\leq \beta$.

\subsection{Perturbations of generalized path spaces}
\label{sec0.3}
We now describe perturbations of generalized path spaces obtained
from a Feynman-Kac-Nelson kernel. Unless stated otherwise, we will
consider the case $\beta<\infty$.
 
Let  $(Q, \Sigma, \Sigma_{0}, U(t), R, \mu)$ be an OS-positive
$\beta$-periodic  path space. Let $V$ be a selfadjoint operator on
$\cH$, which is
affiliated to ${\cal U}$. Using the isomorphism between ${\cal U}$ and
$L^{\infty}(Q, \Sigma_{0}, \mu)$ we can view $V$ as a real
$\Sigma_{0}$-measurable function on $Q$, which we still denote by $V$.

Assume that $V\in L^{1}(Q, \Sigma_{0}, \mu)$ and $\exp ({-\beta V})\in L^{1}(Q,
\Sigma_{0}, \mu)$. Then (see \cite{KL1} or \cite[Proposition~6.2]{GeJ})
the function $F:=\exp \bigl({-\int_{-\beta/2}^{\beta/2}U(t)V\d t}\bigr)$
belongs to $L^{1}(Q, \Sigma, \mu)$. One can hence define the perturbed
measure $\d\mu_{V}:= (\int_{Q}F\d\mu)^{-1}F\d\mu$. 
The perturbed path space $(Q, \Sigma, \Sigma_{0}, U(t), R, \mu_{V})$ is 
OS-positive and $\beta$-periodic (see \cite{KL1}). Hence we can associate to
this perturbed path space a stochastically positive $\beta$-KMS system~$(\cB_{V}, \cU_{V}, \tau_{V} , \omega_{V})$.

The following concrete realization of the perturbed  
$\beta$-KMS system~$(\cB_{V}, \cU_{V}, \tau_{V} , \omega_{V})$ has been  obtained  in~\cite{KL1} (with
some improvements in \cite{GeJ}):
\begin{itemize}
\item[--] the physical Hilbert space $\cH_{V}$ 
obtained from the  reconstruction theorem outlined in the previous subsection
is equal to the physical Hilbert space $\cH$
of the unperturbed $\beta$-KMS system~$(\cB, \cU, \tau , \omega)$;
\item[--] the von Neumann algebra ${\cal B}_{V}$ and the abelian algebra ${\cal
U}_{V}$ are equal to ${\cal B}$ and ${\cal U}$, respectively;
\item[--] the operator sum $L+V$ is essentially selfadjoint on ${\cal D} (L) \cap {\cal D} (V)$ and if
$H_{V}:=\overline{L+V}$, then the perturbed time-evolution~$\tau_{V}$ on
${\cal B}$ is given by \[ \tau_{V, t}(B)=\e^{\i t
H_{V}}B\e^{-\i tH_{V}}, \: \: B \in \cB ;\]
\item[--] the distinguished vector $\Omega$ of the unperturbed KMS system belongs to
$\cD\bigl(\e^{-\frac{\beta}{2}H_{V}}\bigr)$ and the perturbed KMS state $\omega_{V}$
is given by  $\omega_{V}(B)= (\Omega_{V},
B\Omega_{V})$, where 
\[Ê\Omega_{V}:= \frac{\e^{-\frac{\beta}{2}H_{V}}\Omega}{ 
\|\e^{-\frac{\beta}{2}H_{V}}\Omega\| }. \]
\end{itemize}
The following result is shown in \cite[Theorem 6.12]{GeJ}: If $\e^{-\beta V}\in L^{1}(Q, \Sigma_{0}, \mu)$ and
\[
V\in L^{p}(Q, \Sigma_{0}, \mu),\quad
\e^{-\frac{\beta}{2}V}\in L^{q}(Q, \Sigma_{0}, \mu) \quad \hbox{for} \quad p^{-1}+ q^{-1}=
\frac{1}{2} , \quad 2\leq p, q\leq \infty,
\]
then the operator sum $H_{V}-JVJ$ is essentially selfadjoint and 
the {\em Liouvillean\/} $L_{V}$ (for a general definition of 
Liouvilleans see, e.g., \cite{DJP}) for the perturbed $\beta$-KMS system
$({\cal B}_{V}, \tau_{V}, \omega_{V})$ is equal to
$\overline{H_{V}- JVJ}$. Here $J$ denotes the modular conjugation associated to the pair $({\cal B},\Omega)$. 

\subsection{Perturbed dynamics associated to FKN kernels}
\label{fkn}
Let us describe in more details the construction of
$H_{V}= \overline{L+V}$ given in \cite{KL1} which is based on the
Feynman-Kac-Nelson  formula. Note that the results  of this
subsection are  also valid in the
$0$-temperature case $\beta=+\infty$. Let $V$ be a real
$\Sigma_{0}$-measurable function such that $V\in L^{1}(Q,\Sigma_{0},
\mu)$ and $\e^{-TV}\in L^{1}(Q, \Sigma_{0}, \mu)$ for some $T>0$ if
$\beta=\infty$ and for $T=\beta$ if~$\beta<\infty$.
Set
\[
F_{[0,s]}:=\e^{-\int_{0}^{s}U(t)V\d t},\: \: 0\leq s\leq \inf(T, \beta)/2,
\]
which belongs to $L^{2}(Q, \Sigma_{[0, \inf(T, \beta)/2]}, \mu)$. The family
$\{F_{[0,s]}\}_{0\leq s\leq \inf{(T, \beta)}/2}$ is called a {\em
Feyn\-man-Kac-Nelson kernel}.

For $0\leq
t\leq \inf(T, \beta)/2$ we set 
\[
{\cal M}_{t}:=\hbox{ linear span of } \bigcup_{0\leq s\leq \inf(T, \beta)/2-t}F_{[0,s]}L^{\infty}(Q, \Sigma_{[0,
\inf(T, \beta)/2 -t]}, \mu)
\] 
and  
\[
\matrix{ U_{V}(s)\colon &
{\cal M}_{t} & \to & L^{2}(Q, \Sigma_{+}, \mu)\cr
& \psi & \mapsto & F_{[0,s]}U(s)\psi,\: 
\cr} \quad 0\leq s\leq t.
\]
Setting finally 
\beq
\label{spaces}
\cD_{t}={\cal V}({\cal M}_{t}),
\eeq
one can show that 
\[
\matrix{P_{V}(s)\colon &
\cD_{t} & \to & \cH \cr
&\cV(\psi) & \mapsto & \cV(F_{[0,s]}U(s)\psi) 
\cr} 
\]
is a well defined linear operator, and that $\bigl(P_{V}(t), \cD_{t}, \inf(T, \beta)/2 \bigr)$ is
a local symmetric semigroup on~$\cH$. Now let $H_{V}$ be the unique
selfadjoint operator associated to the local symmetric semigroup~$(\cD_{t}, P_{V}(t), \inf(T, \beta)/2)$.
It follows (see \cite{KL1}) that $H_{V}= \overline{L+V}$.

In the sequel we will need the following result.
\begin{proposition}
\label{approx}
Let $V\in L^{2}(Q, \Sigma_{0}, \mu)$ be a real function such that
$\e^{-T
V}\in L^{1}(Q, \Sigma_{0}, \mu)$ for some $T>0$ and $V_{n}:=
V\one_{\{|V|\leq n\}}$ for $n\in \nn$. Denote by $L$ the
selfadjoint operator on~$\cH$ associated to the OS-positive
path space $(Q, \Sigma, \Sigma_{0}, U(t), R, \mu)$. Let $H_{n}$ be the
closure of $L+ V-V_{n}$.
Then 
\[
\e^{-\i tL}=\slim_{n\to \infty}\e^{-\i tH_{n}},\: \: t\in \rr.
\]
\end{proposition}

Note that the selfadjoint operators  $H_{n}$  are  associated to
local symmetric semigroups $\bigl(P_{n}(t), \cD^{(n)}_{t}, T/2 \bigr)$ obtained from the FKN kernels
\[
F_{[0, s]}^{(n)}:=\e^{-\int_{0}^{s}U(t)(V-V_{n})\d t},
\] and the operator  $L$ is associated to the local symmetric
semigroup $\bigl(P_{\infty}(t), \cD_{t}, T/2 \bigr)$ obtained from  the FKN
kernels $F^{(\infty)}_{[0, s]}=1$. 

\noindent
\proof
We first claim that 
\beq
\sup_{0\leq s\leq T/2}\|F_{[0,s]}^{(n)}-1\|_{L^{1}(Q, \Sigma,
\mu)}\to 0  \: \: \hbox{for} \: \: n\to \infty.
\label{idiot}
\eeq
In order to prove (\ref{idiot}), we recall  the following bound from \cite[Theorem~6.2 (i)]{KL0}:
\beq
\|\e^{-\int_{a}^{b} U(t)V\d t}\|_{L^{p}(Q, \Sigma, \mu)}\leq
\|\e^{-(b-a)V}\|_{L^{p}(Q,\Sigma, \mu)},\: \: 1\leq p<\infty.
\label{bound1}
\eeq
Now let $W$ be a real measurable function on $Q$. Using $1-\e^{-a}=a\int_{0}^{1}\e^{- \theta a}\d \theta$
we find
\[
1-\e^{-\int_{0}^{s}U(t)W\d t}= \int_{0}^{s}U(t)W\d
t\int_{0}^{1}\e^{-\theta\int_{0}^{s} U(t)W\d t}\d\theta.
\]
This yields
\[
\begin{array}{rl}
\|1-\e^{-\int_{0}^{s}U(t)W\d t}\|_{L^{1}}\leq
&|s| \: \|W\|_{L^{2}}\int_{0}^{1}\|\e^{-\theta\int_{0}^{s} U(t)W\d
t}\|_{L^{2}}\d\theta\\[2mm]
\leq & |s| \:  \|W\|_{L^{2}}\int_{0}^{1}\|\e^{- \theta
sW}\|_{L^{2}}\d\theta\\[2mm]
\leq & |s| \:  \|W\|_{L^{2}} \bigl(1+\int_{0}^{1}\|\e^{\theta
sW_{-}}\|_{L^{2}}\d\theta \bigr)\\[2mm]
\leq &\frac{T}{2} \|W\|_{L^{2}} \bigl(1+\|\e^{T W_{-}}\|_{L^{1}} \bigr),
\end{array}
\]
where $W_{-}=\sup(0, -W)$ denotes the negative part of $W$.
In the first line we have used the Cauchy-Schwarz inequality and the
fact that $U(t)$ is unitary on 
$L^{2}(Q, \Sigma, \mu)$, in the second line the estimate
(\ref{bound1}).

By assumption $V\in L^{2}(Q, \Sigma, \mu)$ and $\e^{-T V}\in L^{1}(Q, \Sigma, \mu)$. Thus  $V-
V_{n}\to 0$ in $L^{2}(Q, \Sigma, \mu)$ and~$\e^{T(V-V_{n})_{-}}\to 0$ in $L^{1}(Q, \Sigma, \mu)$.
Applying the above bound for $W= V-V_{n}$, we obtain~(\ref{idiot}).

Before we finish the proof, we extract a Lemma.

\begin{lemma}
\label{semigroupconv}
Let $\bigl(P_{n}(t), {\cal D}^{(n)}_{t}, T \bigr)$ for $n\in \nn\cup \{\infty\}$ be a family
of local symmetric semigroups on a Hilbert space $\cH$. Let $H_{n}$,
$n\in \nn\cup \{\infty\}$, denote the associated
selfadjoint operators.

 Assume that there exists a family $\{{\cal L}_{t}\}$ for $0< t\leq T'\leq T$ of subspaces of $\cH$ with 
\beq
\label{semigroup0}
{\cal L}_{t}\subset {\cal D}^{(n)}_{t}, \: \bigcup_{0<t\leq T'}{\cal
L}_{t}\hbox{ dense in }\cH.
\eeq Assume moreover that
\beq
 \lim_{n\to \infty}(\Psi, P_{n}(s)\Psi)= (\Psi, P_{\infty}(s)\Psi), \: \Psi\in {\cal
L}_{t}, \:0\leq s\leq t\leq T',
\label{semigroup1}
\eeq
\beq \sup_{n}\sup_{0\leq s\leq t}(\Psi, P_{n}(s)\Psi)<\infty, \: \Psi\in {\cal
L}_{t}, \: 0\leq t\leq T'.
\label{semigroup2}
\eeq
Then
$\: \: \slim_{n\to \infty}\e^{-\i tH_{n}}= \e^{-\i tH_{\infty}} \: \: $ for all $t\in \rr$.
\end{lemma}
\proof
Let us fix $0<t\leq T'$ and $\Psi\in {\cal L}_{t}$. From \cite[Lemma
1]{KL3}, we know that there exist positive measures $\{ \nu_{n} \}$ on $\rr$
such that
\[
 (\Psi, P_{n}(s)\Psi)= \left\|P_{n} \left(\frac{s}{2}\right)\Psi \right\|^{2}= \int_{\rr}\e^{-s a}\d\nu_{n}(a), \: \: 0\leq
s\leq t.
\]
Moreover, one has  (see \cite[Lemma 1]{KL3})

\[
 (\Psi, \e^{-\i yH_{n }}\Psi)=\int_{\rr}\e^{-\i y a}\d \nu_{n}(a).
\]
Set
\[
 f_{n}(z):=\int_{\rr}\e^{-z a}\d\nu_{n}(a), \: \: z\in ]0, t[+\i \rr. 
\]
The family $\{f_{n}\}$ is uniformly bounded on $]0, t[+\i \rr$ by
(\ref{semigroup2})  and
converges pointwise to~$f_{\infty}$ on~$]0, T[$ by (\ref{semigroup1}).
Applying Lemma \ref{T4a} we conclude  that $f_{n}(z)$ converges to
$f_{\infty}(z)$ for all~$z\in \i\rr$. This implies that
on ${\cal L}_{t}$ 
\[
\wlim_{n\to \infty }\e^{-\i yH_{n}}= \e^{-\i yH_{\infty}} \: \: \forall y\in \rr.
\] 
Since by hypothesis $\bigcup_{0<t\leq T}{\cal L}_{t}$ is dense in $\cH$ and
for unitary operators weak convergence implies strong convergence,
this completes the proof of the lemma \qed .

\medskip
\noindent 
{\bf Proof of Proposition \ref{approx} (second part).}
Let us now fix  a convenient family of subspaces~${\cal L}_{t}$. For $0<t<T/4$ 
we set~${\cal L}_{t}= {\cal V}{\cal R}_{t}$, where~${\cal R}_{t}$ equals $L^{\infty}(Q,
\Sigma_{[0, \beta/2-t]}, \mu)$ if $\beta<\infty$ and ${\cal R}_{t} $ equals $
L^{\infty}(Q,\Sigma_{[0, +\infty[}, \mu)$ if $\beta=+\infty$. Clearly~${\cal
L}_{t}$ is included in the spaces $\cD^{(n)}_{t}$ defined in
(\ref{spaces}) (with $V$ replaced by $V-V_{n}$). Moreover, 
$\bigcup_{0<t\leq T/4}{\cal L}_{t}$ is dense in $\cH$,
hence hypothesis (\ref{semigroup0}) of Lemma \ref{semigroupconv} is
satisfied. Let us now fix some~$\Psi\in {\cal L}_{t}$, i.e., $\Psi= {\cal V}\psi$ for
some~$\psi\in {\cal R}_{t}$. Using
(\ref{idiot})  we obtain that 
\[\lim_{n\to
\infty} (\Psi,P_{n}(s)\Psi)= (\Psi,P_{\infty}(s)\Psi) \hbox{ for } 0\leq s\leq t 
\]
and 
$\sup_{n}\sup_{0\leq s\leq t}(\Psi, P_{n}(s)\Psi) <\infty$. Hence hypotheses
(\ref{semigroup1}) and (\ref{semigroup2}) of Lemma \ref{semigroupconv} are
satisfied. Thus we can apply Lemma \ref{semigroupconv}  and this completes the proof of Proposition \ref{approx} \qed .

\goodbreak

\section{Gaussian measures}
\init\label{sec1}
In this Section we recall some standard facts about Gaussian measures
on distribution spaces.
\subsection{Distribution spaces}
\label{sec1.1}
Let $S_{\beta}=[-\beta/2,\beta/2]$ (with end points identified) be the
circle of length $\beta>0$. Points in~$S_{\beta}\times \rr^{d}$,
$d\geq 1$, will be denoted by $(t,x)$. 

The Fr\'echet space of Schwartz functions 
 on $\rr^{d}$ will be denoted by $\cS(\rr^{d})$. For
coherence of notation, the
Fr\'echet space $\cD(S_{\beta})$ 
of smooth periodic functions on $S_{\beta}$ will also be denoted 
by~$\cS(S_{\beta})$.

In addition, we denote by $\cS(S_{\beta}\times \rr^{d})$ the Fr\'echet space of Schwartz
functions on $S_{\beta}\times \rr$, i.e., the space of smooth
functions on $S_{\beta}\times \rr^{d}$, which are $\beta$-periodic in
$t$ and  such that for all~$p\in \nn$ and~$\alpha\in \nn^{d}$
$$\bigl|(1+|x|)^{| \alpha | }\p_{t}^{p}\p^{\alpha}_{x}f(t,x) \bigr|\leq C_{p, \alpha} \,  .$$

We will denote by $\cS'(\rr^{d})$, $\cS'(S_{\beta})$ and
$\cS'(S_{\beta}\times \rr^{d})$ the  duals of
$\cS(\rr^{d})$, $\cS(S_{\beta})$ and $\cS(S_{\beta}\times\rr^{d})$.
The spaces of real elements in these spaces will be denoted by
$\cS'_{\rr}(\rr^{d})$, $\cS'_{\rr}(S_{\beta})$ and
$\cS'_{\rr}(S_{\beta}\times \rr^{d})$.

We set $D_{t}=\i^{-1}\p_{t}$ and $D_{x}= \i^{-1}\p_{x}$,
and we will denote by $D_{t}^{2}$ the selfadjoint operator on~$L^{2}(S_{\beta})$
defined by
\[
D_{t}^{2}:= -\p_{t}^{2}, \quad \cD(D_{t}^{2}):= \bigl\{u\in L^{2}(S_{\beta})
\mid \p_{t}^{2}u\in L^{2}(S_{\beta}), \: u(0)= u(\beta) \bigr\}. 
\]

We denote by $D_{t}^{2}+ D_{x}^{2}$ the selfadjoint operator on
$L^{2}(S_{\beta}\times \rr^{d})$ with domain 
\[
\cD(D_{t}^{2}+ D_{x}^{2}):= \bigl\{
u\in L^{2}(S_{\beta}\times \rr^{d}) \mid (D_{t}^{2}+ D_{x}^{2})u\in
L^{2}(S_{\beta}\times \rr^{d}), \: u  \hbox{ is }\beta\hbox{-periodic in }t \bigr\}.
\]
We denote by $\cS(\zz\times \rr^{d})$ the Fr\'echet space of sequences
$\{u_{n}\}_{n \in \nn}$ with values in $\cS(\rr^{d})$ such that
\[
\sum_{n\in \zz}|n|^{p}\|(D^2_{x}+
x^{2})^{p/2}u_{n}\|_{L^{2}(\rr^{d})}<\infty \quad \forall p\in \nn.
\]
We now fix the notation concerning  partial Fourier transforms.
We first define the (unitary) partial Fourier transform with respect to $t$:
\[
\matrix{
{\cal F}_{t}\colon &\cS(S_{\beta}\times \rr^{d}) & \to 
&\cS(\zz\times \rr^{d}) \cr
&u&\mapsto &\{ \hat{u}_{n}\} \cr} \, , 
\]
where $
\hat{u}_{n}(x)= \beta^{-\frac{1}{2} }\int_{S_{\beta}}\e^{-\i
\nu_n t}u(t,x)\d t $. (The coefficients $\nu_n = 2\pi n/\beta$, $n \in
\nn$, are called in physics {\em Matsubara frequencies}). Its inverse is
\[
 u(t,x)= \beta^{-\frac{1}{2} }\sum_{n\in \zz} \e^{\i \nu_n t}\hat{u}_{n}(x).
\]
The (unitary) partial Fourier transform with respect to $x$ is
\[
\matrix{{\cal F}_{x}\colon &\cS(S_{\beta}\times \rr^{d}) & \to 
& \cS(S_{\beta}\times \rr^{d}) \cr
&u & \mapsto &\hat{u}\cr} \, , 
\]
where $
\hat{u}(t,p)= (2\pi)^{-d/2}\int_{\rr^{d}}\e^{-\i
x.p}u(t,x)\d x$. Its  inverse is
\[
 u(t,x)= (2\pi)^{-d/2}\int_{\rr^{d}} \e^{\i x.p}\hat{u}(t, p)\d p.
\]
For later use we fix two approximations of the Dirac $\delta$
functions in $t$ and $x$. We set, for~$k\in \nn$,
\[
 \delta_{k}(s):=\beta^{-1}\sum_{|n|\leq k}\e^{i \nu_n s } \quad \hbox{and} \quad
\delta_{k}(x):= k \chi(kx),
\] where $\chi$ is a function in $C^{\infty}_{0\:\rr} (\rr^{d})$ with $\int
\chi(x)\d x=1$.

\subsection{Gaussian measures}
\label{sec1.2}
We set
\beq
\label{c}
C(f,g)=\bigl(f,(D_{t}^{2}+ D_{x}^{2}+  m^{2})^{-1}g \bigr),\: \: f,g\in
\cS(S_{\beta}\times \rr^{d}),
\eeq
where $(.,.)$ is the scalar product on
$L^{2}(S_{\beta}\times \rr^{d})$.

Let  $Q:=\cS'_{\rr}(S_{\beta}\times \rr^{d})$ and let $\Sigma$ be the
Borel $\sigma$-algebra on $Q$. If $f\in \cS_{\rr}(S_{\beta}\times\rr^{d})$, 
then~$\phi(f)$ denotes the coordinate function  
\[
\matrix{ \phi(f) \colon & Q & \to & \cc 
\cr
& q & \mapsto &  \langle q, f\rangle \cr}.
\]
Let $F$ be a Borel function on~$\rr$. Then $F(\phi(f)) $ denotes the function
\[
\matrix{ F(\phi(f)) \colon & Q & \to & \cc 
\cr
& q & \mapsto & F \bigl( \langle q, f\rangle \bigr) \cr}.
\]
We denote by $\d \phi_{C}$ the
Gaussian measure  on~$(Q, \Sigma)$ with
covariance $C$ defined by  
\beq
\label{e1.00}
\int_{Q}\e^{\i \phi(f)}\d\phi_{C}= \e^{- C(f,f)/2}, \quad f\in
\cS_{\rr}(S_{\beta}\times \rr^{d}).
\eeq
We have
\beq
\int_{Q}\phi(f)^{p}\d\phi_{C}=\left\{
\begin{array}{l}
0, \: p\hbox{ odd},\\
(p-1)!!C(f,f)^{p/2},\: p\hbox{ even},
\end{array}
\right.
\label{e1.0}
\eeq
where $n!!= n(n-2)(n-4)\cdots 1$. One easily deduces from (\ref{e1.0})
that $\e^{\phi(f)}\in L^{1}(Q, \Sigma, \d\phi_{C})$
if~$f\in \cS_{\rr}(S_{\beta}\times \rr^{d})$.

The cylindrical functions $F \bigl(\phi(f_{1}), \dots, \phi(f_{n}) \bigr)$, $f_{i}\in
\cS_{\rr}(\rr\times S_{\beta})$, $F$ a Borel function on~$\rr^{n}$ and $n\in
\nn$, are dense
in $L^{p}(Q, \Sigma, \d\phi_{C})$ for  $1\leq p<\infty$.

\subsection{Sharp-time fields}
\label{sec1.3}
We now recall some standard results about the existence of {\em sharp-time
fields}.
We will make use of the following well known identity (see \cite{KL2}):
\beq
\frac{1}{\beta}
 \sum_{n\in \zz}\frac{\e^{i \nu_n t}}{ \nu_n^{2}+
\epsilon^{2}}= \frac{\e^{-|t|\epsilon}+
\e^{-(\beta-|t|)\epsilon}}{ 2 \epsilon(
1-\e^{- \beta \epsilon})}  \: \hbox{ for }\epsilon>0,  \: \nu_n = \frac{2\pi n}{\beta}, \: \: 0\leq |t|\leq \beta.
\label{e1.1}
\eeq
For $h_{1}, h_{2}\in \cS_{\rr}(\rr^{d})$, $0\leq t_{1}, t_{2}\leq \beta$, and
$k\in \nn$ 
\[
\begin{array}{rl}
 &C \bigl(\delta_{k}(.-t_{1})\otimes h_{1}, \delta_{k}(.-t_{2})\otimes
h_{2} \bigr)\\[2mm]
= &\beta^{-1}\sum_{|n|\leq k} \e^{i \nu_n (t_{1}- t_{2})}
\bigl(\hat{h}_{1n}, ( \nu_n^{2} + D^{2}_{x}+ m^{2})^{-1}\hat{h}_{2n} \bigr)_{L^{2}(\rr^{d})} .
\end{array}
\]
Using (\ref{e1.1}) we see that
\[
 \lim_{k\to \infty} C \bigl(\delta_{k}(.-t_{1})\otimes h_{1},
\delta_{k}(.-t_{2})\otimes h_{2} \bigr)
= \Bigl( h_{1}, \frac{\e^{-|t_{2}-
t_{1}|\epsilon}+ \e^{-(\beta-|t_{2}-t_{1}|)\epsilon}}{
2\epsilon(1-\e^{- \beta \epsilon})}h_{2} \Bigr)_{L^{2}(\rr^{d})},
\]
where $\epsilon:= (D_{x}^{2}+ m^{2})^{\frac{1}{2} }$.

Using (\ref{e1.0}) this implies that, for $h\in \cS_{\rr}(\rr^{d})$ and
$t\in S_{\beta}$ fixed, the sequence of functions~$\{ \phi( \delta_{k}(.-t)\otimes h) \}_{k \in \nn}$
is Cauchy in 
$\bigcap_{1\leq p<\infty}L^{p}(Q, \Sigma, \d \phi_{C})$.

We set
\beq
 \phi (t, h):=\lim_{k\to \infty}\phi(\delta_{k}(.-t)\otimes h)
\label{e1.4}
\eeq
and  
\beq
C_0  ( t_{1},h_{1}, t_{2} , h_{2}) := \Bigl( h_{1}, \frac{\e^{-|t_{2}-
t_{1}|\epsilon}+ \e^{-(\beta-|t_{2}-t_{1}|)\epsilon}}{
2\epsilon(1-\e^{- \beta \epsilon})}h_{2} \Bigr)_{L^{2}(\rr^{d})}.
\label{e1.4b}
\eeq
We note that  $\phi (t, h)$ belongs to $\bigcap_{1\leq p<\infty}L^{p}(Q, \Sigma, \d \phi_{C})$. 
For later use we define the {\em temperature~$\beta^{-1}$ covariance on
$\rr^{d}$}:
\beq
C_{0}(h_{1}, h_{2}):=  \Bigl( h_{1}, \frac{(1+\e^{-\beta\epsilon})}{
2\epsilon(1-\e^{- \beta \epsilon})}h_{2} \Bigr)_{L^{2}(\rr^{d})},  \: \: h_{1}, h_{2}\in
\cS(\rr^{d}).
\label{c0}
\eeq

\subsection{Sharp-space fields}
\label{sec1.4}
If $d=1$, then it is possible to define similarly {\em sharp-space
fields}.
We first recall another well-known identity, which is analogous  to (\ref{e1.1}):
\beq
(2\pi)^{-1}\int_{\rr}\frac{\e^{\i p x}}{p^{2}+ b^{2}}\d p=
\frac{\e^{- b|x|}}{ 2b} \hbox{ for } b>0, \: x\in \rr .
\label{e1.2}
\eeq
For $g_{1}, g_{2}\in \cS_{\rr}(S_{\beta})$ and $x_{1}, x_{2}\in \rr$  one
has
\beq
\label{e1.4c}
\begin{array}{rl}
&C\bigl(g_{1}\otimes \delta_{k}(.-x_{1}), g_{2}\otimes
\delta_{k}(.-x_{2}) \bigr)\\[2mm]
=&\int_{\rr} \hat{\chi}^{2}(\frac{p}{k})\e^{\i p(x_{1}- x_{2})}
\bigl( g_{1}, (D_{t}^{2}+ p^{2}+
m^{2})^{-1}g_{2} \bigr)_{L^{2}(S_{\beta})}\d p.
\end{array}
\eeq
Using (\ref{e1.2}) and  $\hat{\chi}(0)= (2\pi)^{-\frac{1}{2} }$ we find
\beq
\label{e1.4d}
\lim_{k\to \infty}C\bigl(g_{1}\otimes \delta_{k}(.-x_{1}), g_{2}\otimes
\delta_{k}(.-x_{2}) \bigr)= \Bigl(g_{1}, \frac{\e^{- |x_{1}- x_{2}|b}}{
2b}g_{2} \Bigr)_{L^{2}( S_{\beta})},
\eeq
where $b:= (D_{t}^{2}+ m^{2})^{\frac{1}{2} }$. Now we can use  (\ref{e1.0}) again:
for $g\in \cS_{\rr}(S_{\beta})$ and $x\in \rr$ fixed, the sequence of functions $\bigl\{ \phi \bigl(g\otimes\delta_{k}( .-x) \bigr) 
\bigr\}_{k \in \nn}$
is Cauchy in $\bigcap_{1\leq p<\infty}L^{p}(Q, \Sigma, \d \phi_{C})$.

We set
\[
 \phi (g, x):= \lim_{k\to \infty}\phi \bigl(g\otimes\delta_{k}(.-x) \bigr)
\]
and
\beq
C_\beta (g_{1}, x_1, g_{2}, x_2)  :=  \Bigl(g_{1}, \frac{\e^{- |x_{1}- x_{2}|b}}{
2b}g_{2} \Bigr)_{L^{2}( S_{\beta})}.
\label{e1.5b}
\eeq
We note that $ \phi (g, x)$ belongs to $\bigcap_{1\leq k<\infty}L^{p}(Q, \Sigma, \d\phi_{C})$. 
For later use we define the {\em $0$-temperature covariance on
$S_{\beta}$}:
\beq
\label{cbeta}
C_{\beta}(g_{1}, g_{2}):=\Bigl(g_{1}, \frac{1}{2b}g_{2}\Bigr)_{L^{2}( S_{\beta})}, \: g_{1}, g_{2}\in
\cS(S_{\beta}).
\eeq

\subsection{Some elementary properties}
\label{sec1.5}
From (\ref{e1.0}), (\ref{e1.4b}) and (\ref{e1.5b}) we deduce   that the
maps
\beq
\label{e1.6bb}
\matrix{
& H^{-1}_{\rr}(S_{\beta} \times\rr^{d}) & \to  & \bigcap_{1\leq
p<\infty}L^{p}(Q, \Sigma, \d\phi_{C}) \cr
&f & \mapsto & \phi(f) \cr}  \, ,
\eeq
\beq
\matrix{ & S_{\beta}\times H_{\rr}^{-\frac{1}{2} }(\rr^{d}) &
\to & \bigcap_{1\leq p<\infty}L^{p}(Q, \Sigma, \d\phi_{C})
\cr
& (t,h) & \mapsto & \phi ( t, h)\cr}  
\eeq
and
\beq
\matrix{ 
& H_{\rr}^{-\frac{1}{2} }(S_{\beta})\times \rr & \to & \bigcap_{1\leq p<\infty}L^{p}(Q, \Sigma, \d\phi_{C})
\cr
& (g, x) & \mapsto & \phi (g, x) 
\cr}  
\label{e1.6b}
\eeq
are continuous.

\goodbreak
For $f\in \cS_{\rr}(S_{\beta}\times \rr)$, $t\in S_{\beta}$ and $x\in \rr$  we
set
\[
\matrix{ 
f_{t} \colon & \rr & \to & \cc 
\cr
& x & \mapsto & f(t,x) \cr} \: ,\qquad \qquad
\matrix{ 
f_{x} \colon & S_{\beta} & \to & \cc
\cr
& t & \mapsto & f(t, x) \cr} \: .
\]
We note that $f_{t}\in \cS_{\rr}(\rr)$ and $f_{x}\in \cS_{\rr}(S_{\beta})$.
\begin{lemma}
\label{1.0b}
If $f\in \cS_{\rr}(S_{\beta}\times \rr)$, then the following identity holds on
$\bigcap_{1\leq p<\infty}L^{p}(Q, \Sigma, \d\phi_{C})$:
\[
\int_{\rr}\phi (f_{x}, x )\d x= \int_{S_{\beta}}\phi  (t, f_{t})\d t=
\phi(f).
\]
\end{lemma}
\proof
Let $f\in \cS_{\rr}(S_{\beta}\times \rr)$ and $k\in \nn$. The map
\[
\matrix {&  \rr & \to &  H^{-1}(S_{\beta}\times
\rr)
\cr
& x & \mapsto & f_{x}\otimes \delta_{k}(.-x)
\cr}
\]
is continuous.
Since $f\in \cS_{\rr}(S_{\beta}\times \rr)$, the bound $\|f_{x}\otimes
\delta_{k}(.-x)\|_{H^{-1}(S_{\beta}\times \rr)}\in O \bigl(|x|^{-\infty} \bigr)$ holds true.
Hence by (\ref{e1.6b}) the map
\[
\matrix { & \rr & \to & \bigcap_{1\leq p<\infty}L^{p}(Q, \Sigma, \d\phi_{C})
\cr
& x & \mapsto & \phi \bigl(f_{x}\otimes \delta_{k}(.-x) \bigr) \cr}
\]
is continuous and $\bigl\| \phi \bigl(f_{x}\otimes \delta_{k}(.-x) \bigr) \bigr\|_{L^{p}(Q, \Sigma,
\d\phi_{C})}\in O\bigl(|x|^{-\infty} \bigr)$.  Therefore $\int_{\rr}\phi \bigl(f_{x}\otimes
\delta_{k}(.-x) \bigr)\d x $
is well defined as an element of $\bigcap_{1\leq
p<\infty}L^{p}(Q, \Sigma, \d\phi_{C})$. Moreover,  
\[
 \int_{\rr}\phi \bigl(f_{x}\otimes \delta_{k}(.-x) \bigr)\d x= \phi \Bigl(\int_{\rr}
f_{x}\otimes \delta_{k}(.-x)\d x \Bigr)= \phi(f* \delta_{k}),
\]
where the convolution product $*$ acts only in the space variable $x$.
Since $\lim_{k\to \infty} f*\delta_{k}= f$ holds in~$H^{-1}(S_{\beta}\times
\rr)$, we obtain from (\ref{e1.6bb}) 
\[
 \lim_{k\to \infty} \int_{\rr}\phi \bigl(f_{x}\otimes \delta_{k}(.-x) \bigr)\d x=
\phi(f) \quad \hbox{in} \:  \bigcap_{1\leq p<\infty}L^{p}(Q, \Sigma, \d\phi_{C}).
\]
It follows from (\ref{e1.4c}) and  (\ref{e1.4d}) 
that 
\[ \lim_{ k\to \infty}\sup_{x\in \rr}
|x|^{N} \bigl\|\phi \bigl(f_{x}\otimes \delta_{k}(.-x) \bigr)- \phi (f_{x},
x) \bigr\|_{L^{p}(Q, \Sigma, \d\phi_{C})} = 0 
\]
for $f\in \cS_{\rr}(S_{\beta}\times \rr)$ and $N\in \nn$. Hence
\[
\lim_{k\to \infty} \int_{\rr}\phi \bigl(f_{x}\otimes \delta_{k}(.-x) \bigr)\d
x=\int_{\rr}\phi  (f_{x}, x)\d x.
\]
This proves the first identity of the lemma. The second one can be
shown by similar  arguments \qed .

\medskip

\section{Path spaces supported by $(\cS'_{\rr}(S_{\beta}\times \rr),\Sigma, \d\phi_{C})$}
\init\label{sec2}
In this section we recall two well known path spaces supported by
$(Q,\Sigma,
\d\phi_{C})$. The first is associated  to the free neutral scalar field  of
mass $m$ on
$S_{\beta}$ at  temperature $0$; the second is associated to
the free neutral scalar field of
mass $m$ on $\rr$ at temperature~$\beta^{-1}$. 

We recall that $(t,x)$ denotes a point in $S_{\beta}\times \rr$, and refer to $t$ as the (euclidean) time and 
to~$x$ as space variable. The time translation induced on $Q$ by the map $(t,x) \mapsto (t +s, x)$ will be denoted 
by ${\scriptstyle \frak T}_s \colon
Q\to Q$
and the spatial translations induced on $Q$  by the map $(t,x) \mapsto (t,x + y)$ will be denoted by ${\frak a}_y
\colon Q\to Q$.

\subsection{The free  massive euclidean field on the circle at $0$-temperature}
\label{sec2.1}
In this subsection we identify the generalized path space on
$(Q,\Sigma,\d\phi_{C})$ corresponding to the free
massive scalar field on the circle $S_{\beta}$ at temperature $0$.

Let $\Sigma_{0}^{\tt C}$ be the sub $\sigma$-algebra of
$\Sigma$ generated by the functions $\{ \phi(g,0) \mid g\in
\cS_{\rr}(S_{\beta}) \}$.

We denote by  $\{U_{\tt C}(x) \}_{x \in \rr}$ the $1$-parameter group  
 generated by the spatial translations~$\{{\frak a}_{x}\}_{x \in \rr}$.  
More precisely, if $F\colon Q\to \cc$ is a function on~$Q$, then $U_{\tt C}(x)F(q):=
F\bigl({\frak a}_{-x}(q) \bigr)$ for $q\in Q$. Applying  (\ref{e1.00}) we see that 
$x \mapsto U_{\tt C}(x)$ is a strongly continuous unitary group on $L^{2}(Q, \Sigma,
\d\phi_{C})$, and hence extends to  a group of measure-preserving
automorphisms of $L^{\infty}(Q, \Sigma, \d \phi_{C})$ which is
continuous in measure.

Let $r_{\tt C} \colon Q\to Q$ be the space reflection around $x=0$. We
denote by $R_{\tt C}$ the measure preserving transformation of $(Q,
\Sigma, \d\phi_{C})$ generated by $r_{\tt C}$.

For $g\in \cS_{\rr}(S_{\beta})$ we have 
\beq
\label{tony}
U_{\tt C} (x)\phi(g, 0)= \phi(g,x).
\eeq 
Using  then Lemma \ref{1.0b}, we see that $\Sigma=\bigvee_{x\in
\rr}U_{\tt C}(x)\Sigma_{0}^{\tt C}$. 

Hence 
$(Q, \Sigma, \Sigma_{0}^{\tt C},U_{\tt C}(x), R_{\tt C}, \d \phi_{C})$ 
is a generalized path space. Moreover, it
is
OS-positive (see e.g.~\cite{KL2}). 

It describes the {\em free neutral scalar euclidean field of mass $m$ on the circle $S_{\beta}$ at
temperature $0$}.

Let us now briefly describe a well-known concrete form of the
physical objects associated to this path space by the reconstruction
theorem.
Let $H^{-\12}(S_{\beta})$ be the Sobolev space of order~$-\frac{1}{2}
$ equipped with its canonical complex structure $\i$ and 
scalar product $(h_{1}, (2b)^{-1}h_{2})_{L^{2}(S_{\beta})}$, where  $b=(D_{t}^{2}+ m^{2})^{\frac{1}{2} }$.
Then the physical Hilbert space  can be unitarily identified with the bosonic Fock space $\Gamma
\bigl(H^{-\frac{1}{2} }(S_{\beta}) \bigr)$
over $H^{-\frac{1}{2} }(S_{\beta})$. The distinguished unit vector $\Omega^\circ_{\tt C} :={\cal V}1$ 
is identified with the  Fock vacuum $\Omega$ in $\Gamma \bigl(H^{-\frac{1}{2}
}(S_{\beta}) \bigr)$. The (free) Hamiltonian is 
\[
H^\circ_{\tt C}=\d\Gamma(b).
\]
The abelian von Neumann algebra $\cU_{\tt C}$ obtained from the reconstruction
theorem can be identified  with the von Neumann algebra generated by $\{W_F(g) \mid  g\in
H^{-\12}_{\rr}(S_{\beta}) \}$. In fact, if~$A=\e^{\i \phi(g, 0)}$ for~$g\in \cS_{\rr}(S_{\beta})$, then the
operator $\tilde{A}$ defined in (\ref{identi}) is identified with  the Fock
Weyl operator $W_F(g)= \e^{\i \phi_F (g)}$ on~$\Gamma \bigl(H^{-\12}(S_{\beta}) \bigr)$.

\subsection{The free massive euclidean field  on $\rr$ at temperature
$\beta^{-1}$} 
\label{sec2.2a}
We now identify the  generalized path space on
$(\cS_{\rr}'(S_{\beta}\times \rr), \Sigma, \d\phi_{C})$ corresponding to the free
massive scalar euclidean field on $\rr$ at temperature $\beta^{-1}$.

Let $\Sigma_{0}$ be the sub $\sigma$-algebra of
$\Sigma$ generated by the functions $\{ \phi(0,h) \mid h\in \cS_{\rr}(\rr) \}$.
We denote by~$\{ U(t)\}_{t\in S_{\beta}}$  the one parameter
group  generated by $\{ {\scriptstyle \frak T}_t \}_{t\in S_{\beta}}$. If $F\colon Q\to \cc$ is a
function on $Q$, then $U(t)F(q) := F\bigl({\scriptstyle \frak T}_{-t}(q)\bigr)$ for $q\in Q$.
 Using~(\ref{e1.00}) we see 
that $t \mapsto U(t)$ is a strongly continuous
$\beta$-periodic unitary
group on $L^{2}(Q, \Sigma,  \d\phi_{C})$. Hence it extends to a group of measure-preserving
automorphisms of $L^{\infty}(Q, \Sigma, \d \phi_{C})$ which is
continuous in measure.

Let $r$ be the (euclidean) time reflection around $t=0$. We denote 
by $R$ the measure preserving transformation of $(Q,
\Sigma, \d\phi_{C})$ generated by $r$.

For $h\in \cS_{\rr}(\rr)$ we have 
\beq
\label{zeke}
U(t)\phi(0, h)= \phi(t, h).
\eeq
Again by  Lemma \ref{1.0b}, we see that $\Sigma=\bigvee_{t\in
S_{\beta}}U(t)\Sigma_{0}$. 
Hence 
$(Q, \Sigma, \Sigma_{0},U(t), R, \d \phi_{C})$ 
is a generalized path space. Moreover, it
is $\beta$-periodic and 
OS-positive (see e.g.~\cite{KL2}). It describes the {\em free neutral scalar field of mass $m$ on $\rr$ at
temperature~$\beta^{-1}$}.

We now describe a well known  concrete form of the $\beta$-KMS system
associated to the generalized path space  $(Q, \Sigma,
\Sigma_{0},U(t), R, \d \phi_{C})$.
Let  $\ch:= H^{-\frac{1}{2} }(\rr)$ be the Sobolev
space of order $-\12$, equipped with its canonical complex
structure $\i$ and scalar product $(h_{1}, h_{2})=(h_{1},
(2\epsilon)^{-1}h_{2})_{L^{2}(\rr)}$, where $\epsilon= (D_{x}^{2}+
m^{2})^{\frac{1}{2} }$. On $\ch$ we consider the unitary dynamics
$\e^{-\i t\epsilon}$.

On the Weyl algebra $\cal {W}(\ch)$ we define a 
state $\omega^\circ_\beta $ and a one-parameter group of automorphisms
$\{\tau^{\circ}_{t}\}_{t\in \rr}$
by
\beq
\label{dyn}
\omega^{\circ}_\beta \bigl(W(h) \bigr):= \e^{-\frac{1}{4}(h, (1+ 2\rho)h)}, \quad 
\tau^{\circ}_{t}\bigl(W(h) \bigr):= W(\e^{\i t\epsilon}h), \:h\in \ch,\: t\in \rr,
\eeq
where $\rho:= ( \e^{\beta {\rm \epsilon}}-1)^{-1}$, $\beta>0$. It can
be easily seen that $\omega^{\circ}_\beta$ is a quasi-free
$(\tau^{\circ}, \beta)$-KMS state on~${\cal W}(\ch)$.

Let us now recall some terminology. If $\ch$ is a
complex vector space, then the {\em conjugate vector space} $\chbar$ is the
real vector space $\ch$ equipped with the complex structure $-\i$. We
will denote by $\ch\ni h\mapsto \overline{h}\in \ch$ the (anti-linear)
identity operator. If $a\in {\cal L}(\ch)$, then we denote by
$\overline{a}\in {\cal L}(\chbar)$ the operator
$\overline{a}\overline{h}:= \overline{ah}$. If $\ch$ is a Hilbert
space, then $\chbar$ is equipped with the Hilbert space 
structure~$(\overline{h_{1}}, \overline{h_{2}}):= (h_{2}, h_{1})$. 

\goodbreak

We recall a  convenient realization of the GNS representation associated to $\bigl( {\cal W}(\ch),
\omega^\circ_\beta \bigr)$, which is called the {\em  right Araki-Woods
representation\/}. It is specified by setting
\[
\begin{array}{l}
\cH_{\scriptscriptstyle AW}:= \G(\ch\oplus \chbar),\\[3mm]
\Omega_{\scriptscriptstyle AW}:= \Omega, \\[3mm]
\pi_{\scriptscriptstyle AW}(W(h))= W_{\scriptscriptstyle AW}(h):= W_{\rm F} \bigl( (1+ \rho)^{\frac{1}{2}}h\oplus
\overline{\rho}^{\frac{1}{2}}\overline{h} \bigr),\quad h\in 
\ch.\\[3mm]
\end{array}
\] 
Here  $W_{F}(.)$ denotes the Fock Weyl operator  on
$\G(\ch\oplus\chbar)$ and $\Omega\in \G(\ch\oplus \chbar)$ is the Fock
vacuum.

The physical Hilbert space associated to the path space $(Q, \Sigma,
\Sigma_{0},U(t), R, \d \phi_{C})$ can be
unitarily identified with $\G(\ch\oplus\overline{\ch})$. The distinguished vector ${\cal V}1$
 is identified with the Fock vacuum vector $\Omega$ in
$\G(\ch\oplus\overline{\ch})$. The Liouvillean $L_{\scriptscriptstyle AW}$
satisfies
\[ {\rm e}^{i L_{\scriptscriptstyle AW} t} \pi_{\scriptscriptstyle AW} (A) \Omega_{\scriptscriptstyle AW} =
\pi_{\scriptscriptstyle AW} \bigl(\tau_t^\circ(A)\bigr) \Omega_{\scriptscriptstyle AW} \quad \hbox{\rm and} \quad
L_{\scriptscriptstyle AW} \Omega_{\scriptscriptstyle AW} =0, \]
and can be identified with  $\d\Gamma(\epsilon\oplus
-\overline{\epsilon})$. 

The abelian von Neumann algebra
$\cU_{\scriptscriptstyle AW}$ obtained by the reconstruction theorem  can be identified  with
the abelian von Neumann algebra generated by
$\{W_{\scriptscriptstyle AW}(h) \mid  h\in
H^{-\12}_{\rr}(\rr) \}$. In fact, if $A=\e^{\i \phi(0, h)}$ for
$h\in \cS_{\rr}(\rr)$, then the 
operator $\tilde{A}$ defined in (\ref{identi}) is identified with  the
Weyl operator~$W_{\scriptscriptstyle AW}(h)= \e^{\i \phi_{\scriptscriptstyle AW}(h)}$ on~$\G(\ch\oplus\chbar)$.

The von Neumann algebra
$\cB_{\scriptscriptstyle AW}$ generated by $\bigcup_{t \in \rr}  \tau_t^\circ (\cU_{\scriptscriptstyle AW})$ can be identified  with
the von Neumann algebra $\cRb$ generated by
$\{W_{\scriptscriptstyle AW}(h) \mid  h\in
H^{-\12} (\rr) \}$.

\section{Perturbations of path spaces}

\init\label{perturbations}
In this section we describe perturbations of the two path spaces
defined in Subsects.~\ref{sec2.1} and~\ref{sec2.2a}, obtained from FKN
kernels corresponding to $P(\phi)_{2}$ interactions. 
\subsection{Interaction terms}
\label{sec2.2}
We recall some well known facts concerning  the Wick ordering of
Gaussian random variables. Let~$(K, \nu)$ be a probability space and   $X$ 
a real vector space equipped with a positive quadratic form~$f\mapsto
c(f,f)$ called a {\em covariance}. Let $f\mapsto \phi(f)$ be a $\rr$-linear map from $X$
into the space of real measurable functions on $K$.
 
The {\em Wick ordering} $\:  : \! \phi(f)^{n}\!:_c$ with respect to the
covariance $c$ is defined by the following generating series:
\beq
\label{wickdef}
:\!\e^{\alpha\phi(f)}\!:_{c} \: :
=\sum_{n=0}^{\infty}\frac{\alpha^{n}}{n!}:\!\phi(f)^{n}\!:_{c}=
\e^{\alpha\phi(f)}\e^{-\frac{\alpha^{2}}{2}c(f,f)}.
\eeq
Thus
\beq
\label{wick}
:\!\phi(f)^{n}\!:_{c}=
\sum_{m=0}^{[n/2]}\frac{n!}{m!(n-2m!)}\phi(f)^{n-2m} \Bigl(-\frac{1}{2}  c(f,f) \Bigr)^{m},
\eeq
where $[.]$ denotes the integer part. 

\goodbreak

\begin{lemma}
\label{wicko}
\quad
\vskip .3cm
\halign{  \indent \indent \indent #  \hfil & \vtop { 
\parindent =0pt 
\hsize=12cm                           \strut # \strut} \cr {\rm (i)}  &  For $f\in L^{1}(S_{\beta} \times \rr )\cap L^{2}(S_{\beta} \times \rr)$
the following  limit exists  
in $\kern -.2cm \bigcap \limits_{1\leq p<\infty} \kern -.2cm L^{p}(Q, \Sigma, \d\phi_{C})$:
\[
\: \lim_{(k, k')\to \infty}\int_{S_{\beta}\times \rr}f(t,x):\!
\phi \bigl(\delta_{k}(.-t)\otimes \delta_{k'}(.-x) \bigr)^{n}\!:_{C}\d t \d x .
\]
It will be denoted by $
\int_{S_{\beta}\times \rr}f(t,x):\!
\phi(t,x)^{n}\!:_{C}\d t \d x$.
\cr
{\rm (ii)}  &  For $h\in L^{1}(\rr)\cap L^{2}(\rr)$
the following limit exists  
in $\bigcap_{1\leq p<\infty}L^{p}(Q, \Sigma, \d\phi_{C})$:
\[
 \lim_{k\to \infty}\int_{\rr}h(x):\!\phi(0, \delta_{k}(.-x))^{n}\!:
_{C_{0}}\d x  . \]Ê
It will be denoted by $\int_{\rr} h(x) :\!\phi(0, x)^{n}\!:_{C_{0}}\d x $.
\cr
{\rm (iii)} &  For $g\in L^{1}(S_{\beta})\cap L^{2}(S_{\beta})$
the following limit exists  
in $\bigcap_{1\leq p<\infty}L^{p}(Q, \Sigma, \d\phi_{C}) $:

\[ \lim_{k' \to \infty}\int_{S_{\beta}}g(t):\!\phi(\delta_{k}(.-t), 0)^{n}\!:
_{C_{\beta}}\d t  . \]
It will be denoted by $\int_{S_{\beta}} g(t) :\!\phi(t, 0)^{n}\!:_{C_{\beta}}\d t $. \cr}
\end{lemma}
We recall that the covariances $C$, $C_{0}$ and $C_{\beta}$ have been
defined in (\ref{c}), (\ref{c0}) and (\ref{cbeta}), respectively. In Lemma
\ref{wicko} the probability space is $(Q, \Sigma, \d\phi_{C})$ and the real
vector spaces are equal to $\cS_{\rr}(S_{\beta}\times
\rr)$, $\cS_{\rr}(\rr)$ and  $\cS_{\rr}(S_{\beta})$, respectively.

\proof The proof is straightforward, adapting standard arguments (see e.g.~\cite{Si1}, \cite[Section 9]{GeJ}) used for the spatially cutoff $P(\phi)_{2}$ model at 
$0$-temperature  \qed . 
\begin{remark}
If $P=P(\lambda)$ is a polynomial, then the functions
\[
\int_{S_{\beta}\times \rr} \kern -.5cm f(t,x):\!
P(\phi(t,x))\!:_{C}\d t \d x, \: \:Ê
\int_{\rr} \kern -.1
cm h(x) :\!P(\phi(0,x))\!:_{C_{0}}\d
x \: 
\hbox{ and }\int_{S_{\beta}} \kern -.2cm g(t) :\!P(\phi(t,0))\!:_{C_{\beta}}\d t
\]
are well defined, by linearity. It can be easily shown (see \cite[Proposition~8.4]{GeJ}) using the
so-called {\em Wick
reordering identities} that there exists a linear  invertible map between
polynomials 
\[
 P\mapsto \tilde{P}
\]
with ${\rm deg}P={\rm deg}\tilde{P}$, ${\rm deg}(P-\tilde{P})\leq {\rm deg}(P)-1$ such that 
\[
\int_{\rr} h(x) :\!P(\phi(0,x))\!:_{C_{0}}\d x= \int_{\rr} h(x):\!
\tilde{P}(\phi(0, x))\!:_{\rm vac}\d x .
\]
Here $: $ $:_{\rm vac}$ denotes Wick ordering with respect to the
$0$-temperature covariance $(h, \frac{1}{2\epsilon}h)_{L^{2}(\rr)}$.
\end{remark}
\begin{lemma}
\label{2.2}
Let $P$ be a polynomial, $h\in L^{1}(\rr)\cap L^{2}(\rr)$ and $g\in L^{1}(S_{\beta})\cap
L^{2}(S_{\beta})$. Set 
\beq
\begin{array}{l}
V_0 (h) := \int_{\rr} h(x):\!P(\phi(0,x))\!:_{C_{0}}\d x,\\[3mm]
V_{\beta} (g) := \int_{S_{\beta}} g(t):\!P(\phi(t,0))\!:_{C_{\beta}}\d t,
\end{array}
\label{e2.10}
\eeq
as functions on $Q$. 
\goodbreak
Then
\beq \: \:
 \int_{S_{\beta}} \kern -.1cm g(t)U(t)V_0 (h) \d t = \int_{S_{\beta}\times \rr}
\kern -.5cm \bigl( g(t)\otimes h(x) \bigr) :\!P(\phi(t,x))\!:_{C}\d t \d
x
=\int_{\rr} \kern -.1cm h(x)U_{\tt C} (x)V_{\beta} (g) \d x \\
\label{e1.11}
\eeq
as functions on  Q.
\end{lemma}
\proof Let $W$ be a function in $L^{p}(Q, \Sigma, \d \phi_{C})$ for some
$1\leq p<\infty$. The one parameter groups~$\{ÊU(t) \}_{t \in S_{\beta}}$
and~$\{ U_{\tt C}(x) \}_{x \in \rr}$ are strongly continuous groups of isometries of
$\bigcap_{1\leq p<\infty}L^{p}(Q, \Sigma, \d\phi_{C})$. Therefore the
functions $\int_{\rr}h(x)U_{\tt C}(x)W\d x$ and
$\int_{S_{\beta}}g(t)U(t)W \d t$ belong to $L^{p}(Q,\Sigma,
\d\phi_{C})$.

Together with Lemma \ref{wicko} this implies that all three functions given in
(\ref{e1.11}) belong to~$L^{p}(Q, \Sigma, \d\phi_{C})$. Let us now prove that they are identical. By linearity, we
may assume that $P(\lambda)= \lambda^{n}$.
Using Lemma \ref{wicko} and the Wick identity (\ref{wick}), it follows that 
\[
\int_{S_{\beta}\times \rr}
\bigl( g(t)\otimes h(x) \bigr)  :\!P(\phi(t,x))\!:_{C}\d t \d x=\lim_{(k, k')\to \infty}
F(k, k')\hbox{ in }L^{p}(Q, \Sigma, \d\phi_{C}),
\]
where
\[
F(k, k')= \sum_{m=0}^{[n/2]} \frac{n! \bigl( -\frac{1}{2} C(\delta_{k,
k'}, \delta_{k, k'}) \bigr)^{m}}{m!(n-2m)!} \int_{S_{\beta}\times \rr}  \bigl( g(t)\otimes h(x) \bigr)
\phi \bigl(\delta_{k}(.-t)\otimes \delta_{k'}(.-x) \bigr)^{m}\d t \d x 
\]
and $\delta_{k, k'}(t,x):= \delta_{k}(t)\otimes \delta_{k'}(x)$. Since
\[
\lim_{k\to \infty}C(\delta_{k,k'}, \delta_{k, k'})= C_{0}(
\delta_{k'}, \delta_{k'}), 
\]
the definition given in (\ref{e1.4}) of sharp-time fields implies that 
\[
 \lim_{k\to \infty}F(k, k')=  \int_{S_{\beta}} g(t)V_{k'}(t,h)\d
t \: \hbox{ in } \: L^{p}(Q, \Sigma, \d\phi_{C}), 
\]
where
\[
V_{k'} (t, h) = \sum_{m=0}^{[n/2]} \frac{n!}{m!(n-2m)!} \Bigl(-\frac{1}{2} 
C_{0}(\delta_{k'}, \delta_{k'}) \Bigr)^{m}\int_{\rr}h(x)
\phi \bigl(t, \delta_{k'}(.-x) \bigr)^{m}\d x.
\]
Note that (\ref{zeke}) implies $V_{k'}(t, h)=U(t)V_{k'}(0, h)$.
By Lemma \ref{wicko} {\rm (ii)} we know that 
\[
\lim_{k'\to \infty}V_{k'}(0, h)= \int_{\rr}h(x):\!P(\phi(0, x))\!:
_{C_{0}}\d
x  \: \hbox{ in } \: L^{p}(Q, \Sigma, \d \phi_{C})
\]
and hence 
\[
 \lim_{k\to \infty}\int_{S_{\beta}} g(t)V_{k'}(t,h)\d t=
\int_{S_{\beta}} g(t)U(t)V_{0}(h)\d t  \: \hbox{ in } \: L^{p}(Q, \Sigma,  \d \phi_{C}).
\]
Applying Lemma \ref{a1} with $E= L^{p}(Q, \Sigma, \d \phi_{C})$ we
obtain the first identity in (\ref{e1.11}).  The second identity follows by the
same argument, taking first the limit $k'\to \infty$ and using then
that
\[
\lim_{k'\to \infty}C(\delta_{k,k'}, \delta_{k, k'})= C_{\beta}(
\delta_{k}, \delta_{k})\: \Box.  
\]

\subsection{The $P(\phi)_{2}$ model on the circle $S_{\beta}$ at  temperature $0$}
\label{sec2.4a}
Let $P(\lambda)$ be a real valued polynomial, which is bounded from below. The {\em $P(\phi)_{2}$
model on the circle~$S_{\beta}$} is specified by the formal interaction term
\[
V_{\tt C}:= V_{\beta} \bigl(1_{[- \beta/2, \beta/2]} \bigr) =\int_{S_{\beta}}:\!P(\phi(t, 0))\!:_{C_{\beta}}\d t.
\]
This expression can be given two equivalent meanings: first of all, as
recalled in Lemma \ref{wicko}, it can be viewed as a
$\Sigma_{0}^{\tt C}$ measurable function $V_{\tt C}\in \bigcap_{1\leq
p<\infty}L^{p}(Q, \Sigma_{0}^{\tt C}, \d\phi_{C})$. Secondly, $V_{\tt C}$~can be considered
as a selfadjoint operator on $\Gamma\bigl(H^{-\12}(S_{\beta}) \bigr)$ affiliated to the
abelian algebra $\cU_{\tt C}$. More precisely, for $t\in S_{\beta}$ and
$\Lambda\gg 1$ an UV cutoff parameter, we define an approximation 
$ h_{\Lambda, t} \in H^{-\12}(S_{\beta})$ of the Dirac delta-function $\delta (.-t) \in  H^{-\12}(S_{\beta})$ by
\[
 h_{\Lambda, t}:=\one_{[0, \Lambda]}(b)\delta(.-t)\in H^{-\12}(S_{\beta}),
\]
where  $b=(D_{t}^{2}+ m^{2})^{\12}$. Setting
$\phi_{\Lambda}(t,0):= \phi_{F}(h_{\Lambda, t})$  one obtains by well-known
arguments that
\[
V_{\tt C}=\lim_{\Lambda \to \infty}\int_{S_{\beta}}:
\!P(\phi_{\Lambda}(t, 0))\!:_{C_\beta}\d t
\]
on a dense set of vectors in $\Gamma \bigl(H^{-\12}(S_{\beta}) \bigr)$. Since
$h_{\Lambda, t}\in H^{-\12}_{\rr}(S_{\beta})$ is a real valued function, it is easy to see that
$V_{\tt C}$ is a selfadjoint operator affiliated to $\cU_{\tt C}$.

It is then easy to verify, by adapting well-known results for the
spatially cutoff~$P(\phi)_{2}$ model on the real line $\rr$ at $0$-temperature (see
\cite{SHK}) that $V_{\tt C}\in \bigcap_{1\leq p<\infty}L^{p}(Q,
\Sigma_{0}^{\tt C}, \d\phi_{C})$ and $\e^{-TV_{\tt C}}\in L^{1}(Q,
\Sigma_{0}^{\tt C}, \d\phi_{C})$ for all $T>0$. Now consider, for $0\leq b-a<\infty$,
\beq
\label{e3.1}
G_{[a,b]}:=\e^{-\int_{a}^{b}U_{\tt C}(x)V_{\tt C} \: \d x}
\eeq
as a function on $Q$. It
follows from Jensen's inequality (see \cite[Theorem 6.2]{KL0}) that
\beq
\|G_{[a,b]}\|_{L^{p}(Q, \Sigma,  \d \phi_{C})}\leq
\|\e^{-(b-a)V_{\tt C}}\|_{L^{p}(Q, \Sigma,  \d \phi_{C})}
\label{e6.1b}
\eeq
and hence $G_{[a,b]}\in \bigcap_{1\leq p<\infty}L^{p}(Q, \Sigma, \d
\phi_{C})$.
From the results recalled in Subsection~\ref{sec0.3}, we obtain a
selfadjoint operator 
\[
H_{\tt C}=\overline{\d\Gamma(b)+ V_{\tt C}}\] 
on $\Gamma \bigl(H^{-\12}(S_{\beta}) \bigr)$
associated to the FKN kernel $\{G_{[0,s]}\}$. The Hamiltonian
$H_{\tt C}$ is called the~{\em $P(\phi)_{2}$ Hamiltonian on the circle
$S_{\beta}$}. 

\begin{proposition}
The Hamiltonian $H_{\tt C}$ is bounded from below
and has a  unique normalized ground state  such that
$(\Omega_{\tt C}, \Omega)>0$. We set
\[
\omega_{\tt C} ( \: . \: )= (\Omega_{\tt C}, \: . \: \Omega_{\tt C}).
\]
Moreover, for $c\gg 1$,
\beq
\label{e3.1b}
\|\phi_F(g)(H_{\tt C}+ c)^{-\12}\|\leq C\|g\|_{H^{-\12}(S_{\beta})},
\eeq
\beq
\label{e3.1c}
\pm \phi_F(g)\leq C \|g\|_{H^{-\12}(S_{\beta})}(H_{\tt C}+c)^{\12}
\eeq
and
\beq
\label{e3.1d}
 \pm \phi_F(g)\leq C\|g\|_{H^{-1}(S_{\beta})}(H_{\tt C}+c)
\eeq
for all $g\in H^{-\12}(S_{\beta})$. As before, $W_F(g)= \e^{\i \phi_F (g)}$ 
is the Fock
Weyl operator on~$\Gamma \bigl(H^{-\12}(S_{\beta}) \bigr)$.
\label{3.1}
\end{proposition}

\proof
The existence and uniqueness of the vacuum state  can be shown by following the
proofs of the corresponding results for spatially cutoff~$P(\phi)_{2}$
models. For example, one easily obtains (see e.g.~\cite[Theorem~V.20]{Si1} or \cite[Theorem~6.4 (ii)]{DG}) that
\beq
\label{e3.1bb}
(\dG(b)+1)\leq C (H_{\tt C}+ c)  \: \: \hbox{for} \: \: c\gg 1.
\eeq 
Since $\dG(b)$ has compact resolvent on $\Gamma \bigl(H^{-\12}(S_{\beta}) \bigr)$,
it follows that $H_{\tt C}$ is bounded from below with a compact resolvent and hence has a
ground state. The uniqueness of the vacuum  (i.e., the ground state of
$H_{\tt C}$) follows from 
a Perron-Frobenius argument (see e.g.~\cite[Theorem~V.17]{Si1}). 
Since $b\geq m>0$, we  see  that
it suffices to check (\ref{e3.1b}) and (\ref{e3.1c}), with
$H_{\tt C}$ replaced by the number operator $N$, which is immediate.
To prove (\ref{e3.1d}) we use   (\ref{e3.1bb}) and the well known 
bound (see e.g. \cite[Appendix]{Ge})
\[
 \pm \phi_F (g)\leq \|b^{-\frac{1}{2}
}g\|_{H^{-\12}(S_{\beta})}(\d\Gamma(b)+ 1) \: \Box .
\]

\goodbreak
Without proof we quote the following result (see \cite{HO}).

\begin{theoreme} Let $H_{\tt C}^{\rm ren}:=
H_{\tt C}-E_{\tt C}$, where  $E_{\tt C}:=\inf(\sigma(H_{\tt C}))$ and let $P_{\tt C}$ denote the generator of the 
 translations along the circle $S_{\beta}$.
The joint spectrum of $H_{\tt C}^{\rm ren} $ and $P_{\tt C}$ is purely discrete and is contained in the forward light cone.
\end{theoreme}

Consequently the correlation function
\[ (t,x) \mapsto \bigl(\Omega_{\tt C},  A {\rm e}^{i x H_{\tt C}^{\rm ren}+ i tP_{\tt C} } B \Omega_{\tt C} \bigr), 
\quad A, B \in \cB \bigl(\Gamma (H^{-\12}(S_{\beta})) \bigr), \]
allows an analytic continuation to the tube $\rr^2 + i V_+$, where $V_+
:= \{Ê(t, x) \mid | t | < x; \:  x>0 \} $ denotes the forward light cone (with $t$ and $x$
reversed, due to our conventions).

\subsection{The spatially cutoff $P(\phi)_{2}$  model on $\rr$ at temperature
$\beta^{-1}$}
\label{sec2.4}
Let $P(\lambda)$ be a real valued polynomial, which is bounded from below (as in Subsection~\ref{sec2.4a}), and let $l\in
\rr^{+}$ be a spatial cutoff parameter. The {\em spatially cutoff
$P(\phi)_{2}$ model on $\rr$} is  specified by the formal interaction term (see (\ref{e2.10}))
\[
V_{l}:= V_0 \bigl (\one_{[-l, l]} \bigr) =\int_{-l}^{l} :\! P(\phi(0, x))\!:_{C_{0}}\d x.
\]
Again this formal expression can be given two  equivalent meanings: 
first
of all, as recalled in Lemma~\ref{wicko}, it can be viewed as a
$\Sigma_{0}$-measurable
function $V_{l} \in
\bigcap_{1\leq p<\infty}L^{p}(Q, \Sigma_{0}, \d\phi_{C})$. 
Secondly, $V_{l}$~can be considered as a selfadjoint operator on
$\Gamma(\ch\oplus\overline{\ch})$ affiliated to the abelian von
Neumann algebra $\cU_{\scriptscriptstyle AW}$. As in Subsection~\ref{sec2.4a}  
we define an approximation 
$ h_{\Lambda, x} \in H^{-\12}(\rr)$ of the Dirac delta-function $\delta (.-x) \in  H^{-\12}(\rr)$.
For $x\in \rr$ and $\Lambda\gg 1$ we set
\[
h_{\Lambda, x}:= \one_{[0, \Lambda]}(\epsilon)\delta(.-x)\in
H^{-\frac{1}{2} }(\rr)
\]
and introduce cutoff fields $\phi_{\Lambda}(0,x):= \phi_{\scriptscriptstyle AW}(h_{\Lambda, x})$, where 
$\phi_{\scriptscriptstyle AW}(h)$ is the selfadjoint field operator
associated to $W_{\scriptscriptstyle AW}(h)$,
$h\in \ch$.

As before, the limit
\beq
\label{interact}
V_{l}= \lim_{\Lambda \to \infty}\int_{-l}^{l}:\!P(\phi_{\Lambda}(0,x))\!:
_{C_{0}}\d x
\eeq
exists on a dense set of vectors in $\Gamma(\ch\oplus\overline{\ch})$. Since $h_{\Lambda,
x}\in H^{-\frac{1}{2} }_{\rr}(\rr)$, one obtains that  $V_{l}$ is a
selfadjoint operator affiliated
to ${\cal U}_{\scriptscriptstyle AW}$. 

Adapting well known arguments (see
\cite[Section~8.2]{GeJ})  it can be shown that~$\e^{-TV_{l}}\in L^{1}(Q, \Sigma_{0}, \mu)$ for
all~$T>0$. Consequently, we can associate to $V_{l}$ the FKN kernel
\[
F_{[a,b]}^{l}:=\e^{-\int_{a}^{b} U(t)V_{l}\d t}, \quad 0\leq b-a\leq
\beta,
\]
and the measure 
\[ \d\mu_{l}:=\frac{F^l_{[-\beta/2,\beta/2]}\d\phi_{C} }{\int_{Q}F^{l}_{[-\beta/2,
\beta/2]}\d\phi_{C}}. \] 
The
generalized path space $(Q, \Sigma, \Sigma_{0}, U(t), R,
\mu_{l})$ is $\beta$-periodic and OS-positive. The associated
$\beta$-KMS system is called the {\em spatially cutoff~$P(\phi)_{2}$ model
on $\rr$ at temperature $\beta^{-1}$}. Applying the abstract results
recalled in Subsection~\ref{sec0.3}, we obtain the following facts:
\begin{itemize}
\item[--] the physical Hilbert space  $\cH_{V_{l}}$ is equal to $\cH_{\scriptscriptstyle AW}Ê= \Gamma(\ch\oplus
\overline{\ch})$;
\item[--] the $W^{*}$-algebra ${\cal B}_{V_{l}}$ and the abelian algebra ${\cal
U}_{V_{l}}$ are equal to $\cRb$ and ${\cal U}_{\scriptscriptstyle AW}$, respectively;
\item[--] the operator sum $L_{\scriptscriptstyle AW}+V_{l}$ is essentially selfadjoint on $\cD(L_{\scriptscriptstyle AW}) \cap \cD (V_{l})$ 
and  if
$H_{l}:=\overline{L_{\scriptscriptstyle AW}+V_{l}}$, then the perturbed time-evolution on
${\cal B}$ is given by $\tau^{l}_{t}(B):=\e^{\i t
H_{l}}B\e^{-\i tH_{l}}$, $B \in \cB$;
\item[--] the GNS vector  $\Omega_{\scriptscriptstyle AW}Ê\in \Gamma(\ch\oplus\overline{\ch})$ belongs to
$\cD\bigl(\e^{-\frac{\beta}{2}H_{l}}\bigr)$ and the perturbed KMS state~$\omega_{l}$
is given by  $\omega_{l}(B)= (\Omega_{l} ,
B\Omega_{l} )$, where $\Omega_{l} :=
\|\e^{-\frac{\beta}{2}H_{l}}\Omega_{\scriptscriptstyle AW}Ê\|^{-1}
\e^{-\frac{\beta}{2}H_{l}}\Omega_{\scriptscriptstyle AW}Ê$.
\end{itemize}

\noindent
The following consequence of Lemma \ref{2.2} will be important in Section~\ref{sec3}:
\beq
F_{[-\beta/2,\beta/2]}^{l}= G_{[-l, l]},
\label{e3.2}
\eeq
where $G_{[a,b]}$ was defined in (\ref{e3.1}).  The analog identity in the temperature zero case  is called {\it Nelson symmetry}.

\section{The thermodynamic limit}
\init\label{sec4}
In this section we prove that the limits
\[
\lim_{l\to +\infty}\tau^{l}_{t}(A)=: \tau_{t}(A) \quad \hbox{and} \quad \lim_{l\to
+\infty}\omega_{l}(A)=: \omega_\beta (A)
\]
exist for $A$ in the $C^{*}$-algebra  of local observables $\cA$ and that
$(\cA, \tau , \omega_\beta)$ is a $\beta$-KMS system, describing
the {\em translation invariant $P(\phi)_{2}$ model at temperature
$\beta^{-1}$}. 
\subsection{Preparations}
\label{sec4.1}
 We first recall a well known relationship between $\e^{-\i
t \epsilon}$ and the Klein-Gordon equation: let
\beq
\matrix{
U\colon & H^{-\frac{1}{2} }(\rr) & \to & H^{-\frac{1}{2} }_{\rr}(\rr)\oplus H^{\frac{1}{2} }_{\rr}(\rr).  
\cr
&  h & \mapsto & ({\rm Re}h, \epsilon^{-1}{\rm
Im}h)=(\varphi, \pi).\cr} 
\eeq
(Note that $U$ is $\rr$-linear but not $\cc$-linear).
Then 
\beq
\label{ekg2}
U\e^{-\i t\epsilon}= T(t)U, \hbox{ where }T(t)(\varphi, \pi)= (\varphi_{t},
\p_{t}\varphi_{t}),
\eeq
 and $\varphi_{t}$ is the solution of the Klein-Gordon
equation
\[
\left\{
\begin{array}{l}
\bigl(\p_{t}^{2}-\p_{x}^{2} + m^{2} \bigr) \varphi_{t}=0,\\[2mm]
\varphi_{t=0}= \varphi, \quad
\bigl( \p_{t}\varphi \bigr)_{t=0}=-\epsilon^{2}\pi.
\end{array}
\right.
\]
Moreover if $h_{i}\in H^{-\frac{1}{2} }(\rr)$ and $Uh_{i}= (\varphi_{i}, \pi_{i})$
for $i=1,2$, then 
\beq
\label{ekg3}
\sigma(h_{1}, h_{2}):={\rm Im}(h_{1}, h_{2})_{H^{-\frac{1}{2} }(\rr)}=
\int_{\rr}\varphi_{1}(x)\pi_{2}(x)- \pi_{1}(x)\varphi_{2}(x)\d x.
\eeq
For $I\subset \rr$ a bounded open interval we define the  real vector
subspace $\ch_{I}$ of $\ch$ 
\beq
\ch_{I}:=\{h\in \ch \mid  \supp Uh\subset I \times I\}.
\label{ekg4}
\eeq
It follows from (\ref{ekg2}) that $\i \epsilon \colon \cD(\epsilon)\cap
\ch_{I}\to \ch_{I}$, and hence $(1+ \alpha\epsilon^{2} )^{-1} \colon\ch_{I}\to
\ch_{I}$ for~$\alpha>0$. In particular $\cD(\epsilon)\cap \ch_{I}$ is
dense in $\ch_{I}$. Moreover  (\ref{ekg3}) shows that $\ch_{I}$
and $\ch_{J}$ are orthogonal for the symplectic form $\sigma$ if
$I\cap J=\emptyset$.
\subsection{The net of local algebras}
\label{sec4.1a}
We start by recalling  a result of Araki \cite[Thm.~1]{Ar} which will be
useful later on. Let us recall a standard notation: If $\cH_{1}$,
$\cH_{2}$ are two vector subspaces of a Hilbert space $\cH$, then $\cH_{1}\vee \cH_{2}$ denotes
$\overline{\cH_{1}+\cH_{2}}$. 
If ${\cal R}_{1}$,
${\cal R}_{2}$ are two 
$^{*}$-sub-algebras of $\cB(\cH)$, then ${\cal R}_{1}\vee {\cal R}_{2}$ denotes
the von Neumann algebra generated by ${\cal R}_{1}\cup {\cal R}_{2}$. 
\begin{proposition}
\label{araki}
Let $X$ be a
Hilbert space and let $Z$ be a real vector subspace of $X$. Let 
${\cal W}(Z)\subset {\cal W}(X)$ denote the $C^*$-algebra generated by $\{W(x) \mid
x\in Z\}$  and let $\pi_{F}: {\cal W}(X)\to {\cal B}(\Gamma(X))$ be
the Fock representation.  Then
\beq
\bigcap_{\alpha} \pi_F  \bigl( {\cal W}(Z_{\alpha}) \bigr)''= \pi_F  \bigl({\cal W}\left(\cap_{\alpha}
Z_{\alpha} \right) \bigr)'', \:\: \:\:
\bigvee_{\alpha}\pi_F  \bigl({\cal W}(Z_{\alpha})\bigr)''= \pi_F 
\bigl({\cal W}\left(\vee_{\alpha} Z_{\alpha} \right) \bigr)''
\label{et.2}
\eeq
and
\beq
\pi_F  \bigl({\cal W}(Z)\bigr)'= \pi_F  \bigl({\cal W}(Z^{\perp})\bigr)'',
\label{et.1}
\eeq
where $Z_{\alpha}$ is a family of real vector subspaces of $X$ and $Z^{\perp}$ is the vector space 
orthogonal to~$Z$ 
for the symplectic
form $\sigma(x_{1}, x_{2})={\rm Im}(x_{1}, x_{2})$.
\end{proposition}

We now define the {\em net of local von Neumann algebras} $I \to \cRb (I)$ describing free thermal 
scalar bosons. Let $I\subset \rr$ be a bounded open interval. We denote by $\cRb (I)$ the von Neumann 
algebra generated by
\[
\{W_{\scriptscriptstyle AW}(h) \mid h\in \ch_{I}\}. 
\]

\begin{lemma} \quad
\vskip 0cm
\label{l6.1}
\halign{ \indent \indent \indent #  \hfil & \vtop { 
\parindent =0pt 
\hsize=12cm
                            \strut # \strut} \cr 
{\rm (i)}  & The local von Neumann algebras for the free thermal field are regular from the inside and regular from the outside:
\[ \bigcap_{J\supset \overline{I}}\cRb(J)=\cRb(I)= \bigvee_{\overline{J}\subset
I}\cRb(J); \]
\vskip -.3cm
\cr
{\rm (ii)}  & The net of local von Neumann algebras for the free thermal field is {\rm additive}:
\[ \cRb(I)= \bigvee_{J_i} \cRb(J_i) \hbox{ if }  I = \cup_i J_i ;  \]
\vskip -.3cm
\cr
{\rm (iii)}  & For each open and bounded interval $I$, the local observable algebra $\cRb(I)$ is $*$-isomorphic to the
unique hyper-finite factor of type~III$_1$. \cr}
\end{lemma}
\proof
Recalling the definition of the Araki-Woods representation we see that, with
the notation introduced above,
\[
 \cRb(I)= \pi_{\scriptscriptstyle F} \bigl({\cal W}(Z_{I}) \bigr)'',
\]
where $Z_{I}\subset \ch\oplus\overline{\ch}$ is the
vector subspace
\[
Z_{I}=\{(1+\rho)^{\12}h\oplus\overline{\rho}^{\12}\overline{h} \mid
h\in \ch_{I}\}.
\]
Clearly $\bigcap_{J\supset \overline{I}}Z_{J}=\bigvee_{\overline{J}\subset I}Z_{J}=Z_{I}$,
which using (\ref{et.2}) implies (i). Part (ii) is a direct consequence of (\ref{et.2}). To prove (iii) we use
(\ref{et.1}) and (\ref{et.2}) which implies that 
\[
\cRb(I)\cap \cRb(I)'=  \pi_{\scriptscriptstyle F} \bigl({\cal W}(Z_{I}\cap Z_{I}^{\perp} \bigr)'',
\]
where $Z^{\perp}$ is the orthogonal space to $Z$ in
$\ch\oplus\overline{\ch}$ for the symplectic form $\sigma(f,g)={\rm
Im}(f,g)$ on~$\ch\oplus\overline{\ch}$. We claim that
\beq
Z_{I}\cap Z_{I}^{\perp}=\{0\},
\label{et.3}
\eeq
which will imply that $\cRb(I)$ is a factor. To prove our claim we pick $h\in \ch_{I}$ such that
$(1+\rho)^{\12}h\oplus\overline{\rho}^{\12}\overline{h}\in
Z_{I}^{\perp}$. This implies that ${\rm Im}(h, g)=0$ for all $g\in
\ch_{I}$. Hence to prove (\ref{et.3}) it suffices to check that
\beq
\ch_{I}\cap \ch_{I}^{\perp}=\{0\}.
\label{et.3b}
\eeq
But if $h\in \ch_{I}\cap \ch_{I}^{\perp}$, we have ${\rm Im}(h,
\i \epsilon (1+ \alpha \epsilon^{2})^{-1}h)=0$ for $\alpha>0$, 
since $\i \epsilon  (1+ \alpha\epsilon^{2} )^{-1}h \in \ch_{I}$
for $ h \in \ch_{I}$.  
Letting $\alpha\to 0$ this yields ${\rm Re}(h,\epsilon h)= (h,
\epsilon h)=0$, since $\epsilon$  is selfadjoint. Using that
$\epsilon\geq m>0$ this implies that $h=0$, which proves (\ref{et.3b})
and hence (\ref{et.3}).
Thus $\cRb(I)$ is a factor, 
if $I$ is bounded. Note that (\ref{et.3b}) shows that $\pi_{F}({\cal
W}(\ch_{I}))''$ is a factor, and it is well known 
(see e.g.~\cite{BD'AF}\cite{L} and lit.\ cit.) that  $\pi_{F}({\cal
W}(\ch_{I}))''$
is $*$-isomorphic to the
unique hyper-finite factor of type~III$_1$.
Thus Lemma~\ref{localfock} below completes the proof of the lemma  \qed .

We now recall an easy fact about the restriction of the free KMS state
$\omega_{\beta}^{\circ}$ to the local algebras ${\cal W}(\ch_{I})$. 
\begin{lemma}
\label{localfock}
Let $I\subset \rr$ be a bounded open interval. Then the representations
$\pi_{\scriptscriptstyle AW}$ and~$\pi_{F}$ of~${\cal W}(\ch_{I})$ are
quasi-equivalent.
\end{lemma}
\proof 
Let $\ch$ be a Hilbert space and let $\epsilon\geq m>0$ be a  positive
selfadjoint operator on~$\ch$. Let~$\omega_{\beta}^{\circ}$ be the quasi free state on
${\cal W}(\ch)$ defined by $\omega_{\beta}^{\circ} \bigl(W(h) \bigr)= \e^{-\frac{1}{4}(h, (1+
2\rho)h)}$, where $\rho=(\e^{\beta\epsilon}-1)^{-1}$. Then it is well known that $\omega^{\circ}_{\beta}$ is
normal with respect to the Fock representation~$\pi_{F}$ of~${\cal
W}(X)$ iff ${\rm Tr}\:\e^{-\beta\epsilon}<\infty$ (see e.g. \cite[Prop.
5.2.27]{BR}).

This fact implies  that if $\ch_{1}\subset\ch$ is a complex vector
subspace, then the restriction of $\omega^{\circ}_{\beta}$ to ${\cal
W}(\ch_{1})$ is $\pi_{F}$-normal iff ${\rm
Tr}(E\e^{-\beta\epsilon}E)<\infty$, where $E$ is the orthogonal
projection onto~$\ch_{1}$.

We will apply this remark to $\ch=H^{-\12}(\rr)$, $\rho=(\e^{\beta
\epsilon}-1)^{-1}$ and $\ch_{1}=\cc\ch_{I}$. 
 Let $E_{I}$ denote the orthogonal projection on $\cc\ch_{I}$. Let  $\chi\in
C^{\infty}_{0\:\rr}$ such that $\chi\equiv 1$ near $I$ and  $x=\i \p_{k}$. If $h\in
\ch_{I}$, then ${\rm Re}h= \chi(x){\rm Re}h$ and ${\rm Im}h=
\epsilon\chi(x)\epsilon^{-1}{\rm Im}h$. Using pseudodifferential
calculus, we see that the operators $(1+|x|)^{N}\chi(x)$ and
$(1+|x|)^{N}\epsilon \chi(x)\epsilon^{-1}$  are bounded on
$H^{-\12}(\rr)$ for all~$N\in \nn$. This implies that
 \beq
\label{stup}
\|(1+|x|)^{N}h\|_{H^{-\12}(\rr)}\leq C\|h\|_{H^{-\12}(\rr)},  \: \: h\in
\ch_{I}.
\eeq
Clearly (\ref{stup}) extends to $\cc\ch_{I}$,
which implies that $(1+|x|)^{N}E_{I}$ is bounded for all $N\in \nn$.
Since $\e^{-\beta\epsilon}(1+|x|)^{-N}$ is trace
class for $N$ large enough we see that $E_{I}\e^{-\beta \epsilon} E_{I}$ is trace
class. 
Using the arguments given above we obtain that $\omega^{\circ}_{\beta}$ restricted to
${\cal W}(\cc \ch_{I})$ (and hence also to ${\cal W}(\ch_{I})$) is
$\pi_{F}$-normal. 

Finally we have seen in the proof of Lemma \ref{l6.1} that
$\pi_{F}({\cal W}(\ch_{I}))''$ is a factor, hence $\pi_{F}$ is a
factor representation of ${\cal W}(\ch_{I})$. It is shown in
\cite[Prop. 10.3.14]{KR} that if ${\cal R}$ is a $C^{*}$-algebra and
$\pi$ is a factor representation of ${\cal R}$, then $\pi$ is
quasi-equivalent to the GNS representation of any $\pi$-normal state
$\omega$. Since the restriction of $\pi_{\scriptscriptstyle AW}$ to
${\cal W}(\ch_{I})$ is the GNS representation for the quasi-free state
$\omega^{\circ}_{\beta}$, this completes the proof of the lemma \qed .
 
\goodbreak

\subsection{Existence of the limiting dynamics}
\label{sec4.2}
The $C^*$-algebra of local observables $\cA$ is defined as follows:
\[
{\cal A}:=\overline{\bigcup_{I \subset \rr} \cRb(I)}^{(*)} ,
\]
where the union is over all open bounded intervals $I\subset \rr$ and
the symbol $\overline{\bigcup_{I \subset \rr} \cRb(I)}^{(*)}$ denotes the $C^{*}$-inductive limit (see e.g.~\cite[Proposition 11.4.1.]{KR}).

We denote by $\{\alpha_{x}\}_{x\in \rr}$ the group of
space translations on~$\cA$, defined by
\[
\alpha_{x} \bigl(W_{\scriptscriptstyle AW}(h) \bigr):= W_{\scriptscriptstyle AW}(\e^{\i x.k}h), \:  \: x\in \rr,
\]
where $k$ is the momentum operator acting on $\ch=H^{-\12}(\rr)$.

\begin{theoreme}
\label{existencedyn} 
{\rm (Existence of limiting dynamics).}
Let $I \subset \rr $ be a  bounded open interval. For~$t\in \rr$ fixed,
the norm limit
\[
\lim_{l\to \infty}\tau_{t}^{l}(B)=: \tau_{t}(B)\]
exists
for all $B\in \cRb(I)$. The map $\tau \colon t \mapsto \tau_t$  
defines a group of $*$-automorphisms of ${\cal
A}$ such that $\tau_{t} \circ \alpha_{x}=\alpha_{x} \circ \tau_{t} $ for all $t, x\in \rr$. Moreover,  
\beq
 \tau_{t} \colon  \cRb(I)\to \cRb \bigl( I+]-t, t[ \bigr).
\label{e6.1f}
\eeq
\end{theoreme}
\proof 
The proof follows the well-known proof in  
the $0$-temperature case, which is based on finite propagation speed
(see \cite[Theorem 4.1.2]{GJ2}). 
To prove the  existence of
the limit and the group property, it suffices to  show that
$\tau^{l}_{t}(B)$, for $B\in \cRb(I)$ and $|t|\leq T$,
is independent of $l$ for $l>|I|+ T$.
 
It follows from (\ref{ekg2}) and Huygens principle 
that \beq
\label{e6.1d}
\tau^{\circ}_{t} \colon \cRb(I)\to \cRb( I+]-t, t[).
\eeq
Moreover (\ref{ekg3}) implies that $\cRb(I_{1})\subset \cRb(I_{2})'$, if $I_{1}\cap I_{2}= \emptyset$.

The dynamics
$\tau^{l}_{t}$ is unitarily implemented by $\e^{\i tH_{l}}$, where
$H_{l}=\overline{L_{\scriptscriptstyle AW}+ V_{l}}$ for
\[
 V_{l}=\int_{]-l, l[}:\!P(\phi(0,x))\!:_{C_0} \d x.
\]
Trotter's formula yields $\e^{\i
tH_{l}}=\slim_{n\to\infty}(\e^{\i tL_{\scriptscriptstyle AW}/n}\e^{\i tV_{l}/n})^{n}$ and
hence
\beq
 \tau^{l}_{t}(A)=\slim_{n\to \infty} \bigl(\tau^{\circ}_{t/n}\circ
\gamma^{l}_{t/n} \bigr)^{n}(A), \: \: A\in \cB(\cH_{\scriptscriptstyle AW}),
\label{e6.1e}
\eeq
where $\gamma^{l}_{t}(A) := \e^{\i  tV_{l}}A\e^{-\i tV_{l}}$. Note
that for $l'>l$  
\[
V_{l'}= V_{l}+ \int_{]-l', l'[ \: \setminus \: ]-l, l[}:\!P(\phi(0,x))\!:_{C_0}\d x.
\]
Since $V_{l'}- V_{l}$ is affiliated to $\cRb \bigl( ]-l', l'[ \: \setminus \: [-l,
l] \bigr)$, we see that $\gamma^{l}_{t}= \gamma^{l'}_{t}$ on $\cRb(I)$ for
$l, l'>|I|$. Using (\ref{e6.1d}) and (\ref{e6.1e}), this implies that
$\tau^{l}_{t}= \tau^{l'}_{t}$ on $\cRb(I)$ for $|t|\leq T$ and $l,
l'>|I|+T$.  This proves our claim. The same argument using again
(\ref{e6.1d}) proves (\ref{e6.1f}). 

It remains to check that
$\tau$ and $\alpha$ commute. Let $T>0$ and $I$ a
bounded interval. For~$|t|\leq T$ the time evolution is locally (i.e., applied to elements in  $\cRb(I)$) generated 
by~$H_{l}$ if $l>|I|+t$. Now~$\alpha_{x}$ is
implemented by $\e^{\i xP}$ with $P=\d\Gamma(k\oplus \overline{k})$.
It follows that $\alpha_{x} \circ \tau_{t}  \circ \alpha_{x}^{-1}$ is implemented by~$\e^{\i
tH_{l, x}}$ with $H_{l,x}=\e^{\i xP}H_{l}\e^{-\i xP}$. It is
easy to see that 
\[
H_{l,x}= L_{\scriptscriptstyle AW} + \int_{]-l +x , l+ x[}:\!P(\phi(0,x))\!:_{C_0} \d x.
\]
By the same argument as above,
$\tau_{t}$ is implemented by $\e^{\i tH_{l,x}}$ for $|t|\leq T$ if $l>
|I|+|T|+|x|$, which implies that $\alpha_{x} \circ \tau_{t} \circ \alpha_{x}^{-1}=
\tau_{t}$ \qed .

\medskip 

\subsection{An identification of local algebras}
\label{ila}
In order to apply the results of Section~\ref{sec3} to the
algebra of local observables
$\cA$, it is necessary to identify the  local Weyl algebra $\cRb(I)$ with the von Neumann algebra $\cB(I)$ obtained by 
applying the interacting dynamics $\tau$ to the local abelian
algebra of time-zero fields $\cU_{\scriptscriptstyle AW} (I)$. This is done in Proposition \ref{p6.1}
below. Note that by similar arguments the corresponding result holds
also  in the
$0$-temperature case.
 
For $I\subset \rr$ a bounded open interval, we denote by $ \cU_{\scriptscriptstyle AW}(I)$ the
von Neumann algebra generated by 
$\{W_{\scriptscriptstyle AW}(h) \mid h\in \ch_{I},\: h\hbox{ real
valued}\}$.
Note that $\cU_{\scriptscriptstyle AW}(I)\subset \cRb(I)$ is abelian. 
We denote by ${\cal B}_{\alpha}(I)$ the
von Neumann algebra generated by 
\beq
\label{localalg}
\{\tau_{t}(A) \mid A\in \cU_{\scriptscriptstyle AW}(I), \;
|t|<\alpha\}.
\eeq
\begin{proposition} \label{p6.1}
Set ${\cal B}(I):=\bigcap_{\alpha>0} {\cal B}_{\alpha}(I)$. Then
\[
{\cal B}(I)= \cRb(I).
\]
\end{proposition}
\proof 
Let us first prove that ${\cal B}(I)\subset \cRb(I)$. 
Using
(\ref{e6.1f})  and $\cU_{\scriptscriptstyle AW}(I)\subset \cRb(I)$, we see that ${\cal B}_{\alpha}(I)\subset
\cRb \bigl(I+]-\alpha, \alpha[ \bigr)$ for all $\alpha >0$. According
to Lemma~\ref{l6.1} (i) this
implies~${\cal B}(I)\subset \cRb(I)$.

Let us now prove that $\cRb(I)\subset {\cal B}(I)$. Using Lemma~\ref{l6.1} (i) it suffices to show that 
for all~$\overline{J}\subset
I$ and $\alpha\ll 1$ one has 
\beq
\label{e6.02}
\cRb(J)\subset {\cal B}_{\alpha}(I).
\eeq
To this end we fix $I$ and $J$ with $\overline{J}\subset I$ and set
$\delta=\12{\rm dist}(J, I^{c})$. We will first prove that
\beq
\label{e6.03}
\e^{\i tL_{\scriptscriptstyle AW}}A\e^{-\i tL_{\scriptscriptstyle AW}}\in {\cal B}_{\alpha}(I), \: A\in
\cU_{\scriptscriptstyle AW}(J), \: |t|< \alpha,
\eeq
if $\alpha<\delta$.
The proof of 
Theorem~\ref{existencedyn} shows that for $|t|\leq \delta$ the unitary group $\e^{\i
tH_{I}}$, with $H_{I}:=\overline{L_{\scriptscriptstyle AW}+ V_{I}}$ and
\[
 V_{I}:=\int_{I}:\!P(\phi(0, x))\!:\d x, 
\]
induces the correct dynamics $\tau$ 
on $\cRb (J)$. Applying then Proposition \ref{approx}, we obtain 
\[
 \e^{\i tL_{\scriptscriptstyle AW}}=\slim_{n\to \infty}\e^{\i tH_{I}^{(n)}}, \: t\in \rr,
\]
for $H_{I}^{(n)}=\overline{L_{\scriptscriptstyle AW}+ V_{I}- V_{I}^{(n)}}$, where $V_{I}^{(n)}=
V_{I}\one_{\{|V_{I}|\leq n\}}$. Since $V_{I}^{(n)}$ is bounded,
\[H_{I}^{(n)}= \overline{L_{\scriptscriptstyle AW}+V_{I}}- V_{I}^{(n)}= H_{I}- V_{I}^{(n)}, \]
and
hence by Trotter's formula
\[
 \e^{\i tH_{I}^{(n)}}= \slim_{p\to \infty} \bigl(\e^{\i t H_{I}/p}\e^{-\i t
V_{I}^{(n)}/p} \bigr)^{p}.
\]
This yields, for $A\in \cRb(J)$,
\[
 \e^{\i tL_{\scriptscriptstyle AW}}A\e^{-\i tL_{\scriptscriptstyle AW}}= \slim_{n\to \infty}\slim_{p\to \infty}\bigl(\e^{\i t H_{I}/p}\e^{-\i t
V_{I}^{(n)}/p} \bigr)^{p}A \bigl(\e^{\i t
V_{I}^{(n)}/p}\e^{-\i t H_{I}/p}\bigr)^{p}.
\]
Using again Theorem~\ref{existencedyn} we obtain, for
$|t|<\alpha$,
\[
\bigl(\e^{\i t H_{I}/p}\e^{-\i t
V_{I}^{(n)}/p} \bigr)^{p}A \bigl(\e^{\i t
V_{I}^{(n)}/p}\e^{-\i t H_{I}/p}\bigr)^{p}= \bigl(\tau_{t/p}\circ
\gamma^{(n)}_{t/p}\bigr)^{p}(A),
\]
where $\gamma^{(n)}$ is the dynamics implemented by the unitary
group $t \mapsto \e^{-\i t V_{I}^{(n)}}$. Since $V_{I}$ is
affiliated to $\cU_{\scriptscriptstyle AW} (I)$, $\e^{-\i t V_{I}^{(n)}}\in \cU_{\scriptscriptstyle AW}(I)$ and hence
$\bigl(\tau_{t/p}\circ
\gamma^{(n)}_{t/p} \bigr)^{p}(A)\in \cB_{\alpha}(I)$ for $|t|<\alpha$. Since~$\cB_{\alpha}(I)$
is weakly closed, we obtain (\ref{e6.03}).

Let us now prove (\ref{e6.02}). Clearly the operators $W_{\scriptscriptstyle AW}(h)$ for
$h\in \ch_{J}$ and  $h$ real valued belong to $\cU_{\scriptscriptstyle AW}(J)$ and hence to ${\cal
B}_{\alpha}(I)$. Let us now pick $h\in \ch_{J}\cap \cD(\epsilon)$ and
$h$ real valued. 
(This is possible; see  the discussion 
presented at the end of Subsection~\ref{sec4.1}). Applying (\ref{e6.03}) to $A=
W_{\scriptscriptstyle AW}(h)$, we obtain that $W_{\scriptscriptstyle AW}(\e^{\i t \epsilon}h)\in {\cal
B}_{\alpha}(I)$ for $|t|<\alpha$. Hence $W_{\scriptscriptstyle AW}\bigl(t^{-1}(\e^{\i
t\epsilon}h -h) \bigr)\in {\cal
B}_{\alpha}(I)$ for $|t|<\alpha$. Letting $t\to 0$ and using the fact
that the map $\ch\ni h\mapsto W_{\scriptscriptstyle AW}(h)$ is continuous for the strong
operator topology, we obtain that $W_{\scriptscriptstyle AW}(\i \epsilon h)\in
\cB_{\alpha}(I)$. But any vector $h\in \ch_{J}$ can be
approximated in norm by vectors of the form $h_{1}+ \i \epsilon
h_{2}$, with $h_{i}\in \ch_{J}$ real and $h_{2}\in \cD(\epsilon)$. This implies that for
all $h\in \ch_{J}$  the
operators $W_{\scriptscriptstyle AW}(h)$ belong to
$\cB_{\alpha}(I)$ and hence $\cRb(J)\subset \cB_{\alpha}(I)$. This
completes the proof of the proposition \qed .

\subsection{Existence of the limiting state}
\label{sec4.3}
\begin{theoreme}
\label{limitingstate}
{\rm (Existence of limiting state).}
Let $\{\omega_{l}\}_{l>0}$  be the family of $(\tau^{l},\beta)$-KMS states 
for the spatially cutoff~$P(\phi)_{2}$ models constructed in
Subsection~\ref{sec2.4}.

Then
\[
\wlim_{l\to +\infty}\omega_{l}=: \omega_\beta \hbox{ exists on }\cA.
\]
The state $\omega_\beta$ on $\cA$ has the following properties:
\vskip .3cm
\halign{ \indent \indent \indent #  \hfil & \vtop { 
\parindent =0pt 
\hsize=12cm
                            \strut # \strut} 
\cr 
{\rm (i)} & $\omega_\beta$ is a $(\tau, \beta)$-KMS state on $\cA$;
\cr 
{\rm (ii)} &$\omega_\beta$ is {\em locally normal}, i.e., if $I$ is an open and bounded interval, then $\omega_{
\beta |\cRb(I)}$ is normal 
w.r.t.~the Araki-Woods representation;
\cr 
{\rm (iii)} &$\omega_\beta$ is invariant under spatial translations, i.e.,
\[
\omega_\beta (\alpha_{x}(A))= \omega_\beta(A), \: \: \: x\in \rr, \: A\in \cA;
\]
\vskip -.3cm
\cr 
{\rm (iv)} &$\omega_\beta$ has the {\em spatial clustering property}, i.e.,
\[
\lim_{x\to\infty}\omega_\beta(A\alpha_{x}(B))= \omega_\beta(A)\omega_\beta(B) \: \: \: \: \forall  A,
B\in \cA.
\]
\vskip -.3cm
\cr}
\end{theoreme}

\begin{remark}
Let ${\cal R}$ be a $C^{*}$-algebra, $\pi_{i}\colon {\cal R}\to {\cal
B}(\cH_{i})$, $i=1,2$, two
quasi-equivalent representations of ${\cal R}$. Then there exists a
$*$-isomorphism $\tau$ between $\pi_{1}({\cal R})''$ and
$\pi_{2}({\cal R})''$ intertwining the two representations. This isomorphism is automatically weakly continuous.
Therefore the representation $\pi_{2}$ extends uniquely from ${\cal
R}$ to $\pi_{1}({\cal R})''$ and is quasi-equivalent to the concrete
representation of $\pi_{1}({\cal R})''$ in ${\cal B}(\cH_{1})$.

Applying this easy observation to the representations $\pi_{\scriptscriptstyle AW}$
and $\pi_{F}$ of ${\cal W}(\ch_{I})$, which are quasi-equivalent by
Lemma \ref{localfock},  we see that the Fock
representation $\pi_{F}$ extends by weak continuity from $\pi_{\scriptscriptstyle AW}({\cal W}(\ch_{I}))$ to 
$\cRb(I)$ and is quasi-equivalent to the Araki-Woods representation.
Since two quasi-equivalent representations have the same set of normal
states, we obtain that $\omega_{\beta|\cRb(I)}$ is also normal with
respect to the Fock representation.
\end{remark}

\proof
The family $\{\omega_{l}\}_{l>0}$ of states on $\cA$ is weak$^{*}$
compact by the Banach-Alaoglu theorem. Let $\omega_{1}$ be one of the
limit points of $\{\omega_{l}\}_{l>0}$. Then we can find a 
subnet\footnote{A net $\{ y_\beta \}_{\beta \in B}$ is a subnet of a net $\{
x_\alpha \}_{\alpha \in A}$ if there exists a map $B \ni \beta 
\mapsto \alpha(\beta) \in A$ such that: i.) $y_\beta =
x_{\alpha(\beta)}$
for all $\beta \in B$; ii.) for all $\alpha_{0} \in A$ there exists 
some $\beta_{0}$ such that $\alpha(\beta) \ge \alpha_{0}$
whenever $\beta \ge \beta_{0}$.}
$\{\omega^{r}\}_{r \in R}$ such that
$\omega_{1}=\wlim_{r \in R}\omega^{r}$. 

We claim that $\omega_{1}$ is a $(\tau,\beta)$-KMS state. Let $A, B\in
\cA$. Writing \[
\omega_{1}(A\tau_{t}(B))-
\omega^{r}(A\tau^{l_r}_{t}(B))= (\omega_{1}- \omega^{r})(A\tau_{t}(B))+
\omega^{r} \bigl( A\tau_{t}(B)- A\tau^{l_r}_{t}(B) \bigr)
\]
and using that $\lim_{l \to \infty}\| \tau^{l} (A) - \tau_{t} (A) \|=0$ for $A \in \cA$ and $t \in \rr$ fixed, we find
\beq
\omega_{1}(A\tau_{t}(B))=\lim_{r\in R}
\omega^{r} \bigl( A\tau^{l_r}_{t}(B) \bigr),  \: \: t\in \rr.
\label{e7.01}
\eeq
The same argument shows 
\beq
\omega_{1}(\tau_{t}(B)A)=\lim_{r\in R}\omega^{r} \bigl(\tau^{l_r}_{t}(B)A \bigr),  \: \: t\in \rr. 
\label{e7.02}
\eeq
Since the $\omega^{r}$'s are  $(\tau^{l_r}, \beta)$-KMS states there exist
functions $F^{r}(z)$, which are holomorphic in $I_{\beta}^{+}=\{0<{\rm Im}z<\beta\}$
and continuous in $\overline{I^{+}_{\beta}}$,
such that $F^{r}(t)= \omega^{r} \bigl(A\tau^{l_r}_{t}B \bigr)$ and $F^{r}(t+ \i \beta)=
\omega^{r} \bigl(\tau^{l_r}_{t}(B)A \bigr)$. Moreover, one has $\sup_{z\in
I_{\beta}}|F^{r}(z)|\leq \|A\|\|B\|$. Applying Vitali's theorem and
possibly extracting a subnet, we know that $\lim_{r \to
\infty}F^{r}(z)= F(z)$ exists and is holomorphic and bounded in
$I_{\beta}^{+}$. By Lemma \ref{T4a}, we obtain that
$F$ is continuous on $\overline{I_{\beta}^{+}}$ and  
\[
F(t)=\lim_{r\to \infty}F^{r}(t),\: \: \: F(t+\i \beta)= \lim_{r\to
\infty}F^{r}(t+\i \beta).
\]
Using (\ref{e7.01}) and (\ref{e7.02}) this implies that $\omega_{1}$ is a
$(\tau, \beta)$-KMS state. 

We now apply a result of Takesaki and Winnink \cite{TW}: clearly
$I \to \{\cRb(I)\}$  is a net of von Neumann algebras (see \cite[Section~2]{TW}). 
The algebras $\cRb(I)$ are $\sigma$-finite, since the Hilbert
space $\Gamma(\ch\oplus \overline{\ch})$ on which they act is separable.
Moreover, as factors on a separable infinite dimensional Hilbert space they are properly
infinite. Applying \cite[Theorem~1]{TW}, we obtain that the KMS state
$\omega_{1}$ is {\em  normal} on $\cRb(I)$. 

Let us now show that all limit states are identical.
Let us denote by $\cUc (I)$ the abelian $C^{*}$-algebra generated by
\[ \bigl\{F (\phi_{\scriptscriptstyle AW}(h)) \mid h\in C^{\infty}_{0\: \rr}(I), \:F\in
\coinf(\rr) \bigr\} \]
and by $\cBc (I)$ the $*$-algebra generated
by $\bigl\{\tau_{t}(A) \mid A\in \cUc (I), \: |t|<\alpha \bigr\}$.

From Theorem~\ref{existencedyn} and Proposition~\ref{T4} we
deduce that 
\[
\lim_{l\to
\infty}\omega_{l} \bigl(\prod_{1}^{n}\tau_{t_{i}}(A_{i}) \bigr)=
\tilde{\omega} \bigl(\prod_{1}^{n}\tilde{\tau}_{t_{i}}(A_{i}) \bigr)=\omega_{1}
\bigl(\prod_{1}^{n}\tau_{t_{i}}(A_{i}) \bigr),\: \: A_{i}\in
\cUc (I), \: t_{i}\in \rr,
\]
where $\tilde{\omega}$ and $\tilde{\tau}$ are defined in Subsection~\ref{secsec}.
Therefore all weak accumulation points of~$\{\omega_{l}\}_{l>0}$
coincide on the algebras $\cBc(I)\subset
\cRb \bigl(I+]-\alpha, \alpha[ \bigr)$.  We 
note  that $\cBc(I)$ is weakly dense 
in the von Neumann algebra ${\cal B}_{\alpha}(I)$ defined in
(\ref{localalg}). Moreover, we have seen that all limit states
are normal on the local algebras $\cRb(I)$, $I$ open and bounded. Therefore they
coincide on the von Neumann algebras ${\cal B}_{\alpha}(I)$, and hence
by Proposition~\ref{p6.1} on $\cRb(I)$. Consequently, they also coincide 
on the norm closure~$\cA$. Thus the weak$^*$ compact family $\{\omega_{l}\}_{l>0}$ has a
unique accumulation point, which implies that
\[
\omega_{\beta} :=\wlim_{l\to \infty}\omega_{l}\hbox{ exists on }\cA.
\]
We have already seen that $\omega_{\beta}$ is a locally normal $(\tau, \beta)$-KMS state on
$\cA$, which completes the proof of (i) and (ii).
Property (iii) follows from the invariance of the state
$\tilde{\omega}$ under space translations shown in Lemma
\ref{spaceclust} and the same density argument as above. It remains to
prove (iv). Let $(\cH_{\beta}, \pi_{\beta}, \Omega_{\beta})$ denote the
GNS objects associated to $(\cA, \omega_{\beta})$. The group~$\{\alpha_{x}\}_{x \in \rr}$
is implemented in $\cH_{\beta}$ by a strongly continuous group of unitary operators
$\{ \e^{\i xP_\beta} \}_{x \in \rr}$
with $P_\beta \Omega_{\beta}=0$. Lemma \ref{spaceclust} (ii) implies that,
for $A, B\in \cBc(I)$,
\beq
\label{clust}
\lim_{x\to \infty} \bigl(\pi_{\beta}(A)\Omega_{\beta}, \e^{\i
xP_\beta}\pi_{\beta}(B)\Omega_{\beta} \bigr)= \bigl(\pi_{\beta}(A)\Omega_{\beta},
\Omega_{\beta})(\Omega_{\beta},\pi_{\beta}(B)\Omega_{\beta} \bigr).
\eeq
Since $\cRb(I)$ is a factor, the representation $\pi_{\beta}$ provides a weakly continuous
$^{*}$-isomorphism between~$\cRb(I)$ and~$\cR_{\beta}( I)= \pi_{\beta} (\cRb(I)) = \pi_{\beta} (\cRb(I))''$. Hence,
by the same  weak density argument as above,
(\ref{clust}) extends  to all $A,B\in\cRb(I)$. Thus the space clustering property holds on
$\cRb(I)$ for all $I$, $I$ open and bounded, and extends to~$\cA$ by  norm
density.

\section{Construction of the interacting  path space}
\init\label{sec3}
In this section we construct  the interacting path space supported by
${\cal S}_{\rr}' (S_{\beta}\times \rr)$ describing the translation
invariant $P(\phi)_{2}$ model at temperature $\beta^{-1}$ and study
some of its properties.
\subsection{Construction of the interacting  measure}
Let $H_{\tt C}^{\rm ren}= H_{\tt C}^{\rm ren} - E_{\tt C}$ be the renormalized
$P(\phi)_{2}$ Hamiltonian on $S_{\beta}$ defined in
Subsection~\ref{sec2.4a}. 
Let $f\in \cS_{\rr}(S_{\beta}\times \rr)$. For $x\in \rr$ the function $f_{x}$ defined in
Subsection \ref{sec1.5} belongs to~$\cS_{\rr}(S_{\beta})$. 
We will apply the
results of Appendix \ref{td} to the selfadjoint operator  $H=H_{\tt C}^{\rm ren}$, 
$R(x)=\phi_{F}(f_{x})$ (replacing the variable
$t$ in Appendix ~\ref{td} by the variable $x$). 

It follows from the bound (\ref{e3.1b}) in Proposition~\ref{3.1} and the fact that the map 
\[ \matrix{ & \rr & \to & \cB \bigl(\Gamma(H^{-\frac{1}{2} }(S_{\beta})) \bigr)
\cr
& x & \mapsto & \phi(f_{x})(H_{\tt C}^{\rm ren}+1)^{-\frac{1}{2} } \cr}
\] 
is  infinitely
differentiable that the hypothesis~(\ref{e4.3}) in Subsection~\ref{td1} is
satisfied. Similarly, using the bound 
(\ref{e3.1c}) and the fact that  the map $x\mapsto \|f_{x}\|_{H^{-\12}(S_{\beta})} $ is in
$L^{1}(\rr)\cap L^{\infty}(\rr)$, we see that hypotheses
(\ref{e4.12a}) and 
(\ref{e4.12b}) in Subsection~\ref{td3} is satisfied. Therefore we can
apply all the abstract results from Subsections \ref{td1} and
\ref{td3}. In particular there exists a solution $U(b, a)$ of the
time-dependent
heat equation:
\[
\frac{\d}{\d b}U(b,a)= (-H_{\tt C}^{\rm ren} +\i \phi_{F}(f_{b}))U(b, a), \: U(a,a)=\one.
\]
We will set for
$-\infty\leq a\leq b\leq +\infty$:
\[
 W_{[a,b]}(f):= U(b,a)^{*}.
\]

\begin{proposition}
\label{3.2}
Let $f\in \cS_{\rr}(S_{\beta}\times \rr)$ and assume that $\supp f\subset
S_{\beta}\times[-a, a]$. Then
\[
\int_{Q}\e^{\i \phi(f)}G_{[-l, l]}\d\phi_{C}=
\e^{-2lE_{\tt C}} \bigl(\e^{-(l-a) {H}^{\rm ren}_{\tt C}}\Omega_{\tt C}^{\circ},
W_{[-a, a]}(f)\e^{-(l-a) {H}^{\rm ren}_{\tt C}}\Omega_{\tt C}^{\circ} \bigr),
\]
where $\Omega_{\tt C}^{\circ}$ is the free vacuum on $\Gamma \bigl(H^{-\frac{1}{2} }(S_{\beta}) \bigr)$.
\end{proposition}
\proof
Let us first introduce a notation which we will use throughout the proof. If $A$ is a~$\Sigma$-measurable 
function on $Q$, the image of $A$ under $U_{\tt C}(x)$ for $x \in \rr$ will be denoted by 
$U_{\tt C}(x) (A)$. On the other hand, the expression $U_{\tt C}(x) A$ will denote 
the operator product of the operator $U_{\tt C}(x)$
and the operator of multiplication by $A$, acting on $L^{2}(Q, \Sigma,
\d\phi_{C})$.

Using Lemma \ref{1.0b} we find
\[
\e^{\i \phi(f)}=\e^{\i\int_{-a}^{a}\phi(f_{x},x)\d x}=\e^{\i
\int_{-a}^{a}U_{\tt C}(x) (\phi(f_{x}, 0))\d x}.
\]
We will approximate the above integral  using Riemann sums. Let $n,p\in
\nn$ and $0\leq j\leq 2np$. We set $x_{j}= -a + j\frac{a}{np}$ and $z_{j}= -a + [ j/ p]\frac{a}{n}$, where $[ . ]$ denotes the integer part.

\goodbreak
 
It follows from (\ref{e1.6b}) that the map $x\mapsto \phi(f_{x},x)\in\bigcap_{1\leq 
p<\infty}L^{p}(Q,\Sigma, \d\phi_{c})$ is continuous. 
Therefore
\[
\int_{-a}^{a}U_{\tt C}(x)\bigl(\phi(f_{x}, 0) \bigr)\d x = \lim_{n,p \to \infty} \sum_{j=0}^{2np-1}
(x_{j+1} -x_j)  U_{\tt C}(x_j) \bigl(\phi(f_{z_j}, 0) \bigr)  \]
in $\bigcap_{1 \le p \le \infty} L^p (Q, \Sigma,  \d\phi_{C})$ and hence
\beq
\begin{array}{rl}
{\rm e}^{\i  \int_{-a}^{a}U_{\tt C}(x)(\phi(f_{x}, 0))\d x } &= \lim_{n,p \to \infty} \prod_{j=0}^{2np-1}
{\rm e}^{\i (x_{j+1} -x_j)  U_{\tt C}(x_j) (\phi(f_{z_j}, 0))}
\\[3mm]
&= \lim_{n,p \to \infty} \prod_{j=0}^{2np-1} U_{\tt C}(x_j) \bigl(
{\rm e}^{\i (x_{j+1} -x_j)  \phi(f_{z_j}, 0) } \bigr)
\end{array}
\label{e3.3}
\eeq
in $\bigcap_{1 \le p \le \infty} L^p (Q, \Sigma,  \d\phi_{C})$, where in the last line we use the fact that 
$U_{\tt C}(x_j)$ is an automorphism of $L^\infty (Q, \Sigma,  \d\phi_{C})$. Since $G_{[a,b]}$ 
is a FKN kernel,  
\[
\begin{array}{rl}
G_{[-l,l]} &= G_{[-l,-a]} \Bigl[ \prod_{j=0}^{2np-1}
G_{[x_j,x_{j+1}]}   \Bigr] G_{[a,l]}
\\[3mm]
&= G_{[-l,-a]} \Bigl[ \prod_{j=0}^{2np-1}
U_{\tt C}(x_j) \bigl(G_{[0,x_{j+1}- x_j]} \bigr)  \Bigr] G_{[a,l]} .\end{array}  \]
Therefore
\[
\begin{array}{rl}
&G_{[-l,l]}  \prod_{j=0}^{2np-1} U_{\tt C}(x_j) \bigl({\rm e}^{\i (x_{j+1} -x_j)  \phi(f_{z_j}, 0) }  \bigr)
\\[3mm]
=& G_{[-l,-a]}  \Bigl[ \prod_{j=0}^{2np-1} U_{\tt C}(x_j) \bigl({\rm e}^{\i (x_{j+1} -x_j)  \phi(f_{z_j}, 0) } 
G_{[0,x_{j+1}- x_j]} \bigr)  \Bigr] G_{[a,l]}.
\end{array} \]
Next let $A_j$, $0 \le j < 2np -1$, be the multiplication operators by
$\Sigma$-measurable functions. Using the identity $U_{\tt C}(x_j)
(A_j)= U_{\tt C}(x_j) A_j U_{\tt C}(-x_j)$ and the fact that $U_{\tt
C}(x)$ is an automorphism of $L^{\infty}(Q, \Sigma, \d\phi_{C})$, we
obtain
as an operator identity on $L^2 (Q, \Sigma, \d\phi_{C})$:
\[
\prod_{j=0}^{2np-1}U_{\tt C}(x_j)  (A_j) = 
U_{\tt C}(x_0)
\prod_{j=0}^{2np-1}A_j U_{\tt C}(x_{j+1} - x_j)  U_{\tt C}(-x_{2np}).
\]
In the above identity the product on the l.h.s.~is the  operator of
multiplication by the product of the functions
$U_{\tt C}(x_j) (A_j)$ and the product on the r.h.s.~is an operator product.  Using that 
$x_0= - x_{2np} =-a$ and that $U_{\tt C}(-a)^* = U_{\tt C}(a)$ we get
\[
\begin{array}{rl}
& \int_Q G_{[-l,l]}  \prod_{j=0}^{2np-1}U_{\tt C}(x_j)  \bigl({\rm e}^{\i (x_{j+1} -x_j)  \phi(f_{z_j}, 0) } 
 \bigr) \d\phi_{C}
\\[3mm]
= & \int_Q G_{[-l,-a]} U_{\tt C}(-a) \Bigl[ \prod_{j=0}^{2np-1} {\rm e}^{\i (x_{j+1} -x_j)  \phi(f_{z_j}, 0) }
U_{\tt C}(x_{j+1}- x_j) G_{[0,x_{j+1}- x_j]} \Bigr] U_{\tt C}(-a) G_{[a,l ]} \d\phi_{C}
\\[3mm]
= & \int_Q G_{[-l+a,0]} \Bigl[ \prod_{j=0}^{2np-1} {\rm e}^{\i (x_{j+1} -x_j)  \phi(f_{z_j}, 0) }
U_{V}(x_{j+1}- x_j) \Bigr] G_{[0,l-a]} \d\phi_{C} \end{array}   \]
for $U_{V}(s)= G_{[0, s]}U_{\tt C}(s)$. 

Let us now set for $0 \le k \le 2n$, $y_k = -a + k \frac{a}{n}$. We note that $z_j = y_{[j /p]}$ and
that 
$(x_{j+1} -x_j) = \frac {a}{np} = (y_{k+1} - y_{k}) /p$. We obtain
that
\[
\begin{array}{rl}
& \int_Q G_{[-l,l]}  \prod_{j=0}^{2np-1}U_{\tt C}(x_j)  \bigl({\rm e}^{\i (x_{j+1} -x_j)  \phi(f_{z_j}, 0) } 
 \bigr) \d\phi_{C}
\\[3mm]
=&  \int_Q G_{[-l+a,0]}  \prod_{k=0}^{2n-1}  \bigl( {\rm e}^{\i
(y_{k+1} -y_k)  \phi(f_{y_k}, 0) /p}
U_{V}(\frac{y_{k+1}- y_k}{p} )\bigr)^pG_{[0,l-a]} \d\phi_{C} 
\\[3mm]
=&  \int_Q R_{\tt C} (G_{[0,l-a]})  \prod_{k=0}^{2n-1}  \bigl( {\rm
e}^{\i (y_{k+1} -y_k)  \phi(f_{y_k}, 0) / p}
U_{V}(\frac{y_{k+1}- y_k}{p} )\bigr)^pG_{[0,l-a]} \d\phi_{C} . \end{array}   \]
Taking into account  the construction of
$H_{\tt C}$ recalled in Subsection~\ref{sec2.4a} we find
\[
\begin{array}{rl}
&\int_Q G_{[-l,l]}  \prod_{j=0}^{2np-1}U_{\tt C}(x_j)  \bigl({\rm e}^{\i (x_{j+1} -x_j)  \phi(f_{z_j}, 0) } 
 \bigr) \d\phi_{C}
\\ [3mm]
 = &\Bigl(\e^{-(l-a)H_{\tt C}}\Omega_{\tt C}^{\circ},   
\prod_{k=0}^{2n-1} \bigl(
\e^{\i
(y_{k+1}-y_{k}) \phi(f_{y_{k}})/p}  \e^{-(y_{k+1}-y_{k}) {H}_{\tt C}/p} \bigr)^p
 \e^{-(l-a)H_{\tt C}} 
\Omega_{\tt C}^{\circ} \Bigr)\\[3mm]
 =&\e^{-2l E_{\tt C}} \Bigl(\e^{-(l-a) {H}^{\rm ren}_{\tt C}}\Omega_{\tt C}^{\circ}, 
\prod_{k=0}^{2n-1} \bigl(
\e^{\i
(y_{k+1}-y_{k}) \phi(f_{y_{k}})/p}\e^{-(y_{k+1}-y_{k}) {H}^{\rm ren}_{\tt C}/p} \bigr)^p
 \e^{-(l-a) {H}^{\rm ren}_{\tt C} } 
\Omega_{\tt C}^{\circ} \Bigr).
\end{array}
\]
Letting now $n$ and $p$ tend to $\infty$ and using Proposition~\ref{timeprod}, we
obtain the proposition \qed .

\goodbreak

\begin{theoreme}
\label{3.3} \quad
\vskip .3cm
\halign{ \indent \indent \indent #  \hfil & \vtop { 
\parindent =0pt 
\hsize=12cm
                            \strut # \strut} \cr 
{\rm (i)}  & Let $f\in
C^{\infty}_{0\:\rr}(S_{\beta}\times \rr)$. Then 
\[
\lim_{l\to +\infty}\int \e^{\i
\phi(f)}\d\mu_{l}= (\Omega_{\tt C},
W_{[-\infty, \infty]}(f)\Omega_{\tt C}),
\]
where $\Omega_{\tt C}$ is the unique vacuum state of $H_{\tt C}$.
\cr
{\rm (ii)} & The map
\[
\cS_{\rr}(S_{\beta}\times\rr)\ni f\mapsto (\Omega_{\tt C},
W_{[-\infty, \infty]}(f)\Omega_{\tt C})
\]
is the generating functional of a Borel probability measure $\mu$ on $(Q,
\Sigma)$.
\cr
{\rm (iii)} & The measure $\mu$ is invariant under  space
translations $\{{\frak a}_{x}\}_{x\in \rr}$, time 
translations~$\{{\scriptstyle \frak T}_t \}_{t\in S_{\beta}}$ and the
time reflection $r$.
\cr
{\rm (iv)} & The functions $\phi(f)$ belong to $\bigcap_{1\leq
p<\infty}L^{p}(Q,\Sigma, \mu)$ for $f\in \cS_{\rr}(S_{\beta}\times\rr)$. Moreover,
\[
\int_{Q}\phi(f)^{n}\d \mu= n! \kern -1cm \int \limits_{-\infty<x_{1}\leq\cdots\leq
x_{n}<\infty}  \kern -.8cm
\bigl(\Omega_{\tt C},
\bigl[ \: \prod_{1}^{n-1}\phi(f_{x_{k}})\e^{-(x_{k+1}-x_{k})H_{\tt C}^{\rm ren}} \: \bigr] \phi(f_{x_{n}})  \Omega_{\tt C} \bigr)\d
x_{1}\dots\d x_{n}.
\]
\cr
{\rm (v)} & Let $f_{i}\in C^{\infty}_{0\:\rr}(S_{\beta}\times \rr)$ for $1\leq i\leq n$. Then 
\[
\lim_{l\to +\infty}\int_{Q}
\prod_{i=1}^{n}\phi(f_{i})\d\mu_{l}=\int_{Q}\prod_{i=1}^{n}\phi(f_{i})\d\mu.
\]
\cr}
\end{theoreme}
\proof
Note first that applying Proposition~\ref{3.2} for $f=0$, we
obtain $W_{[-a,a]}(0)=\e^{-2aH^{\rm ren}_{\tt C}}$:
\[
\int_{Q}G_{[-l, l]}\d\phi_{C}=\e^{-2lE_{\tt C}} \bigl(\e^{-(l-a) 
{H}^{\rm ren}_{\tt C}}\Omega_{\tt C}^{\circ},
\e^{-(l-a) {H}^{\rm ren}_{\tt C}}\Omega_{\tt C}^{\circ} \bigr)..
\]
Let $f\in C^{\infty}_{0\:\rr}(S_{\beta}\times \rr)$ with $\supp f\subset
S_{\beta}\times [-a, a]$ for some $a\in \rr$. Using Proposition~\ref{3.2} we find
\[
\int \e^{\i
\phi(f)}\d\mu_{l}\\[2mm]
= \frac{
\bigl(\e^{-(l-a)H_{\tt C}^{\rm ren}}\Omega_{\tt C}^{\circ},
W_{[-a, a]}(f)\e^{-(l-a)H_{\tt C}^{\rm ren}}\Omega_{\tt C}^{\circ}\bigr) }{ 
(\e^{-l H_{\tt C}^{\rm ren}}\Omega_{\tt C}^{\circ},
\e^{-l H_{\tt C}^{\rm ren}}\Omega_{\tt C}^{\circ}) } .
\]
Now $\lim_{l\to +\infty}\e^{-(l -a) H_{\tt C}^{\rm ren}}\Omega_{\tt C}^{\circ} 
= (\Omega_{\tt C} ,  \Omega_{\tt C}^{\circ})\Omega_{\tt C}$, where $\Omega_{\tt C}$ is the eigenvector
for the simple eigenvalue $\{0\}$ of  $H^{\rm ren}_{\tt C}$. Thus
\[
\lim_{l\to +\infty}\int \e^{\i
\phi(f)}\d\mu_{l}= (\Omega_{\tt C},
W_{[-a, a]}(f)\Omega_{\tt C}),
\]
Because
$\supp f\subset S_{\beta}\times [-a, a]$, we see that  $\bigl(\Omega_{\tt C},
W_{[s,t]}(f)\Omega_{\tt C}\bigr)$ is constant for $s\leq -a$, $t\geq a$, which proves {\rm (i)}. 

To prove {\rm (ii)} we apply Minlos theorem (see e.g.~\cite{GV}).
As a
limit of functionals of Borel probability measures on $(Q, \Sigma)$ the functional $f\mapsto \bigl(\Omega_{\tt C},
W_{[-\infty, \infty]}(f)\Omega_{\tt C} \bigr) $ is of positive type.
It remains to show that the map
\[
\matrix{ & \cS(S_{\beta}\times \rr) & \to & \cc
\cr & f & \mapsto & (\Omega_{\tt C},
W_{[-\infty, \infty]}(f)\Omega_{\tt C})\cr} 
\]
is continuous.
Using the bound (\ref{e3.1c}) we obtain
\[
\pm \bigl(\phi_{F}(f_{2,x})-\phi_{F}(f_{1, x}) \bigr)\leq C r(x)( {H}^{\rm ren}_{\tt C}+
1)^{\frac{1}{2} }  \hbox{ for } f_{1},\: f_{2}\in \cS(S_{\beta}\times \rr) ,
\]
where $C>0$ is some constant and
\[
 r(x):=\|(f_{2,x}- f_{1,x})\|_{H^{-\frac{1}{2} }(S_{\beta})}.
\]
Clearly $\|r\|_{L^{2}(\rr)}\leq C\|f_{1}-f_{2}\|_{p}$, where
$\|.\|_{p}$  is a Schwartz semi-norm on $\cS(S_{\beta}\times \rr)$.
Applying Lemma~\ref{minlos} for $\delta=\12$, we obtain 
\[
\bigl\| W_{[-\infty, +\infty]}(f_{2})-W_{[-\infty, +\infty]}(f_{1}) \bigr\|\leq
C\|f_{2}- f_{1}\|_{p},
\]
which proves the desired continuity result. 

Let us now verify (iii). The measure $\mu$ is invariant under time
translations and time reflection as the weak limit of the time
translation and time reflection invariant measures~$\mu_{l}$. 
The fact that~$\mu$ is invariant under space translations follows directly from
(i) and Remark~\ref{translationremark}. 

To prove  {\rm (iv)} we apply Lemma~\ref{5.1}, using the estimates
in Proposition \ref{titi}~{\rm (ii)}. We obtain that $\phi(f)\in
\bigcap_{1\leq p<\infty}L^{p}(Q,\Sigma,  \mu)$. The formula in {\rm (iv)}
follows  from Proposition~\ref{titi}~{\rm (iii)}.

It remains to prove {\rm (v)}. Let $f\in C^{\infty}_{0\:\rr}(S_{\beta}\times \rr)$ with $\supp f\subset
S_{\beta}\times [-a, a]$. We consider the family of functions
\[
u_{l}(\lambda)= \int
\e^{\i \lambda
\phi(f)}\d\mu_{l} \hbox{ for } \lambda\in \cc.
\]
Since $\e^{\phi(f)}\in \bigcap_{1\leq p<\infty}L^{p}(Q, \Sigma, 
\d\phi_{C})$ and $F_{[-\beta/2, \beta/2]}^{l}\in L^{1+\epsilon}(Q, \Sigma, 
\d\phi_{C})$, the functions~$u_{l}(\lambda)$ are entire and
\beq
\label{e3.9}
\frac{\d^{n}}{\d
\lambda^{n}}u_{l}(0)=\i^{n}\int
\phi(f)^{n}\d\mu_{l}.
\eeq
Using Proposition \ref{3.2} and $\lambda \phi (f) = \phi (\lambda f) $ for $\lambda \in \rr$ we find
\[
u_{l}(\lambda)= \frac{\bigl(W_{[-a, a]}(\lambda f)   \e^{-(l-a) {H}^{\rm ren}_{\tt C}}\Omega_{\tt C}^{\circ},
\e^{-(l-a) {H}^{\rm ren}_{\tt C}}\Omega_{\tt C}^{\circ} \bigr)}{ \| \e^{- l
 {H}^{\rm ren}_{\tt C}}\Omega_{\tt C}^{\circ}\|} \: \: \hbox{ for }\lambda \in \rr.
\]
The r.h.s.~is an entire function by Lemma \ref{analytic}. Therefore  this
identity extends to $\lambda \in \cc$. Applying (\ref{e3.1c}) and Proposition~\ref{titi}~(i) 
with $\delta = 1/2$ 
we obtain
\beq
|u_{l}(\lambda)|\leq \e^{C|{\rm Im}\lambda|^{2}}, \: l\in \rr^{+}, \: \lambda \in \cc.
\label{e3.8}
\eeq 
We have seen above that
\[
\lim_{l\to \infty}u_{l}(\lambda)=\int_{Q}\e^{\i \lambda \phi(f)}\d \mu \hbox{ for } \lambda \in \rr.
\]
By Vitali's theorem we obtain 
\[
\lim_{l\to \infty}\frac{\d^{n}}{\d
\lambda^{n}}u_{l}(0)=\i^{n}\int_{Q}\phi(f)^{n}\d\mu.
\]
Using (\ref{e3.9})  and multi-linearity, this proves {\rm (v)} \qed .

\medskip

\subsection{Existence and properties of sharp-time fields}
\label{secsec}
\begin{proposition}
\label{T2}
Let $h\in \cS_{\rr}(\rr)$ and $t\in S_{\beta}$. Then the sequence
$\phi \bigl(\delta_{k}(.-t)\otimes h \bigr)$ is Cauchy
in~$\bigcap_{1\leq p<\infty}L^{p}(Q, \Sigma, \mu)$ and hence
\[
 \phi(t, h):= \lim_{k\to \infty}\phi \bigl(\delta_{k}(.-t)\otimes h \bigr)\in
\bigcap_{1\leq p<\infty}L^{p}(Q, \Sigma, \mu).
\]
Moreover, the map
\[
\matrix{ 
& S_{\beta} & \to &  \bigcap_{1\leq p<\infty}L^{p}(Q, \Sigma, \mu)
\cr
&t & \mapsto & \phi(t, h) \cr}
\]
is continuous for each $h\in \cS_{\rr}(\rr)$.
\end{proposition}
\proof For $p\geq 1$ we have
\[
\begin{array}{rl}
&\int_{Q} \bigl(\phi(\delta_{k}(.-t)\otimes h)-\phi(\delta_{k'}(.-t)\otimes
h) \bigr)^{2p}\d\mu\\[2mm]
=&(-\i)^{2p}\frac{\d^{2p} }{\d \lambda^{2p}} \Bigl(\Omega_{\tt C}, W_{[-\infty,
+\infty]}\bigl(\lambda(\delta_{k}(.-t)\otimes h-\delta_{k'}(.-t)\otimes
h)\bigr)\Omega_{\tt C} \Bigr)_{|\lambda=0}.
\end{array}
\]
If $f= \delta_{k}(.-t)\otimes h$, then for $x\in \rr$ the function $f_{x}\in
\cS_{\rr}(S_{\beta})$ is equal to 
$\delta_{k}(.-t)h(x)$.  It follows then from the estimate (\ref{e3.1d}) in Proposition \ref{3.1} that
\[
\pm \bigl(\phi_{F}(\delta_{k}(.-t)h(x))-\phi_{F}(\delta_{k'}(.-t)h(x)) \bigr)\leq \tC \, \|\delta_{k}(.-t)-
\delta_{k'}(.-t)\|_{H^{-1}(S_{\beta})} \: |h(x)| \: \bigl( H^{\rm ren}_{\tt C}+1 \bigr).
\]
Applying now Lemma \ref{T1} we obtain that
\beq
\begin{array}{rl}
&\bigl\|\frac{\d^{2p} }{\d \lambda^{2p}}W_{[-\infty,
+\infty]}\bigl(\lambda(\delta_{k}(.-t)\otimes h-\delta_{k'}(.-t)\otimes
h) \bigr) \bigr\|\\[2mm]
\leq &\tC_{p}\|\delta_{k}(.-t)-
\delta_{k'}(.-t)\|^{2p}_{H^{-1}(S_{\beta})} \: \|h\|_{\infty}^{2p} \: \e^{\|h\|_{1}\|h\|^{-1}_{\infty}}.
\end{array}
\label{e6.6b}
\eeq
Since $\delta_{k}(.-t)$ converges to $\delta(.-t)$ in
$H^{-1}(S_{\beta})$, we see that $\phi \bigl( \delta_{k}(.-t)\otimes h \bigr)$ is
Cauchy in~$L^{2p}(Q,\Sigma, \mu)$.  A similar argument shows that
$t\mapsto \phi(t, h)\in L^{2p}(Q, \Sigma, \mu)$ is continuous, using the
fact that $t\mapsto \delta(.-t)\in H^{-1}(S_{\beta})$ is continuous \qed . 

\medskip 
Using the existence of sharp-time fields, we can  equip the
probability space $(Q, \Sigma, \mu)$ with an OS-positive
$\beta$-periodic path space structure:  
We recall that~$U(t)$ is the group of transformations
generated by the (euclidean) time translations~${\scriptstyle \frak
T}_{t}$ and~$R$ is the 
transformation generated by time reflection,
and~$\Sigma_{0}$ is the sub$-\sigma$-algebra of
$\Sigma$ generated by the functions~$\{\phi(0, h) \mid h\in \cS_{\rr}(\rr) \}$.

\begin{theoreme}
$(Q, \Sigma, \Sigma_{0},U(t), R,  \mu)$ is an OS-positive
$\beta$-periodic generalized path space.
\label{3.5}
\end{theoreme}
\proof
Since the measure $\mu$ is invariant under time translations and
time reflection, we see that $U(t)$ and $R$ are
measure preserving automorphisms of $L^{\infty}(Q, \Sigma, \mu)$. 
Proposition~\ref{T2} implies that the map $S_{\beta}\ni t\mapsto
\e^{\i \phi(t,h)}\in L^{2}(Q, \Sigma, \mu)$ is continuous. 
Hence~$U(t)$ is a strongly continuous group on $L^{2}(Q, \Sigma, \mu)$.
This implies that~$U(t)$ is strongly continuous in measure on
$L^{\infty}(Q, \Sigma, \mu)$. Clearly it is $\beta$-periodic.

The generalized path space $(Q, \Sigma, \Sigma_{0},U(t), R,  \mu)$
is OS-positive, since $\mu$ is the weak limit of the measures
$\mu_{l}$, which are
associated to OS-positive path spaces.
Finally we have already seen that $\Sigma=\bigvee_{t\in S_{\beta}}\Sigma_{t}$. This completes the
proof of the theorem \qed . 

\medskip

By the reconstruction theorem, we obtain a stochastically positive $\beta$-KMS system
\[
(\tilde{\cB}, \tilde{\cU}, \tilde{\tau}, \tilde{\omega})
\] 
which
describes the {\em translation invariant $P(\phi)_{2}$ model at
temperature }$\beta^{-1}$.

\subsection{Properties of the interacting $\beta$-KMS system}
\label{sec7.3}

We first prove the convergence of sharp-time Schwinger functions.
\begin{proposition}
\label{T3}
Let $h_{i}\in C^{\infty}_{0\: \rr}(\rr)$ and $t_{i}\in S_{\beta}$ for
$1\leq i\leq n$.
Then
\[
\lim_{l\to \infty}\int_{Q}\prod_{1}^{n}\e^{\i \phi(t_{j},
h_{j})}\d\mu_{l}= \int_{Q}\prod_{1}^{n}\e^{\i \phi(t_{j},
h_{j})}\d\mu.
\]
\end{proposition}
\proof 
Let $a>0$ such that $\supp h_{j}\subset [-a,a]$. By Proposition \ref{T2}, we
know that
\[
 \phi(t_{j}, h_{j})= \lim_{k\to \infty}
\phi \bigl(\delta_{k}(.-t_{j})\otimes h_{j} \bigr)\hbox{ in }L^{1}(Q, \Sigma,
\mu).
\]
After extracting a subsequence, this implies that
\[
\phi(t_{j}, h_{j})= \lim_{k\to \infty}
\phi \bigl(\delta_{k}(.-t_{j})\otimes h_{j} \bigr) \hbox{ pointwise }\mu\hbox
{ a.e. on }Q,
\]
and hence 
\beq
\begin{array}{rl}
\int_{Q}\prod_{1}^{n}\e^{\i \phi(t_{j},
h_{j})}\d\mu=&\lim_{k\to \infty}\int_{Q}\prod_{1}^{n}\e^{\i
\phi(\delta_{k}(.-t_{j})\otimes h_{j})}\d \mu\\[2mm]
=&\lim_{k\to
\infty} \bigl(\Omega_{\tt C}, W_{[-a,
a]} \bigl(\sum_{1}^{n}\delta_{k}(.-t_{j})\otimes
h_{j} \bigr)\Omega_{\tt C} \bigr),
\end{array}
\label{e6.6c}
\eeq
by Theorem~\ref{3.3} (i).  Note that for all $l>0$
\[
 \phi(t_{j}, h_{j})= \lim_{k\to \infty}
\phi \bigl(\delta_{k}(.-t_{j})\otimes h_{j} \bigr)\hbox{ in }L^{1}(Q, \Sigma,
\mu_{l}),
\]
because this convergence holds in $L^{2}(Q, \Sigma, \d
\phi_{C})$ and 
\[  \d\mu_{l}: = \frac{G_{[-l, l]}\d \phi_{C}}{\int_Q G_{[-l, l]}\d \phi_{C}}, \] 
where $G_{[-l,
l]}\in L^{2}(Q, \Sigma, \d\phi_{C})$ as a consequence of (\ref{e6.1b}). By the same
arguments as above, we obtain
\[
\int_{Q}\prod_{1}^{n}\e^{\i \phi(t_{j},
h_{j})}\d\mu_{l}=\lim_{k\to
\infty}\frac{\bigl(\e^{-(l-a) {H}^{\rm ren}_{\tt C}}\Omega_{\tt C}^{\circ}, W_{[-a,
a]}\bigl(\sum_{1}^{n}\delta_{k}(.-t_{j})\otimes
h_{j} \bigr)\e^{-(l-a) {H}^{\rm ren}_{\tt C}}\Omega_{\tt C}^{\circ} \bigr)}{\|\e^{-l {H}^{\rm ren}_{\tt C}}\Omega_{\tt C}^{\circ}\|^{2} }.
\]
Let us denote by $F(k, l)$ the quantity on the r.h.s..  Applying
(\ref{e6.6b}) we obtain  
\[
\lim_{k\to \infty}F(k, l)=\int_{Q}\prod_{1}^{n}\e^{\i \phi(t_{j},
h_{j})}\d\mu_{l} \hbox{ uniformly w.r.t. }l.
\]
As we have seen, Theorem~\ref{3.3} (i) implies
\[
\lim_{l\to \infty}F(k,l)=\int_{Q}\prod_{1}^{n}\e^{\i
\phi(\delta_{k}(.-t_{j})\otimes h_{j})}\d\mu.
\]
Applying now Lemma \ref{a1} (ii) and using (\ref{e6.6c}) we
obtain the proposition \qed .

\medskip
Let us denote by $\cUc
\subset \cB \bigl(\Gamma(\ch\oplus\overline{\ch}) \bigr)$ the $C^{*}$-algebra generated by 
\[
\bigl\{F \bigl(\phi_{\scriptscriptstyle AW}(h_{1}), \dots, \phi_{\scriptscriptstyle AW}(h_{n}) \bigr) \mid h_{i}\in
C^{\infty}_{0\: \rr}(\rr),\: F\in \coinf(\rr^{n}), \: n\in \nn \bigr\} .
\] 
The isomorphism between $L^{\infty}(Q, \Sigma_{0}, \d\phi_{C})$ and $\cU_{\scriptscriptstyle AW}$, which we recalled in 
Subsection~\ref{sec0.3}, maps the operator $F \bigl(\phi_{\scriptscriptstyle AW}(h_{1}), \dots, \phi_{\scriptscriptstyle AW}(h_{n})\bigr)$ 
onto the function $F\bigl(\phi(0, h_{1}), \dots, \phi(0, h_{n}) \bigr)$.
This function is $\Sigma_{0}$-measurable.  We will still denote by
$A$ the image of such a function $A$ in the abelian algebra~$\tilde{\cU}$ provided by the reconstruction theorem 
for the translation
invariant $P(\phi)_{2}$ model.

\begin{proposition}
\label{T4}
Let $A_{i}\in \cUc$ and $t_{i}\in \rr$, $1\leq i\leq n$. Then 
\[
\lim_{l\to
+\infty}\omega_{l} \Bigl(\prod_{1}^{n}\tau^{l}_{t_{i}}(A_{i})
\Bigr)=\tilde{\omega} \Bigl(\prod_{1}^{n}\tilde{\tau}_{t_{i}}(A_{i}) \Bigr).
\]
\end{proposition}
\proof 
Let us fix $A_{i}\in \cUc$ and set
\[
G^{l}(t_{1}, \dots, t_{n}) :=
\omega_{l} \Bigl(\prod_{1}^{n}\tau^{l}_{t_{i}}(A_{i}) \Bigr) \quad \hbox{and} \quad G(t_{1}, \dots,
t_{n})=\tilde{\omega} \Bigl(\prod_{1}^{n}\tilde{\tau}_{t_{i}}(A_{i}) \Bigr).
\]
Due to the KMS condition, the functions $G^{l}$ and $G$ are holomorphic in
\[
I_{\beta}^{n+} =\bigl\{ (z_{1}, \dots, z_{n})\in \cc^{n} \mid {\rm Im}z_{i}<
{\rm Im}z_{i+1}, \; {\rm Im}z_{n}- {\rm Im}z_{1}<\beta \bigr\},
\]
continuous on $\overline{I_{\beta}^{n+}}$ and
bounded by $\prod_{1}^{n}\|A_{i}\|$.

We first claim that
\beq
\lim_{l\to \infty}G^{l}(\i s_{1}, \dots, \i s_{n})= G(\i s_{1}, \dots,
\i s_{n}) \hbox{ for }s_{1}\leq \cdots\leq s_{n},\: s_{n}-s_{1}\leq
\beta.
\label{eT1}
\eeq
Using Proposition \ref{T3} and the identity (\ref{e0.0}) we
see that (\ref{eT1}) holds for $A_{j}= \e^{\i
\phi_{\scriptscriptstyle AW}(h_{j})}$, $h_{j}\in C_{0\:\rr}^{\infty}(\rr)$. Using functional calculus we
can extend (\ref{eT1}) to arbitrary  $A_{j}\in \cUc$. 

Let us now consider, for $s_{2}\leq \cdots \leq s_{n}$ and $s_{n}-s_{2}\leq
\beta$, the functions 
\[u_{l}(z) := G^{l}(z, \i s_{2}, \dots, \i s_{n}) ,\]
which are holomorphic in $\{0<{\rm Im}z<s_{2}\}$ and continuous on
$\{0\leq {\rm Im}z\leq s_{2}\}$. Since the family $\{u_{l}\}$ is uniformly
bounded, we can apply Lemma \ref{T4a}. It follows that
\[
 \lim_{l\to +\infty}G^{l}(t_{1}, \i s_{2}, \dots, \i s_{n} )= G(t_{1},
\i s_{2}, \dots, \i s_{n})
\]
for $s_{2}\leq \cdots \leq s_{n}$, $s_{n}-s_{2}\leq \beta$ and $t_{1}\in
\rr$. Iterating this argument, we obtain  
\[
\lim_{l\to +\infty}G^{l}(t_{1}, \dots, t_{n})=G(t_{1}, \dots, t_{n}). 
\]
This completes the proof of the proposition  \qed .

Let us denote by $\{\alpha_{x}\}_{x\in \rr}$ the group of space
translations on $\cUc$ defined by $\alpha_{x} \bigl(W_{\scriptscriptstyle AW}(h) \bigr)=
W_{\scriptscriptstyle AW} \bigl(h( \: . \: - x) \bigr)$ for $h\in C_{0\:\rr}^{\infty}(\rr)$.
\begin{lemma}
\label{spaceclust}
Let $A_{j}\in \cUc$ and $t_{j}\in \rr$, $1\leq j\leq n$. Set
$A=\prod_{j=1}^{k}\tau_{t_{j}}(A_{j})$ and
$B=\prod_{j=k+1}^{n}\tau_{t_{j}}(A_{j})$. It follows that
\vskip .3cm
\halign{ \indent \indent \indent #  \hfil & \vtop { 
\parindent =0pt 
\hsize=12cm
                            \strut # \strut} 
\cr 
{\rm (i)}  & $\tilde{\omega}(\alpha_{x}(A))= \tilde{\omega}(A)$ for all $ x\in \rr$;
\cr 
{\rm (ii)}  & $\lim_{x\to \infty}\tilde{\omega}(A\alpha_{x}(B))=\tilde{\omega}(A)\tilde{\omega}(B)$.
\cr}
\end{lemma}
\proof
Property (i) follows directly from the invariance of the measure $\mu$  under the space translations 
$\{ \alpha_{x}\}_{x \in \rr}$ shown in Theorem~\ref{3.3} (iii). It remains to prove (ii). We set
\[
\begin{array}{l}
 G_{x}(t_{1}, \ldots, t_{n}):=
\tilde{\omega} \bigl(\prod_{j=1}^{l}\tau_{t_{j}}(A_{j})\prod_{j=l+1}^{n}\alpha_{x} \circ\tau_{t_{j}}(A_{j}) \bigr),
\\[2mm]
G_{\infty}(t_{1}, \ldots, t_{n}):=
\tilde{\omega} \bigl(\prod_{j=1}^{l}\tau_{t_{j}}(A_{j}) \bigr)\cdot \tilde{\omega} \bigl(\prod_{j=l+1}^{n}\tau_{t_{j}}(A_{j}) \bigr).
\end{array}
\]
Due to the KMS condition, the functions $G_{x}$ and $G_{\infty}$ are holomorphic in
$I_{\beta}^{n+}$ and bounded by~$\prod_{j=1}^{n}\|A_{j}\|$. We 
claim that, for $s_{1}\leq \cdots\leq s_{n}$ and $s_{n}-s_{1}\leq
\beta$, 
\beq
\lim_{x\to \infty}G_{x}(\i s_{1}, \dots, \i s_{n})= G_{\infty}(\i s_{1},
\dots, \i s_{n}).
\label{claim1}
\eeq
Let us prove (\ref{claim1}). Let us first assume that $A_{j}= \e^{\i \phi(0,h_{j})}$ for
$h_{j}\in C_{0\:\rr}^{\infty}(\rr)$. 
Then
\[
\begin{array}{rl}
G_{x}(\i s_{1}, \dots, \i s_{n})&= \int_{Q}\prod_{j=1}^{l}\e^{\i
\phi(\delta(.-s_{j})\otimes h_{j})}\prod_{j=l+1}^{n}\e^{\i
\phi(\delta(.-s_{j})\otimes h_{j}( \: . \: - x))}\d\mu,\\[2mm]
G_{\infty}(\i s_{1}, \dots, \i s_{n})&= \int_{Q}\prod_{j=1}^{l}\e^{\i
\phi(\delta(.-s_{j})\otimes h_{j})}\d\mu\times\int_{Q}\prod_{j=l+1}^{n}\e^{\i
\phi(\delta(.-s_{j})\otimes h_{j})}\d\mu.
\end{array}
\]
By Proposition \ref{T2} we have
\[
\begin{array}{rl}
G_{x}(\i s_{1}, \dots, \i s_{n})&= \lim_{k\to +\infty}\int_{Q}\prod_{j=1}^{l}\e^{\i
\phi(\delta_{k}(.-s_{j})\otimes h_{j})}\prod_{j=l+1}^{n}\e^{\i
\phi(\delta_{k}(.-s_{j})\otimes h_{j}( \: . \: - x))}\d\mu,\\[2mm]
G_{\infty}(\i s_{1}, \dots, \i s_{n})&= \lim_{k\to
\infty}\int_{Q}\prod_{j=1}^{l}\e^{\i
\phi(\delta_{k}(.-s_{j})\otimes h_{j})}\d\mu\times\int_{Q}\prod_{j=l+1}^{n}\e^{\i
\phi(\delta_{k}(.-s_{j})\otimes h_{j})}\d\mu.
\end{array}
\]
From Theorem \ref{3.3} we get 
\[
\begin{array}{l}
\int_{Q}\prod_{j=1}^{l}\e^{\i
\phi(\delta_{k}(.-s_{j})\otimes h_{j})}\prod_{j=l+1}^{n}\e^{\i
\phi(\delta_{k}(.-s_{j})\otimes h_{j}( \: . \: - x))}\d\mu\\[2mm]
= \bigl(\Omega_{\tt C},
W_{[-\infty, +\infty]}(R_{1,k}+{\frak a}_{x} (R_{2, k})) \Omega_{\tt C} \bigr)
\end{array}
\]
and
\[ 
\begin{array}{l}
\int_{Q}\prod_{j=1}^{l}\e^{\i
\phi(\delta_{k}(.-s_{j})\otimes h_{j})}\d\mu\times\int_{Q}\prod_{j=l+1}^{n}\e^{\i
\phi(\delta_{k}(.-s_{j})\otimes h_{j})}\d\mu\\[2mm]
= \bigl(\Omega_{\tt C}, W_{[-\infty,
+\infty]}(R_{1, k})\Omega_{\tt C} \bigr)\cdot \bigl(\Omega_{\tt C}, W_{[-\infty, +\infty]}(R_{2,
k})\Omega_{\tt C} \bigr),
\end{array}
\]
where 
\[
R_{1, k}= \phi \bigl(\sum_{j=1}^{l}\delta_{k}(.-s_{j})\otimes h_{j} \bigr)\: \: \hbox{and} \: \:
R_{2, k}= \phi \bigl(\sum_{j=l+1}^{n}\delta_{k}(.-s_{j})\otimes h_{j} \bigr).
\]
As before (see Section \ref{sec2.1}), the group of spatial translations induced on $Q$ by the map
$(t, y) \mapsto (t, y+x)$ has been denoted by ~$\{{\frak a}_{x}\}_{x \in \rr}$.
Applying Lemma \ref{spaceclustering} we find 
\[
\Bigl|  
\bigl(\Omega_{\tt C} \: , \:
W\bigl(R_{1,k}+  {\frak a}_{x}(R_{2, k}) 
\bigr) \Omega_{\tt C}\bigr)
- \bigl(\Omega_{\tt C} \: , \:
W(R_{1,k}) 
\Omega_{\tt C}\bigr)
\bigl(\Omega_{\tt C}, W(R_{2,
k})\Omega_{\tt C} \bigr)
\Bigr|\leq \e^{- (|x|- C)a},
\]
where $a>0$ is the spectral gap of $H_{\tt C}^{\rm ren}$ and $W (.):=W_{[-\infty, +\infty]}(.)$. Letting $k\to \infty$
and using Proposition \ref{T2} we obtain 
\[
\bigl| G_{x}(\i s_{1}, \dots, \i s_{n})- G_{\infty}(\i s_{1}, \dots, \i
s_{n}) \bigr|\leq \e^{- (|x|- C)a}.
\]
Using functional calculus, we conclude that (\ref{claim1}) holds for all
$A_{j}\in \cUc$. 
To complete the proof of the lemma, we can now argue as in the proof of
Proposition \ref{T4}, using Lemma \ref{T4a} \qed .

\appendix
\section{A time-dependent heat equation}
\init
\label{td}
Let $H\geq 0$ be a selfadjoint operator on a Hilbert space $\cH$ and let $R(t)$, $t\in \rr$, be a
family of  closed operators with $\cD(H^{\gamma})\subset \cD(R(t))$
for some $0\leq \gamma<1$. We consider the following
time-dependent heat equation:
\beq\left\{
\begin{array}{l}
\frac{\d}{ \d t}U (t,s)= - \bigl(H+\i \lambda R(t) \bigr)U(t,s),\; s\leq t, \\
U(s,s)=\one.
\end{array}\right.
\label{e4.1}
\eeq
This equation is  (formally) equivalent to the following integral equation:
\beq
U (t,s)= \e^{-(t-s)H} -\i \lambda \int_{s}^{t}\e^{-(t-\tau)H}R(\tau) U(\tau,
s)\d \tau.
\label{e4.2}
\eeq
In the main text we only use the results of this section in the
{\em dissipative\/} case, i.e., 
when $R(t)$ is selfadjoint for all $t\in \rr$. However,  part of the
results are valid and will be proved in the general case.

The solution of (\ref{e4.1}) will be denoted by $U(t,s)$ or $U_\lambda (t,s)$. 
If we want to display its  dependence on the
family $R(t)$, then the solution of (\ref{e4.1}) will be denoted by $U(t,s ;
R)$.  

\subsection{Existence of solutions}
\label{td1}
We assume that the maps 
\beq
\matrix{ & \rr & \to &\cB(\cH)
\cr
& t & \mapsto & R(t)(H+1)^{-\gamma}} 
\quad
\hbox{ and }
\matrix{& \rr & \to &\cB(\cH)
\cr
& t & \mapsto &R^{*}(t)(H+1)^{-\gamma}}
\label{e4.3}
\eeq
are H\"{o}lder
continuous of some order $\epsilon>0$.

In the sequel 
we will use the following result.

\begin{lemma}
\label{4.1}
Assume (\ref{e4.3}). Then
\beq
\|\e^{-(t-\tau)H}R(\tau)\|\leq
\tC_{\gamma}\|R(\tau)^{*}(H+1)^{-\gamma}\| \: \bigl(|t-\tau|^{-\gamma}+1\bigr) \quad \forall  \tau\leq t.
\label{e4.4}
\eeq
\end{lemma}
\proof We have $\|\e^{-(t-\tau)H}R(\tau)\|\leq
\|R(\tau)^{*}(H+1)^{-\gamma}\|\|(H+1)^{\gamma}\e^{-(t-\tau)H}\|$. This proves
the lemma, using
\beq
|(\lambda+1)^{\gamma}\e^{-s\lambda}|\leq \tC_{\gamma} (s^{-\gamma}+1) \quad \hbox{for} \quad s,
\: \lambda\geq 0 \: \Box \, .
\label{e4.4c}
\eeq
The following result is shown in \cite[Theorem 7.1.3]{H}.
\begin{proposition}
\label{4.2}
There exists a unique solution $U(t,s)$ of (\ref{e4.1})
such that 
\vskip .3cm
\halign{ \indent \indent \indent #  \hfil & \vtop { 
\parindent =0pt 
\hsize=12cm
                            \strut # \strut} \cr 
{\rm (i)}  & $U(s,s)=\one $ and $U(t,r)U(r,s)= U(t,s) \hbox{ for }s\leq r\leq t $;
\cr
{\rm (ii)} & $ t\mapsto U(t,s)\in \cB(\cH)$ is strongly continuous in
$[s, +\infty[ $ and strongly differentiable in $]s,+\infty[$.
\cr}
\end{proposition}
\begin{lemma}
\label{analytic}
The map $ \lambda \mapsto U_\lambda (.,s)\Psi \in C\bigl([s, +\infty[,\cH \bigr)$ 
is entire analytic for each $\Psi \in \cH$. 
\end{lemma}

\noindent
\proof 
Let $\Psi \in \cH$. For $s\leq T<\infty$,  set $V(t) \Psi = \e^{-(t-s)H}\Psi$  and
\[\| U (.,s)\Psi \|:=\sup_{t \in [s, T]}\|U (t,s)\Psi \|_{\cH}.\] 
If we define a map   
\[
\matrix {K\colon & C([s, T], \cH) & \to & C([s, T], \cH)
\cr
& W(.) & \mapsto & -\i\int_{s}^{(.)}\e^{-(.-\tau)H}R(\tau)W(\tau)\d \tau, \cr}
\]
then the integral equation (\ref{e4.2}) can be rewritten as $(1-K) (U (\: . \: ,s) \Psi)=V (\: . \: ) \Psi$. Now
(\ref{e4.4}) implies
\[
 \|KU(t ,s)\Psi \|\leq \|U (.,s)\Psi \|\sup_{\tau\in [s,
T]}\|R^{*}(\tau)(H+1)^{-\gamma}\|\int_{s}^{t} \bigl(|t-\tau|^{-\gamma}+ 1 \bigr) \d \tau,
\]
and hence 
\[  \|KU(\: . \: ,s) \Psi \|\leq \tC \sup_{\tau\in [s,
T]}\|R^{*}(\tau)(H+1)^{-\gamma}\| \:|T-s|^{1-\gamma}\|U(\: . \: ,s) \Psi\| , \]
which shows
that $K\in \cB \bigl(C([s, T], \cH) \bigr)$.  Then $U_{\lambda} (\: . \: ,s) \Psi$ solves $(1- \lambda K) \bigl( U_{\lambda}(\: . \: ,s)\Psi \bigr) = V (\: . \: )\Psi$, which implies
that $\lambda \mapsto U_{\lambda} (\: . \: ,s) \Psi$ is entire analytic \qed .

\medskip

\subsection{The dissipative case}
\label{td3}
We now consider the dissipative case when $R(t)$
is selfadjoint for $t\in \rr$. We first prove a result about
approximation by time-ordered products. 
We will make use of  an extension of Gronwall's inequality to integral
equations, shown in \cite[Lemma 7.1.1]{H}.
\begin{lemma}
\label{gronwall}
Let $b\geq 0$ and $\gamma>0$.  Let $a(t)$ and $u(t)$ be   non negative locally
integrable functions on~$s\leq t\leq T<\infty$ such that
\[
u(t)\leq a(t)+ b\int_{s}^{t}(t-\tau)^{-\gamma} u(\tau)\d \tau \hbox{
for }s\leq t\leq T.
\]
Then
\[
 u(t)\leq a(t)+ \tC b^{(1-\gamma)^{-1}}\int_{s}^{t} E(t-\tau) a(\tau)\d \tau \hbox{ for
}s\leq t\leq T,
\] 
where $|E(r)|\leq \tC _{T}(|r|^{-\gamma})$ on $[0, T-s]$.
\end{lemma}
\begin{proposition}
\label{timeprod}
Assume that $R(t)$ is selfadjoint and that (\ref{e4.3}) holds.

Then for $s\leq t$ there exists a sequence $\{ p_{n}\}_{n \in \nn}$ with
$\lim_{n\to \infty}p_{n}=+\infty$ such that
\[
U(t,s)=\slim_{n\to \infty}\prod_{n-1}^{0} \Bigl(\e^{-(t_{j+1}- t_{j})H/ p_{n}}\e^{-\i
(t_{j+1}- t_{j})R(t_{j})/ p_{n}} \Bigr)^{p_{n}},
\]
where $t_{j}= s+ \frac{(t-s)j}{n}$ for $0\leq j\leq n$.

\end{proposition}
\proof
Let $R_{i}$, $i=1,2$, be two families of closed operators
satisfying~(\ref{e4.3}) and let~$U^{(i)}(t,s)$ be the associated
propagators. Then 
\[
\begin{array}{rl}
U^{(1)}(t,s)- U^{(2)}(t,s) 
=&
-\i\int_{s}^{t}\e^{-(t-\tau)H}R_{1}(\tau) \bigl(U^{(1)}(\tau,s)- U^{(2)}(\tau,
s) \bigr)\d \tau\\[2mm]
&-\i \int_{s}^{t}\e^{-(t-\tau)H} \bigl(R_{1}(\tau)-
R_{2}(\tau) \bigr)U^{(2)}(\tau,s )\d\tau.
\end{array}
\]
This implies, using Lemma \ref{4.1}, that
\[
\begin{array}{rl}
&\|U^{(1)}(t,s)- U^{(2)}(t,s)\|\\[2mm]
\leq & \tC_\gamma \int_{s}^{t}
(|t-\tau|^{-\gamma}+1) \: \|(H+1)^{-\gamma}R_{1}(\tau)\| \: \|U^{(1)}(\tau,s)-
U^{(2)}(\tau,s)\|\d \tau\\[2mm]
&+ \tC_\gamma \int_{s}^{t}
(|t-\tau|^{-\gamma}+1)\: \bigl\|(H+1)^{-\gamma} \bigl(R_{1}(\tau)-R_{2}(\tau) \bigr) \bigr\| \: \|U^{(2)}(\tau,
s)\|\d \tau.
\end{array}
\]
We  now apply Gronwall's inequality, as given  in Lemma \ref{gronwall}, with
\[
\begin{array}{l}
b= \sup_{s\leq t\leq T}\|(H+1)^{-\gamma}R_{1}(t)\|,\\[2mm]
a(t)\equiv a= \tC_{T}\sup_{s\leq t\leq
T} \bigl\|(H+1)^{-\gamma} \bigl(R_{1}(t)-R_{2}(t) \bigr) \bigr\| \times \sup_{s\leq t\leq T}\|U^{(2)}(t,
s)\| .
\end{array}
\] 
We obtain 
\beq
\begin{array}{rl}
&\sup_{s\leq t\leq T}\|U^{(1)}(t,s)-U^{(2)}(t,s)\|\\[2mm]
\leq & \tC_{T}\sup_{s\leq t\leq
T}\|(H+1)^{-\gamma}R_{1}(t)\|^{(1-\gamma)^{-1}}\\[2mm]
&\times\sup_{s\leq t\leq
T} \bigl\|(H+1)^{-\gamma} \bigl(R_{1}(t)-R_{2}(t) \bigr) \bigr\| \times \sup_{s\leq t\leq T}\|U^{(2)}(t,
s)\|.
\end{array}
\label{e4.gronwall}
\eeq
Let us now prove the proposition. For $s<t$ fixed, $n\in \nn$, we set 
\[
t_{j}:= s+ \frac{(t-s)j}{n},\: \:0\leq
j\leq n, \quad   R_{n}(\tau)=\sum_{n=0}^{n-1}\one_{[t_{j}, t_{j+1}[}(\tau)R(t_{j}).
\]
Note that $H+\i R(t_{j})$ with domain $\cD(H)$ is the generator of a
$C_{0}$-semigroup of contractions, since it is 
closed and maximal accretive, using  (\ref{e4.3}).

If $U^{(n)}(t,s)$ is the solution of (\ref{e4.1}) for the piecewise constant family of operators~$\{ R_{n} (t) \}$, then
one can easily verify that:
\[
 U^{(n)}(t,s)= \prod_{n-1}^{0} \e^{-(t_{j+1}- t_{j})(H+\i R_n (t_{j}))}.
\]
Since $\rr\ni t\mapsto (H+1)^{-\gamma}R(t)\in \cB(\cH)$ is continuous, we
conclude that
\[
\lim_{n\to \infty}\sup_{s\leq t\leq T} \bigl\|(H+1)^{-\gamma} \bigl(R(t)-
R_{n}(t) \bigr) \bigr\|=0.
\]
Using (\ref{e4.gronwall}) we get 
\[
\lim_{n\to \infty}\sup_{s\leq t\leq T}\|U(t,s)- U^{(n)}(t,s)\|=0.
\]
Applying next \cite{Ch}
we obtain  
\[
\e^{-(t_{j+1}- t_{j})(H+\i R(t_{j}))}=\slim_{p\to
\infty}\Bigl(\e^{-(t_{j+1}-t_{j})H/p}\e^{-\i
(t_{j+1}-t_{j})R(t_{j})/p} \Bigr)^{p}.
\]
Using the fact that $\e^{-\tau(H+\i R(t_{j}))}$, $\e^{-\tau H}$ and
$\e^{-\i \tau R(t_{j})}$ are all contractions, we conclude that there
exists a sequence $p_{n}\to \infty$ such that
\[
U(t,s)=\slim_{n\to \infty} \prod_{n-1}^{0} \Bigl(\e^{-(t_{j+1}-t_{j})H/p_{n}}\e^{-\i
(t_{j+1}-t_{j})R(t_{j})/p_{n}} \Bigr)^{p_{n}}.
\]
This completes the proof of the proposition \qed .

\medskip

\begin{proposition}
\label{titi}
Assume that $R(t)$ is selfadjoint and satisfies (\ref{e4.3}). Assume moreover that the function
\beq
t \mapsto \|(H+1)^{-\gamma}R(t)\| \hbox{ is in } L^{1}(\rr)\cap L^{\infty}(\rr)
\label{e4.12a}
\eeq
and
\beq
\label{e4.12b}
 \pm R(t)\leq r(t)(H+1)^{\delta}, \quad 0\leq\delta<1 , 
\eeq
for some $r\in L^{1}(\rr)\cap L^{\infty}(\rr)$.  Then the
limit
$U_{\lambda}(+\infty, -\infty):=\wlim_{(t,s)\to (+\infty, -\infty)}U_{\lambda} (t,s)
$ exists for all $\lambda \in \cc$. 

\halign{ \indent \indent \indent #  \hfil & \vtop { 
\parindent =0pt 
\hsize=12cm
                            \strut # \strut} \cr 
{\rm (i)}  & the function $\cc\ni \lambda \mapsto U_{\lambda}(+\infty, -\infty)$ is entire and
satisfies
\[
\|U_{\lambda}(+\infty, -\infty)\|\leq \e^{\tC |{\rm Im}\lambda|^{(1-\delta)^{-1}}} \quad \forall 
\lambda \in \cc;
\]
\vskip -.3cm
\cr
{\rm (ii)} & the derivatives w.r.t.~$\lambda$ are uniformly bounded:
\[ \sup_{\lambda\in \rr}|\p_{\lambda}^{n}U_{\lambda}(+\infty, -\infty)|<\infty  \quad \forall n\in \nn ; \]
\vskip -.3cm
\cr
{\rm (iii)} & for $n\in \nn$ and $\lambda \in \rr$ the derivatives at $\lambda = 0$ are given by the following formula:
\[
\begin{array}{rl}&\frac{\d^{n}}{\d \lambda^{n}}U_{\lambda}(+\infty, -\infty)_{|\lambda=0}\\[2mm]
 =&n!(-\i)^{n} \kern -.5cm \int \limits_{-\infty<t_{1}\leq\cdots\leq
t_{n}<\infty} \kern -.5cm
\one_{\{0\}}(H)\left(\prod_{n}^{2}R(t_{k})\e^{-(t_{k}-t_{k-1})H}\right)R(t_{1})\one_{\{0\}}(H)\d
t_{1}\dots\d t_{n}.
\end{array}
\]
\cr}
\end{proposition}

\noindent
\proof
Let $\Psi\in \cH$. 
Using Proposition \ref{4.2} and the fact that~$R(t)$
is selfadjoint we obtain
\[
\begin{array}{rl}
\frac{\d}{\d t}\|U_{\lambda }(t,s)\Psi\|^{2} & =-2{\rm Re} \bigl(U_{\lambda}(t,s)\Psi, (H+\i
\lambda R(t))U_{\lambda}(t,s)\Psi \bigr)\\
& = -2 \bigl(U_{\lambda}(t,s)\Psi, (H-{\rm Im}\lambda R(t))U_{\lambda}(t,s)\Psi \bigr).
\end{array}
\]
Now 
\[
H-{\rm Im}\lambda R(t)\geq H-|{\rm Im}\lambda | \: r(t)(H+1)^{\delta}\geq \tC \bigl(  |{\rm
Im}\lambda | \: r(t) \bigr)^{(1-\delta)^{-1}},
\]
since $\inf_{s\geq 0} \: \: s-t(s+1)^{\delta}=- \tC \, t^{(1-\delta)^{-1}}$ for
$t\geq 0$. This yields
\[
\frac{\d}{\d t}\|U_{\lambda}(t,s)\Psi\|^{2}\leq \tC |{\rm
Im}\lambda |^{(1-\delta)^{-1}}r(t)^{(1-\delta)^{-1}}\|U_{\lambda}(t,s)\Psi\|^{2},
\]
and hence
\[
\|U_{\lambda}(t,s)\Psi\|\leq \e^{\tC |{\rm
Im}\lambda |^{(1-\delta)^{-1}}\int_{s}^{t}r(\tau)^{(1-\delta)^{-1}}\d\tau}\|\Psi\|.
\]
Since $r\in L^{1}(\rr)\cap L^{\infty}(\rr)$ we have
$r^{(1-\delta)^{-1}}\in L^{1}(\rr)$, which yields
\beq
\sup_{s\leq t}\|U_{\lambda}(t,s)\|\leq \e^{\tC |{\rm Im}\lambda |^{(1-\delta)^{-1}}} \quad \forall \lambda\in \cc.
\label{e4.12}
\eeq
Let us now prove that
\beq
\label{e4.12aa}
\wlim_{(t,s)\to (+\infty, -\infty)}U_{\lambda}(t,s)\hbox{ exists for all }\lambda\in
\rr.
\eeq
For $\Psi \in \cH$, $\Phi \in \cD(H)$ and $0 \le \gamma < 1$ we find
\[
\bigl(\Phi , (U_{\lambda}(t,s)-\e^{-(t-s)H})\Psi \bigr)=
-\i \lambda\int_{s}^{t} \Bigl(\e^{-(t-\tau)H}(H+1)^{\gamma}\Phi,
(H+1)^{-\gamma}R(\tau)U(\tau, s)\Psi \Bigr)\d \tau.
\]
Using dominated convergence and hypothesis (\ref{e4.12a}) we obtain
the existence of
\[
\lim_{(t,s)\to (+\infty, -\infty)} \bigl(\Phi, U_{\lambda}(t,s)\Psi \bigr)\hbox{ for }\Psi\in
\cH\hbox{ and }\Phi \in \cD(H^\gamma).
\]

Applying a density argument and the uniform bound~(\ref{e4.12}) this
proves~(\ref{e4.12aa}).

Now  $\{ \lambda \mapsto U_{\lambda}(t,s) \mid s\leq t \}$ is a locally
uniformly bounded family of entire functions.  Applying Lemma
\ref{T4a} and (\ref{e4.12aa}) we obtain that 
\[
U_{\lambda}(+\infty, -\infty)=\wlim_{(t,s)\to (+\infty, -\infty)}U_{\lambda}(t,s)
\]
exists for all $\lambda \in \cc$. Moreover, the map $\lambda \mapsto U_{\lambda}(+\infty, -\infty)$ 
 is entire  and
\[
\|U_{\lambda}(+\infty, -\infty)\|\leq \e^{\tC |{\rm Im}\lambda |^{(1-\delta)^{-1}}} \quad \forall 
\lambda \in \cc.
\] 
If $f(z)$ is a
bounded holomorphic function in a strip~$\{|{\rm Im}z|<a\}$, then
it follows easily from Cauchy's formula that 
$\sup_{x\in \rr}|\p_{x}^{n}f(x)|<\infty$ for all $n\in \nn$. This
completes the proof of {\rm (ii)}.

Let us now prove {\rm (iii)}. Set, as in Subsection \ref{td1},
\[
\matrix {K\colon & C([s, T], \cH) & \to & C([s, T], \cH)
\cr
& W(.) & \mapsto & -\i\int_{s}^{(.)}\e^{-(.-\tau)H}R(\tau)W(\tau)\d \tau, \cr}
\]
and $V(t)\Psi =\e^{-(t-s)H}\Psi$.  From the integral equation $(\one - \lambda
K)U_{\lambda}( \: . \: , s)\Psi= V( \: . \: ) \Psi$ we deduce that
\[
\begin{array}{rl}
\frac{\d^{n}}{\d \lambda^{n}}U_{\lambda}( t,t_0)_{|\lambda=0}\Psi & =n!K^{n}V(t)\Psi \\
& = n!(-\i)^{n}\int \limits_{t_0 \leq t_{1}\leq\cdots\leq
t_{n}\leq
t}\e^{-(t-t_{n})H} \Bigl[ \prod_{n}^{1}R(t_{k})\e^{-(t_{k}-t_{k-1})H} \Bigr] \Psi\d
t_{1}\dots\d t_{n}. 
\end{array}
\]
The function $\cc\ni \lambda \mapsto U_{\lambda}(t,s)\Psi$
is entire and uniformly bounded in $\{|{\rm Im}\lambda |\leq a \}$ for $-\infty
<s\leq t<+\infty$. Therefore
\[
\frac{\d^{n}}{\d \lambda^{n}}U_{\lambda}(+\infty, -\infty)\Psi=\lim_{(t,s)\to
(+\infty,-\infty)}\frac{\d^{n}}{\d \lambda^{n}}U_{\lambda}(t,s)\Psi.
\]
Setting $t_{n+1}=t$ we find
\[
\left\|\e^{-(t-t_{n})H}\prod_{n}^{1}R(t_{k})\e^{-(t_{k}-t_{k-1})H} \right\|\leq
 \tC  \prod_{n}^{1}\bigl(|t_{k+1}-t_{k}|^{-\gamma}+1 \bigr)\|(H+1)^{-\gamma}R(t_k)\|  .\]
 From Lebesgue dominated convergence we deduce  that
\[
\lim_{s\to-\infty}\frac{\d^{n}}{\d \lambda^{n}}U_{\lambda}(t,s)_{|\lambda=0}=
n!(-\i)^{n}\int \limits_{-\infty< t_{1}\leq\cdots\leq
t_{n}\leq
t} \Bigl[ \prod_{n}^{1}\e^{-(t_{k+1}-t_{k})H}R(t_{k})\Bigr] \one_{\{0\}}(H)\d
t_{1}\dots\d t_{n}.
\]
A similar argument yields
\[
\begin{array}{rl}
&\lim_{t\to +\infty}\lim_{s\to-\infty}\frac{\d^{n}}{\d \lambda^{n}}U_{\lambda}(t,
s)_{|\lambda=0}\\[2mm]
=&
n!(-\i)^{n}\int \limits_{-\infty< t_{1}\leq\cdots\leq
t_{n}\leq
+\infty}\one_{\{0\}}(H) \Bigl[ \prod_{n}^{2}R(t_{k})\e^{-(t_{k}-t_{k-1})H} \Bigr] R(t_{1})\one_{\{0\}}(H)\d
t_{1}\dots\d t_{n}.
\end{array}
\]
Applying Lemma \ref{a1} we obtain {\rm (iii)}  \qed .

We will use the following lemma  to show that the limiting
functional obtained in Theorem~\ref{3.3} defines a Borel measure on
$\cS'(S_{\beta}\times \rr)$.
\begin{lemma}
\label{minlos}
Let $R_{i}(t)$, $i=1,2$, be selfadjoint families satisfying
(\ref{e4.3}) and
 (\ref{e4.12a}). 
Assume that
\[
\pm \bigl(R_{1}(t)- R_{2}(t) \bigr)\leq r(t)(H+1)^{\delta}, \: 0\leq
\delta<1,
\]
for $r\in L^{(1-\delta)^{-1}}(\rr)$. Then
\[
\|U(+\infty, -\infty;R_{2})-U(+\infty, -\infty; R_{1})\|\leq
 \tC \|r\|_{(1-\delta)^{-1}}.
\]
\end{lemma}
\proof
Let us denote by $Z_{\lambda}(t,s)$ the operator $U(t,s; R_{1}+
\lambda(R_{2}-R_{1}))$. By the same arguments as used in the proof of Proposition \ref{titi}, we
see that $\lambda\mapsto Z_{\lambda}(t,s)$ is an entire analytic function, which  satisfies the bound
\[
\|Z_{\lambda}(t,s)\|\leq \e^{\tC |{\rm Im
\lambda}|^{\gamma}\|r\|^{\gamma}_{\gamma}} \quad \hbox{for }  \lambda\in\cc  
\hbox{ and } \gamma=(1-\delta)^{-1}. \]
As in Proposition \ref{titi}, the limit
of $Z_{\lambda}(t,s)$ when $(t,s)\to (+\infty, -\infty)$ exists for $\lambda \in
\rr$ fixed. Applying again Vitali's theorem, we obtain the existence of
$Z_{\lambda}(+\infty, -\infty)$ for all~$\lambda\in \cc$, and the bound
\[
\|Z_{\lambda}(+\infty, -\infty)\|\leq \e^{\tC |{\rm Im
\lambda}|^{\gamma}\|r\|^{\gamma}_{\gamma}} \quad \forall   \lambda \in\cc.
\]
Applying Cauchy's formula on the circle of radius $R$ centered around $\lambda \in \rr$ yields
\[
\left| \frac{\d}{\d \lambda}Z_{\lambda}(+\infty, -\infty) \right|\leq
R^{-1}\e^{\tC R^{\gamma}\|r\|_{\gamma}^{\gamma}}. 
\]
Optimizing this bound w.r.t.\ $R$ we get
\[
\left| \frac{\d}{\d \lambda}Z_{\lambda}(+\infty, -\infty) \right|\leq \tC \|r\|_{\gamma}.
\]
Integrating in $\lambda$ from $0$ to $1$ we obtain the lemma \qed . 

\medskip
\subsection{Some additional results}
\label{td4}
We now prove some bounds on $U(t,s)$, which we use in the main text
to show the existence of sharp-time fields and the convergence of 
sharp-time Schwinger functions.
\begin{lemma}
\label{T1}
Let $R_{i}(t)$, $i=1,2$, be two families of selfadjoint operators
satisfying (\ref{e4.3}) and (\ref{e4.12a}). Assume  that
\beq
\pm \bigl(R_{2}(t)- R_{1}(t) \bigr)\leq r(t)(H+1)\hbox{ for }r\in
L^{\infty}(\rr)\cap L^{1}(\rr).
\label{e4.t1}
\eeq
Set $Z_{\lambda}(t,s):=U \bigl(t,s; R_{1}+
\lambda(R_{2}-R_{1}) \bigr)$ for $-\infty\leq s\leq t\leq +\infty$. Then
\[
\left\|\frac{\d^{n}}{\d \lambda^{n}}Z_{\lambda}(t,s) \right\|\leq n! \:
\|r\|^{n}_{\infty} \: \e^{\|r\|_{1}\|r\|^{-1}_{\infty}}.
\]
\end{lemma}
\proof
Since $R_{i}(t)$ satisfy (\ref{e4.3}) and (\ref{e4.12a}), the function
$\lambda \mapsto Z_{\lambda}(t,s)$ is entire. We still denote by $Z_{\lambda}(t,
s)$ its extension to $\lambda \in \cc$. As in the proof of Proposition \ref{titi},
we find
\[
\frac{\d}{\d t} \|Z_{\lambda}(t,s)\Psi\|^{2}= -2 \Bigl( Z_{\lambda}(t,s)\Psi, \bigl(H - {\rm
Im}\lambda(R_{2}(t)- R_{1}(t))\bigr)Z_{\lambda}(t,s)\Psi \Bigr).
\]
Now
\[
H - {\rm
Im}\lambda \bigl(R_{2}(t)- R_{1}(t) \bigr)\geq H- r(t)  \:  |{\rm Im}\lambda| \:  (H+1) \geq -r(t)  \:  |{\rm
Im}\lambda|
\]
for $|{\rm Im}\lambda|\leq \|r\|_{\infty}^{-1}$. This yields
\[
\frac{\d}{\d t} \|Z_{\lambda}(t,s)\Psi\|^{2}\leq 2 \:  |{\rm Im}\lambda| \: r(t)\|\Psi\|^{2}
 \:  \hbox{ for }  \:  |{\rm Im}\lambda|\leq \|r\|^{-1}_{\infty}.
\]
Hence
\[
 \|Z_{\lambda}(t,s)\|\leq \e^{|{\rm Im}\lambda|\|r\|_{1}}  \:  \hbox{ for }  \: |{\rm
Im}\lambda|\leq \|r\|^{-1}_{\infty}.
\]
We apply Cauchy's formula on a circle of radius
$\|r\|_{\infty}^{-1}$ and obtain
\[
\left\|\frac{\d^{n}}{\d \lambda^{n}}Z_{\lambda}(t,s) \right\|\leq n!  \: 
\|r\|^{n}_{\infty}  \: \e^{\|r\|_{1}\|r\|^{-1}_{\infty}}  \:  \hbox{ for }  \: \lambda\in \rr .
\]
This completes the proof of the lemma \qed . 

\medskip

Finally we prove a lemma which  is used in the main text to prove  spatial clustering.

\medskip
\begin{remark}
\label{translationremark}
Let $t_{0}\in \rr$ and define the time-translated family by
$\xi_{t_{0}}\bigl(R(t)\bigr):=R(t-t_{0})$. Then clearly \[ U\bigl(t, s;
\xi_{t_{0}}(R)\bigr)=U(t-t_{0}, s-t_{0}; R) \hbox{ for } -\infty<s\leq t<+\infty . \]
Letting $(s,t)\to (-\infty, +\infty)$ we obtain that 
$U\bigl(+\infty, -\infty;\xi_{t_{0}}(R)\bigr)=U(+\infty, -\infty; R)$.
\end{remark}
\begin{lemma}
\label{spaceclustering}
Assume that $0$ is a simple eigenvalue of $H$ and that $H$ has a
spectral gap, i.e., \[ ]0, a]\cap \sigma(H)=\emptyset \hbox{ for some } a>0 . \]
Let $\{R_{1}(t) \}$, $\{ R_{2}(t) \}$ be two selfadjoint families of operators
satisfying (\ref{e4.3}) and (\ref{e4.12a}) with~$R_{i}(t)\equiv 0$ for
$|t|\geq T$. If $\Omega$ is a normalized ground state of~$H$, then 
\[
 \bigl| \bigl(\Omega, U^\infty (R_{1}+ \xi_{t}(R_{2}))\Omega \bigr)- \bigl(\Omega,
U^\infty (R_{1})\Omega \bigr) \bigl(\Omega, U^\infty (
R_{2})\Omega \bigr) \bigr|\leq \e^{- (|t|-2T)a}
\]
for $|t|>2T$, where $U^\infty (R):= U(+\infty, -\infty;R)$.
\end{lemma}
\proof
It suffices to consider the case $t>0$.
Using the group property and  considering the supports of
$R_{i}(.)$, we find
\beq
\label{toto1}
\begin{array}{rl}
U \bigl(t,s;R_{1}+ \xi_{t_{0}}(R_{2}) \bigr)=&U \bigl(t, t_{0}-T;
\xi_{t_{0}}(R_{2})\bigr)\e^{-(t_{0}-2T)H}U(T, s;R_{1})\\[2mm]
=&U (t-t_{0}, -T;
R_{2} )\e^{-(t_{0}-2T)H}U(T, s;R_{1}) 
\end{array}
\eeq
for  $s\leq -T$, $t_{0}>2T$ and $t>t_{0}+ T$.
Since $H$ has a spectral gap of length $a$, 
\beq
\label{toto2}
\bigl\|\e^{- (t_{0}-2T)H}-|\Omega\rangle\langle\Omega| \: \bigr\|\leq \e^{-(t_{0}-
2T)a}.
\eeq
Moreover, since $H\Omega=0$ and $\supp R_{i}(.)\subset [-T, T]$, 
\beq
\begin{array}{l}
\bigl(\Omega, U(t-t_{0}, -T; R_{2})\Omega\bigr)=\bigl(\Omega, U(t-t_{0}, s;
R_{2})\Omega \bigr),\\[2mm]
\bigl(\Omega, U(T, s; R_{2})\Omega \bigr)= \bigl(\Omega, U(t, s;R_{2})\Omega \bigr). 
\end{array}
\label{toto3}
\eeq
Combining (\ref{toto1}), (\ref{toto2}), (\ref{toto3}) and letting $(t,
s)\to (+\infty, -\infty)$ we obtain the lemma \qed .

\section{Miscellaneous results}
\label{appB}
\init
\begin{lemma}
\label{a1}
Let $F \colon \rr^{2} \to E$ be a map with value in
a metric space $E$. 
\vskip .3cm\halign{ \indent \indent \indent #  \hfil & \vtop { 
\parindent =0pt 
\hsize=12cm                            \strut # \strut} \cr 
{\rm (i)}  & Assume that
\[
\begin{array}{l}
\lim_{k,k'\to \infty}F(k, k')= F_{\infty}\hbox{ exists},\\[2mm]
\lim_{k'\to \infty}F(k, k')= G(k)\hbox{ exists }\forall k\in \nn,\\[2mm]
\lim_{k\to \infty}G(k)=G_{\infty}\hbox{ exists}.
\end{array}
\]
Then $F_{\infty}= G_{\infty}$.
\cr {\rm (ii)}  & Assume that
\[
\begin{array}{l}
\lim_{k'\to \infty}F(k, k')= G(k)\hbox{ exists},\\[2mm]
\lim_{k\to \infty}F(k, k')= F(k')\hbox{ exists and the convergence is uniform w.r.t. }k',\\[2mm]
\lim_{k\to \infty}G(k)= G_{\infty}\hbox{ exists}.
\end{array}
\]
Then $\lim_{k'\to \infty}F(k')=G_{\infty}$.
\cr}
\end{lemma}
The proof is easy and left to the reader.

\begin{lemma}
\label{5.1}
Let $(Q, \Sigma, \mu)$ be a probability space. Let $f$ be a real measurable
function on $Q$ and set 
\[
 C(t) :=\int_{Q}\e^{\i t f}\d \mu.
\]
Then $f\in \bigcap_{1\leq p<\infty}L^{p}(Q, \Sigma, \mu)$ if and only if
$\sup_{t\in \rr}|\p_{t}^{n}C(t)|<\infty$ for all $n\in \nn$. If this is the case, then
\[
\p_{t}^{n}C(t)=\i^{n}\int_{Q}f^{n}\e^{\i t f}\d \mu.
\]
\end{lemma}
\proof
The $\Rightarrow$ part and the formula for $\p_{t}^{n}C(t)$ 
is obvious by differentiating under the integral
sign. It remains to prove the $\Leftarrow$ part. Let $\chi(\tau)=
\e^{-\tau^{2}/2}$ and let $p\geq 1$. By monotone convergence it suffices to prove that
\[
\sup_{n\in \nn}\int_{Q}f^{2p} \chi \left(\frac{f}{n} \right)\d\mu<\infty 
\]
in order to show that $f\in L^{2p}(Q, \Sigma, \mu)$. We have 
\[
\tau^{2p}\chi \left(\frac{\tau}{n} \right)
= \frac{n^{2p+1}}{2\pi}\int\e^{\i
t\tau} \bigl( \p^{2p}_{t}\hat{\chi}(nt) \bigr) \d t.
\] 
Hence
\[
\begin{array}{rl}
\int_{Q}f^{2p}
\chi\left(\frac{f}{n} \right)\d\mu&=\frac{n^{2p+1}}{2\pi}\int_{Q}\int_{\rr}\e^{\i
tf} \bigl( \p^{2p}_{t}\hat{\chi}(nt) \bigr) 
\d t\d\mu\\[3mm]
&=\frac{n^{2p+1}}{2\pi}\int_{\rr}C(t) \bigl(\p^{2p}_{t}\hat{\chi}(nt) \bigr)\d
t\\[3mm]
&=\frac{(-1)^{2p}}{2\pi} n \int_{\rr} \bigl( \p_{t}^{2p}C(t) \bigr)  \hat{\chi}(nt)\d t,
\end{array}
\]
using  Fubini's theorem and integrating by parts $2p$ times. Since
$\hat{\chi}\in L^{1}(\rr)$ and $\p_{t}^{2p}C$ is uniformly bounded, we
obtain that $\sup_{n\in \nn}\int_{Q}f^{2p}\chi(n^{-1}f)\d\mu<\infty$,
which completes the proof of the lemma \qed . 

\medskip
\begin{lemma}
\label{T4a}
Let $I$ be a directed set and 
$\{u_{\alpha}\}_{\alpha\in I}$ a net of functions
which are holomorphic in an open set $\Omega\subset \cc$. 
\vskip .3cm
\halign{ \indent \indent \indent #  \hfil & \vtop { 
\parindent =0pt 
\hsize=12cm
                            \strut # \strut} \cr 
{\rm (i)}  & Assume that the
family $\{u_{\alpha}\}$ is locally uniformly bounded in $\Omega$ and
that there exists a set $\Gamma\subset \Omega$ having an accumulation
point in $\Omega$ such that 
\[
\lim_{\alpha\in I}u_{\alpha}(z)\hbox{ exists for }z\in \Gamma.
\]
Then $\lim_{\alpha\in I}u_{\alpha}=u$ exists in  the compact-open topology
on $\Omega$ and $u$ is a holomorphic function in $\Omega$.
\cr 
{\rm (ii)}  & Assume moreover that $\Omega$ is bounded with a smooth boundary and that 
\[\sup_{\alpha\in I}\sup_{z\in \Omega}|u_{\alpha}(z)|<\infty. \] 
Then $u$ is continuous
on $\overline{\Omega}$ and
$\lim_{\alpha\in I}\sup_{z\in \p\Omega}|u_{\alpha}(z)- u(z)|=0$.
\cr}
\end{lemma}
\proof 
Let us first prove (i). 
By Vitali's theorem the family $\{u_{\alpha}\}$ is compact for the
compact-open topology. Let $\{u_{\beta}\}_{\beta \in J}$ be a subnet
converging to  a continuous function $u$. Assume that the net
$\{u_{\alpha}\}_{\alpha\in I}$ does not converge to $u$. Then there
exists a bounded open set $\Omega_{1}\subset\Omega$ and a subnet
$\{u_{\gamma}\}_{\gamma \in J_{1}}$ such that $\sup_{z\in
\Omega_{1}}|u_{\gamma}(z)- u(z)|\geq \epsilon_{0}>0$ for $\gamma \in
J_{1}$. Applying again Vitali's theorem to the net
$\{u_{\gamma}\}_{\gamma \in J_{1}}$, we obtain another subnet
$\{u_{\delta}\}_{\delta \in J_{2}}$ such that $\lim_{\delta \in
J_{2}}u_{\delta}= v$, with $v\neq u$. But $u$ and $v$ are holomorphic
in $\Omega$, as limits of holomorphic functions for the compact-open 
topology and
coincide on $\Gamma$ by hypothesis. Since $\Gamma$ has an accumulation
point in $\Omega$, we have $u=v$ which gives a contradiction. 

Let us now prove (ii). Assume the contrary and let $\{u_{\beta}\}_{\beta\in J}$ be a
subnet such that
\[
\inf_{\beta \in J}\sup_{z\in \p \Omega}|u_{\beta}(z)-u(z)|\geq \epsilon>0.
\]
Since $\Delta u=0$ in $\Omega$, we see that $u$ belongs
to the Sobolev space $H^{2}(\Omega)$. Using that $\Delta u_{\beta}=0$
in~$\Omega$ and the
fact that the family $\{u_{\beta}\}_{\beta \in J}$ is uniformly bounded in
$\Omega$, we obtain similarly that $\{u_{\beta}\}_{\beta \in J}$ is a bounded
family in $H^{2}(\Omega)$. Hence (i) implies $\lim_{\beta \in
J}u_{\beta}=u$ in $\cD'(\Omega)$. Finally we note that the injection $H^{2}(\Omega)\to H^{3/2}(\Omega)$
is compact. Extracting again a subnet, we obtain $\lim_{\gamma \in J_1}u_{\gamma}= u$ in $H^{3/2}(\Omega)$.
Together with the trace theorem this implies that~$\lim_{\gamma \in J_1}u_{\gamma}=u$ in~$H^{1}(\p \Omega)$ and hence in $C(\p \Omega)$. This gives a
contradiction \qed .

\end{document}